\documentclass[12pt]{article}
\pdfoutput=1
\usepackage[nosort]{cite}
\usepackage{amsmath,amssymb}

\usepackage{color}

\usepackage{cite}
\usepackage{hyperref}
\usepackage{tikz}
\usepackage{slashed}
\usepackage{subcaption}

\usepackage{draft}

\def\({\left(}
\def\){\right)}

\begin{document}

\begin{titlepage}

\begin{center}

\hfill \\
\hfill \\
\vskip 1cm

\title{FCC, Checkerboards, Fractons, and QFT}

\author{Pranay Gorantla$^{1}$, Ho Tat Lam$^{1}$, Nathan Seiberg$^{2}$, and Shu-Heng Shao$^{2}$}

\address{${}^{1}$Physics Department, Princeton University, Princeton NJ, USA}
\address{${}^{2}$School of Natural Sciences, Institute for Advanced Study, Princeton NJ, USA}

\end{center}

\vspace{2.0cm}

\begin{abstract}

\noindent
We consider XY-spin degrees of freedom on an FCC lattice,  such that the system respects some subsystem global symmetry.  We then gauge this global symmetry and study the corresponding $U(1)$ gauge theory on the FCC lattice.  Surprisingly, this $U(1)$ gauge theory is dual to the original spin system.  We also analyze a similar $\bZ_N$ gauge theory on that lattice. All these systems are fractonic.  The $U(1)$ theories are gapless and the $\bZ_N$ theories are gapped. We  analyze the continuum limits of all these systems and present free continuum Lagrangians for their low-energy physics.

Our $\bZ_2$ FCC gauge theory is the continuum limit of the well known checkerboard model of fractons.  Our continuum analysis leads to a straightforward proof of the known fact that this theory is dual to two copies of the $\bZ_2$ X-cube model.

We find new models and new relations between known models.  The $\bZ_N$ FCC gauge theory can be realized  by coupling three copies of an anisotropic model of lineons and planons to a certain exotic $\bZ_2$ gauge theory.    Also, although for $N=2$ this model is dual to two copies of the $\bZ_2$  X-cube model, a similar statement is not true for higher $N$.

\end{abstract}

\vfill

\end{titlepage}

\eject

\tableofcontents

\section{Introduction and summary}\label{sec:intro}

The pioneering work of \cite{PhysRevLett.94.040402,PhysRevA.83.042330} has stimulated a lot of work on fracton phases of matter. There is an extensive literature on this subject and it is reviewed nicely in \cite{Nandkishore:2018sel,Pretko:2020cko}.  These reviews include also many references to the original papers.    These phases are exciting for several different reasons.  In particular, their low-energy dynamics cannot be captured by a standard continuum quantum field theory.

Here, we continue the exploration in \cite{Seiberg:2019vrp, paper1, paper2, paper3, Gorantla:2020xap}, trying to describe the long-distance behavior of these models in the context of continuum field theory.
The goal of this paper is to find such a description for some of the models in \cite{Vijay:2016phm}.
More specifically, we will be interested in a fracton model known as the checkerboard model, as well as its related models.
See \cite{Shirley:2019wdf,Shirley:2019xnp,Shirley:2018vtc} for more  discussions on these models. Along the way, we will find new models and new dualities.  These dualities relate different models some of them are known and others are new.

\subsection{General strategy}

We will follow the general strategy of \cite{Seiberg:2019vrp, paper1, paper2, paper3, Gorantla:2020xap} for analyzing such models.  (See these references for more detail.)  The analysis progresses along the following steps.
\begin{itemize}
\item We start with a lattice model with XY-spins on its sites and a particular Hamiltonian.  We study it in a gapless phase and we look for a description of its continuum limit using periodic fields $\phi\sim \phi+2\pi$.  We refer to that theory as the $\phi$-theory.

    The global symmetries of the system include symmetries that are present in the lattice model, which we refer to as ``momentum symmetries'' and some emergent, accidental global symmetries, which we refer to as ``winding symmetries''.\footnote{This terminology is the standard terminology used in the string theory literature of the compact boson theory.  The momentum and winding here are momentum and winding in the target space and should not be confused with momentum in the coordinate space.}  We analyze the spectrum of excitations of the system and in particular, the states charged under the momentum and winding symmetries.
\item We gauge the momentum symmetry of the $\phi$-theory both on the lattice and in the continuum.  This leads us to a pure gauge theory with gauge fields $A$.  We refer to it as the $A$-theory.  Both the lattice system and its continuum version have a certain electric global symmetry.  In addition, the low-energy theory has an emergent magnetic global symmetry.  We analyze the symmetries and the spectrum of excitations charged under them.
\item We look for dual descriptions of the $\phi$-theory and the $A$-theory.
\item We use the $\phi$-fields to Higgs the $A$-theory to a $\mathbb{Z}_N$ version of the $A$-theory, find its global symmetry, the ground state degeneracy, the observables, etc..  We also look for dual descriptions of these theories.
\end{itemize}

It is important to emphasize that this three-step process is completely determined by the first step.  The $\phi$-theory and its global symmetry uniquely specify the other theories.  Although this is not a complete classification of all possible theories, it does organize, at least some of them, in a sensible way.

This sequence of steps is completely standard in the analysis of quantum field theories.  However, in the context of these exotic theories, there are some new subtle elements.  These systems have very large global symmetries -- subsystem symmetries.  These are global symmetries that act separately on various subspaces, e.g., planes.  Such symmetries had figured earlier in \cite{PhysRevB.70.165310,PhysRevB.66.054526} and many subsequent papers.  As we said above, these can be present either in the underlying lattice system, and therefore also in the continuum theory, or they can arise as accidental symmetries in the low-energy theory.

These subsystem symmetries have many consequence.  One of them is that the standard derivative expansion of the low-energy Lagrangian might not be valid \cite{paper1,paper2}.  Still, for most purposes this expansion is valid, and below we will limit ourselves to the leading order terms in that expansion.

Another consequence of the subsystem symmetry is the need to study discontinuous field configurations in the low-energy field theory.  Here we face a hierarchy of situations:
\begin{itemize}
\item Lattice configurations are totally discontinuous.  They have infinite action in the continuum limit and the typical such configuration does not carry any global symmetry charge.  Such configurations should definitely be excluded.
\item Some discontinuous field configurations have infinite action in the continuum limit, but they are the lowest energy states carrying some global symmetry charge.  A conservative approach discards these configurations, because of their infinite action.  Yet, we find it interesting to study them.
\item Some discontinuous field configurations have finite action, but they lead to infinite energy states in the quantum theory.  Such states are the lowest energy states carrying a conserved charge.  Again, a conservative approach discards these states, because strictly, they are not part of the low-energy theory.  However, we still study them.
\item Some discontinuous field configurations have finite action and lead to finite energy states in the continuum limit.  Here there is no option.  We must consider these configurations.  They are the source of the large ground state degeneracy.
\end{itemize}

\subsection{Some $(3+1)$d anisotropic models}\label{anisI}

Here we demonstrate the general strategy above in a sequence of models, which are very similar to those in
\cite{paper1}.  These models will end up being closely related to the models we are most interested in.  The authors of  \cite{Bulmash:2018knk,Shirley:2019wdf,Gromov:2018nbv,2019arXiv190411530Y,You:2019bvu}  discussed  similar theories with different number of spatial derivatives in the Lagrangian or in the gauge transformation laws.

In \cite{paper1}, space was parameterize by $(x,y)$ with a $\bZ_4$ rotation symmetry.  Here we ``lift'' these models to $(3+1)$d by adding another direction $z$ without enlarging the $\bZ_4$ rotation symmetry to the cubic group $S_4$.  Unlike the dependence on the $(x,y)$ coordinates, the dependence on $z$ will be standard.

Following our general strategy, we start with the $\phi$-theory.
The $2+1$d $\phi$-theory is lifted to
\ie\label{eq:aniso_phi_periodI}
{\cal L} = {\bar\mu_0^{}\over 2} (\partial_0\phi)^2
- {\bar\mu_1^{} \over 2}  (\partial_x \partial_y \phi)^2-{\bar \mu^{}\over 2} (\partial_z\phi )^2\,,
\fe
(The bars on the coupling constants are for later convenience.)
The first and second terms are as in \cite{paper1}, and the third term, which is absent in \cite{paper1}, is standard in field theory. This model has a subsystem symmetry
\ie\label{subIphi}
\phi \to \phi + f_x(x)+f_y(y)~,
\fe
with two arbitrary functions $f_x(x)$ and $f_y(y)$.

We take the field $\phi$ to be $2\pi$-periodic, i.e., $\phi \sim \phi +2\pi n$ with integer $n$.  In a standard field theory, such an integer is position independent.  But in our case, it can depend on the spatial coordinates.  This dependence should be such that the Lagrangian \eqref{eq:aniso_phi_periodI} is invariant under the identification.  In other words, the identification should be a subgroup of the global symmetry  \eqref{subIphi}.  This constrains the dependence to satisfy
\ie\label{anisoperiodicityI}
\phi(t,x,y,z)\sim\phi(t,x,y,z)+2\pi n^x(x)  +2\pi n^y(y)~,
\fe
where $n^x(x), n^y(y)\in \bZ$ are integer-valued functions.  This identification has the effect of turning the subsystem symmetry \eqref{subIphi} from $\mathbb{R}$ to $U(1)$.

The more detailed analysis of this system is very similar to that in \cite{paper1}.

Continuing along the lines of our general strategy, we consider an $A$-theory, in which the global symmetry \eqref{subIphi} is promoted to a gauge symmetry.  The gauge fields are $(A_0, A ,\mathbf{A}')$ and their gauge transformations are
\ie\label{anisoAI}
&A_ 0 \sim A_0 +\partial_0 \alpha\,,\\
&A \sim A +\partial_z \alpha\,,\\
&\mathbf{A}' \sim \mathbf{A}' +\partial_x \partial_y \alpha\,,
\fe
where the gauge parameter $\alpha$ is subject to the same identifications as $\phi$:
\ie\label{anisoalphaperiodicityI}
\alpha(t,x,y,z)\sim\alpha(t,x,y,z)+2\pi n^x(x)  + 2\pi n^y(y)~.
\fe
Here  the prime in $\mathbf{A}'$ means that the field is in the spin-2 representation of the $\bZ_4$ rotation symmetry on the $yz$-plane.
All dimension two gauge fields and dimension three field strengths are in bold face. The prime means that the field is in the spin-2 representation of the spatial $\bZ_4$ symmetry.

The gauge fields $A_0$ and $\mathbf{A}'$ are similar to $A_0$ and $A_{xy}$ of \cite{paper1}.  The gauge field $A$ is a standard gauge field, usually denoted by $A_z$.

The gauge invariant field strengths are
\ie\label{aniEBI}
&E = \partial_0 A - \partial_z A_0\, ,
\\
&\mathbf{E}' = \partial_0 \mathbf{A}' - \partial_x \partial_y A_0\, ,
\\
&\mathbf{B}' = \partial_z \mathbf{A}' - \partial_x \partial_y A\, .
\fe
Here, $E$ is a standard electric field in the $z$ direction (usually denoted $E_z$ or $F_{0z}$), $\mathbf{E}'$ is similar to $E_{xy}$ of \cite{paper1}, and $\mathbf B'$ is a new gauge invariant combination. The Lagrangian is
\ie
{\cal L} &=  {1\over (\bar g_{e1}^{})^2}    E ^2  + {1\over (\bar g_{e2}^{})^2 }   (\mathbf{E}')^2 - {1\over (\bar g_m^{})^2}(\mathbf{B}')^2\,.
\fe
(The bars on the coupling constants are for later convenience.)

It is straightforward to dualize this system and to find that it is dual to the $\phi$-theory of \eqref{eq:aniso_phi_periodI} with parameters\footnote{The details of this duality are similar to those of Section \ref{sec:duality}, however, as explained there, the parameters there differ by a factor of $4$.}
\ie
\bar \mu_0 = {(\bar g_m)^2 \over 8\pi^2}~,\qquad \bar\mu = {( \bar g_{e2})^2 \over 8\pi^2}~,\qquad \bar\mu_1 = { (\bar g_{e1})^2 \over 8\pi^2}~.
\fe

Following our general strategy, we  now Higgs the $A$-theory by a charge $N$ $\phi$-theory to a $\bZ_N$ gauge theory.
We can dualize the Higgs Lagrangian to obtain a ``BF-type Lagrangian''
\ie\label{anisoZNI}
\mathcal L =
\frac{N}{2\pi} \left[ \mathbf A' ( \partial_0 \tilde A' - \partial_z \tilde A_0' )
+A ( \partial_0 \tilde {\mathbf A} - \partial_x \partial_y \tilde A_0' )
- A_0 ( \partial_z \tilde{\mathbf  A}- \partial_x \partial_y \tilde A' ) \right]~.
\fe
Recall that the prime means that the field is in the spin-2 representation of the spatial $\bZ_4$ rotation on the $xy$-plane.
Here the gauge fields $(\tilde A_0  ,  \tilde A' , \tilde {\mathbf A})$ are dual to $\phi$ and have the following gauge transformation laws
\ie
&\tilde A'_ 0 \sim\tilde  A'_0 +\partial_0\tilde\alpha'\,,\\
&\tilde A' \sim\tilde A' +\partial_z\tilde \alpha'\,,\\
&\tilde {\mathbf A} \sim \tilde {\mathbf A} +\partial_x \partial_y \tilde\alpha'\,,
\fe
where $\tilde \alpha'$ has the same periodicity as $\alpha$.
This is a gapped fracton model, whose lattice formulation and  continuum field theory have been studied in \cite{Shirley:2019wdf}.

As we focus on the low-energy theory, our systems do not have particle-like excitations.  Instead, such massive charged particles are described in the low-energy theory using defects.  Let us demonstrate them in this $\bZ_N$ theory.

This theory has lineons that can move only in $z$ direction.  They are described in the low energy theory by the defect
\ie\label{ZNaniso-lineon}
\exp \left[ i \oint_\mathcal{C} (dt~ A_0 + dz~ A) \right]~,
\fe
where $\mathcal C$ is a curve in the spacetime $(t,z)$ plane.  There is also a similar defect for the $\tilde A$ gauge fields. A pair of lineons separated in the $x$ direction can move in the $yz$-plane.  Hence, they form a planon, which is described by the defect
\ie\label{ZNaniso-planon}
\exp \left[ i \int_{x_1}^{x_2} dx\oint_\mathcal{C} (dt~ \partial_x A_0 + dz~ \partial_x A + dy~ \mathbf A')\right]~,
\fe
where $\mathcal C$ is a curve in the $(t,y,z)$ spacetime.  Again, there is a similar defect for $\tilde A$ gauge fields. In addition, similar defects describe planons in the $xz$-plane.

The theory has two $\mathbb{Z}_N$ global symmetries, which can be thought of as  the electric symmetry for $A$ and $\tilde A$ gauge fields. The corresponding symmetry operators are special cases of the defects \eqref{ZNaniso-lineon} and \eqref{ZNaniso-planon}, their variant in the other plane, and their $\tilde A$ counterparts, where the curve $\mathcal C$ is at a fixed time.

\subsection{Outline and summary of continuum models}

Here we outline the organization of the paper.  As we do that, we review the main features of the various continuum theories that we will be discussing.

\bigskip\bigskip\centerline{\it The $\phi$-theory}\bigskip

In Section \ref{sec:XYtetra}, we will study a lattice XY-model, the tetrahedral model, which is  the $U(1)$ analog of the $\mathbb{Z}_2$ tetrahedral model of \cite{Vijay:2016phm}.  In particular, we will derive its continuum limit in its gapless phase.  Then, in Section \ref{sec:tetrphitheory}, we will explore this continuum theory.  We will analyze its global symmetries and its spectrum.

This theory ends up being closely related to three copies of the $\phi$-theory of  \eqref{eq:aniso_phi_periodI} -- one for $x$, one for $y$, and one for $z$.
We denote these field by $\phi^i$ with $i=x,y,z$.
The Lagrangian is
\ie\label{philag}
{\cal L} = {\mu_0\over 2} (\partial_0\phi^x)^2
-{ \mu\over 2}  (\partial_x\phi^x )^2
- {\mu_1 \over 2}  (\partial_y \partial_z \phi^x)^2 +\cdots \,,
\fe
where the ellipses represent similar terms in the other directions.
Note that the Lagrangian does not mix the three fields $\phi^i$.
There are no bars on the coupling constants because the fields $\phi^i$ have different periodicities than those in \eqref{eq:aniso_phi_periodI} (see below).

We suppress here higher derivative terms that for some purposes are as important as those that are included. See \cite{paper1,paper2} for a discussion of such terms, why they are important, and in what sense they can be neglected.

As in \eqref{subIphi} the system has a continuous global subsystem symmetry, which we refer to as momentum symmetry
\ie\label{phiglobals}
&\phi^x(x,y,z,t)\to \phi^x(x,y,z,t) + f^x_y(y) + f^x_z(z)~,\\
&\phi^y(x,y,z,t)\to \phi^y(x,y,z,t) + f^y_x(x) + f^y_z(z)~,\\
&\phi^z(x,y,z,t)\to \phi^z(x,y,z,t) + f^z_x(x) + f^z_y(y)~,
\fe
with arbitrary functions $f_i^j(x^i)$.  In addition, the continuum theory also has similar winding symmetries, which we will analyze in detail below.

Locally, the fields $\phi^i$ take values in $\mathbb{R}^3$.  The analysis of the underlying lattice model tells us that we should impose an identification on that space by a Face-Centered Cubic (FCC) lattice\footnote{Note that the underlying XY lattice model was on an FCC lattice.  This should not be confused with this lattice of identifications.}
\ie\label{FCCident}
\phi^i \sim \phi^i +2\pi n^i \qquad {\rm with}  \qquad   n^i \in\bZ~, \qquad n^x+n^y+n^z \in 2\bZ\,.
\fe

The three fields $\phi^i$ are coupled only through the condition on the sum of the integers.  This condition is the only difference between three decoupled copies of the anisotropic $\phi$-theory of \eqref{eq:aniso_phi_periodI} and this theory.  The relation between them will be discussed in more detail in Section \ref{sec:anisotropic}.
There we will show that we can convert one model to the other by coupling it to a $\bZ_2$ gauge theory.

As in the discussion around \eqref{anisoperiodicityI}, the integers in \eqref{FCCident} can be position dependent.  They are constrained to satisfy
\ie\label{ndepcon}
\partial_xn^x=\partial_y\partial_zn^x=0~,
\fe
and similarly for the other directions.  The effect of this identification is to turn the $\mathbb{R}^3$ global symmetry \eqref{phiglobals} into $U(1)^3$.

In addition to the continuous symmetry \eqref{phiglobals}, our system also has some discrete symmetries.  Its space symmetries include the cubic rotation symmetry $S_4$ and parity and time reversal symmetry.  There are also three $\mathbb{Z}_2$ internal symmetries that flip the sign of one of the three fields $\phi^i$.  They are called internal, because they do not act on the coordinates.  The diagonal $\mathbb{Z}_2$ of these three symmetries $\mathbb{Z}_2^C$ is overall charge conjugation.  The other two $\mathbb{Z}_2$ symmetries are internal, but they do not commute with the spatial rotation group $S_4$.  Together they form a group
\ie
G=(\mathbb{Z}_2\times \mathbb{Z}_2)\rtimes S_4~,
\fe
which we analyze in Appendix \ref{app:G}.

One of the unusual facts about this symmetry is that we can assign the three fields $\phi^i$ to different three-dimensional representations of $S_4$
\ie\label{phirepo}
\mathbf{ 3}: ~ &\phi^i\\
\mathbf{ 3}': ~&\phi^{i'}\\
\mathbf{ 1}\oplus \mathbf{ 2}: ~ & \phi^{ii}\\
\mathbf{ 1}'\oplus \mathbf{ 2}: ~ &\phi^{ii'}
\fe
where we used our notation for these $S_4$ representations.  This will be discussed in detail in Section \ref{sec:phidiscrete}.  The notation in \eqref{philag} and the following discussion used the first of these options, but the other options are also possible, and they turn out to be useful.  It is important to stress, though, that these are not different theories.  The options \eqref{phirepo} are merely different ways of describing the same global symmetry of the same system.

\bigskip\bigskip\centerline{\it The $A$-theory}\bigskip

In Section \ref{sec:Alattice}, we will study a $U(1)$ gauge theory, which we refer to as an FCC lattice gauge theory.  It is the theory obtained by gauging the global subsystem symmetry  of the lattice XY-model of Section \ref{sec:XYtetra}.  We take its continuum limit in its gapless phase.  The resulting continuum Lagrangian is the same as gauging the subsystem global symmetry \eqref{phiglobals}.  As a result, the gauge theory has  three gauge parameters $\alpha^i$, which are subject to the same conditions as the three fields $\phi^i$ in the $\phi$-theory \eqref{FCCident}\eqref{ndepcon}.

This continuum gauge theory is studied in detail in Section \ref{sec:tetrgaugetheory}.  Here we review some of its main features.  The gauge fields are as follows.  The temporal gauge fields $A_0^i$ are in $\mathbf{3}$ of $S_4$ with $i=x,y,z$.  The spatial gauge fields are of two kinds $A^{ii}$ in $\mathbf{1} \oplus \mathbf{2} $ of $S_4$ and $\mathbf A^{ii'}$  in $\mathbf{1}' \oplus \mathbf{2} $ of $S_4$.  (The notation is as in \eqref{phirepo}, and it will be discussed in Section \ref{sec:tetrgaugetheory} and in Appendix \ref{usefulgroup}.)

This gauge theory is closely related to three copies of the $A$-theory of \eqref{anisoAI}, one for $x$, one for $y$, and one for $z$.  Just as the difference between three copies of the anisotropic $\phi$-theory \eqref{eq:aniso_phi_periodI} and the $\phi$-theory is in the identifications of $\phi^i$, the same is true for these gauge theories.  The difference between three copies of the anisotropic $A$-theory of \eqref{anisoAI} and this gauge theory is only in the identifications of the gauge parameters $\alpha^i$.  This will be discussed in more detail in Section \ref{sec:anisotropic}. There we will show that we can convert one model to the other by coupling it to a $\bZ_2$ gauge theory.

For simplicity, let us focus on the gauge fields with $i=x$.  Their gauge transformations are
\ie\label{gaugealpha}
\mathbf{3}: ~ &A_0^x \sim A_0^x + \partial_0 \alpha^x\, ,
\\
\mathbf{1} \oplus \mathbf{2}: ~&A^{xx} \sim A^{xx} + \partial_x \alpha^x\, ,
\\
\mathbf{1}' \oplus \mathbf{2}: ~&\mathbf A^{xx'} \sim \mathbf A^{xx'} + \partial_y \partial_z \alpha^x\, .
\fe
Unlike the discussion of the anisotropic model \eqref{eq:aniso_phi_periodI}, which was not $S_4$ invariant, here we keep track of the $S_4$ representation. The first column in \eqref{gaugealpha} denotes the $S_4$ representation and it reminds us that there are similar gauge fields and similar gauge transformations in the two other directions.
The gauge parameters $\alpha^i$ have the same periodicities as $\phi^i$ in \eqref{FCCident}.

The gauge invariant electric and magnetic fields are
\ie
\mathbf{1} \oplus\mathbf{2}: ~&E^{xx} = \partial_0 A^{xx} - \partial_x A^x_0\, ,
\\
\mathbf{1}' \oplus\mathbf{2}:  ~&\mathbf E^{xx'} = \partial_0 \mathbf A^{xx'} - \partial_y \partial_z A^x_0\, ,
\\
\mathbf{3}': ~&\mathbf B^{x'} = \partial_x \mathbf A^{xx'} - \partial_y \partial_z A^{xx}\, .
\fe
Again, we wrote the $S_4$ representation, but we gave the explicit expression for the field strengths only for one component.

In Section \ref{sec:tetrgaugetheory}, we will also discuss the global symmetries of this theory, its spectrum, including the states charged under the global symmetries, and its restricted mobility defects.

Since in the $\phi$-theory we had four options for the $S_4$ representation of $\phi$ \eqref{phirepo} and since the gauge parameter $\alpha$ in \eqref{gaugealpha} behaves like $\phi$, we can also assign it to four different $S_4$ representations.  When we do that, the $S_4$ representations of the gauge fields and the field strengths are also modified.  Again, these are not four different theories, but four different ways of describing the global symmetry of the same system.  These issues will be crucial below.

Just as for the anisotropic models in Section \ref{anisI}, this $\phi$-theory is dual to this $A$-theory.  This duality will be discussed in Section \ref{sec:duality}.

\bigskip\bigskip\centerline{\it The $\mathbb{Z}_N$-theory}\bigskip

Section \ref{ZNgaugetheory} will discuss a $\mathbb{Z}_N$ version of the FCC gauge theory. We will do it both on the lattice and in the continuum.
The lattice model will be identified with a  $\mathbb{Z}_N$ version of the checkerboard model of \cite{Vijay:2016phm} (see also \cite{Shirley:2019wdf,Shirley:2019xnp,Shirley:2018vtc}).

In the continuum, we start with the fields $(A_0^i,A^{ii},\mathbf A^{ii'})$ of the $U(1)$  $A$-theory and couple them to the fields $\phi^i$ of the $\phi$-theory such that they Higgs the gauge group to $\mathbb{Z}_N$.  The Lagrangian can be taken to be
\ie\label{eq:ZNLag_Higgsi}
  \frac 1 {4\pi} \tilde {\mathbf B}^x \left( \partial_0 \phi^x - N A_0^x \right)- \frac 1 {4\pi} \tilde {\mathbf E}^{xx} \left( \partial_x \phi^x - N A^{xx} \right)-\frac 1 {4\pi}  \tilde E^{xx'} \left(\partial_y \partial_z \phi^x - N \mathbf A^{xx'} \right) +\cdots
\fe
where, for simplicity, we wrote explicitly only the terms involving $\phi^x$ and we need to add similar terms for the other fields.   The fields $\tilde E^{ii'}$, $\tilde {\mathbf E}^{ii}$, and $\tilde {\mathbf B}^i$ are Lagrangian multipliers implementing the Higgsing and their arbitrary coefficients were set for later convenience.

Again, the gauge theory \eqref{eq:ZNLag_Higgsi} is closely related to three decoupled copies of the anisotropic $\bZ_N$ gauge theory in Section \ref{anisI}.  The difference between these two theories is only in the identifications of the gauge parameters, and it can be implemented by coupling to $\bZ_2$ gauge theories.  This will be discussed in detail in Section \ref{sec:anisotropic}.

A standard duality transformation maps the Lagrangian \eqref{eq:ZNLag_Higgsi} to a $BF$-type Lagrangian
\ie\label{eq:ZNLag_BF_oldbasisi}
&\frac{N}{4\pi} \sum_{i=x,y,z} \left(  \mathbf A^{ii'} \tilde E^{ii'}+  A^{ii} \tilde {\mathbf E}^{ii}- A_0^i \tilde {\mathbf B}^i \right)
=&\frac{N}{4\pi} \left(  \mathbf A^{xx'} \tilde E^{xx'}+  A^{xx} \tilde {\mathbf E}^{xx}- A_0^x \tilde {\mathbf B}^x \right) +\cdots~\\
&\tilde E^{xx'} = \partial_0 \tilde A^{xx'} - \partial_x \tilde A_0^{x'}~,
\\
&\tilde {\mathbf E}^{xx} = \partial_0 \tilde {\mathbf A}^{xx} - \partial_y\partial_z \tilde A_0^{x'}~,
\\
&\tilde {\mathbf B}^x = \partial_x\tilde A^{xx} - \partial_y \partial_z \tilde A^{xx'}~.
\fe
The ellipses represent similar terms involving the other directions.  We see here that in addition to the gauge fields \eqref{gaugealpha}, we also have another such set of gauge fields
\ie\label{eq:ZN_gauge_trans_tildeAi}
\mathbf{3}': ~&\tilde A_0^{x'} \sim\tilde A_0^{x'}+ \partial_0 \tilde\alpha^{x'}\, ,
\\
\mathbf{1}'\oplus\mathbf{2}: ~&\tilde A^{xx'} \sim \tilde A^{xx'} + \partial_x\tilde  \alpha^{x'}\, ,
\\
\mathbf{1}\oplus\mathbf{2} :  ~&\tilde {\mathbf A}^{xx} \sim \tilde {\mathbf A}^{xx} + \partial_y \partial_z\tilde\alpha^{x'}\, .
\fe
Note that the $S_4$ representations of the gauge parameters $\tilde \alpha$, their gauge fields, and the field strengths differ from those of \eqref{gaugealpha}.  They are related to those in \eqref{gaugealpha} by our regular freedom in the assignment of these representations.  However, it is important that while we have our freedom in changing the $S_4$ assignment of these gauge fields, the two sets of gauge fields are necessarily in different $S_4$ representations.

In Section  \ref{ZNgaugetheory}, we will also analyze the ground state degeneracy and the spectrum of defects of this $\mathbb{Z}_N$ theory.

Section \ref{sec:anisotropic} will explain the  operations that turn the anisotropic models of Section \ref{anisI} into the other models in this paper.  We take three decoupled copies of the models of Section \ref{anisI} along the three different directions.  And we couple the three systems to appropriate gauge fields.

In Section \ref{Xcuberel}, we will relate our $\bZ_N$ gauge theory to the X-cube model \cite{Vijay:2016phm} (see also \cite{PhysRevB.81.184303}).
As shown in \cite{Shirley:2019xnp}, for $N=2$ the checkerboard model is equivalent to  two copies of the X-cube model on the lattice.
Here we demonstrate this equivalence using the continuum field theory descriptions of these models.
We also explain why there is no such equivalence  for our $\bZ_N$ checkerboard model with higher $N$.

In Appendix \ref{usefulgroup}, we review some useful group theory facts.

First, in Appendix \ref{sec:app_reps_S4}, we review the representations of the cubic group $S_4$ and present our notation.

Then, in Appendix \ref{app:fcc}, we review the symmetries of the FCC lattice.  These are essential for understanding the symmetries of our lattice models and how they are realized in the continuum theories. In particular, we discuss the global symmetry group $G=(\mathbb{Z}_2\times \mathbb{Z}_2)\rtimes S_4$ of our continuum models.  Of particular interest are its outer automorphisms, which are crucial for understanding our various dualities.

Finally, in Appendix \ref{sec:X-cube}, we review the continuum field theory for the X-cube model in \cite{Slagle:2017wrc,paper3}.

\section{(3+1)d XY-tetrahedral model}\label{sec:XYtetra}

In this section we study the XY version of the tetrahedral model in \cite{Vijay:2016phm}.

\subsection{The lattice model}

\begin{figure}
\begin{subfigure}{.5\textwidth}
\centering
\includegraphics[scale=0.2]{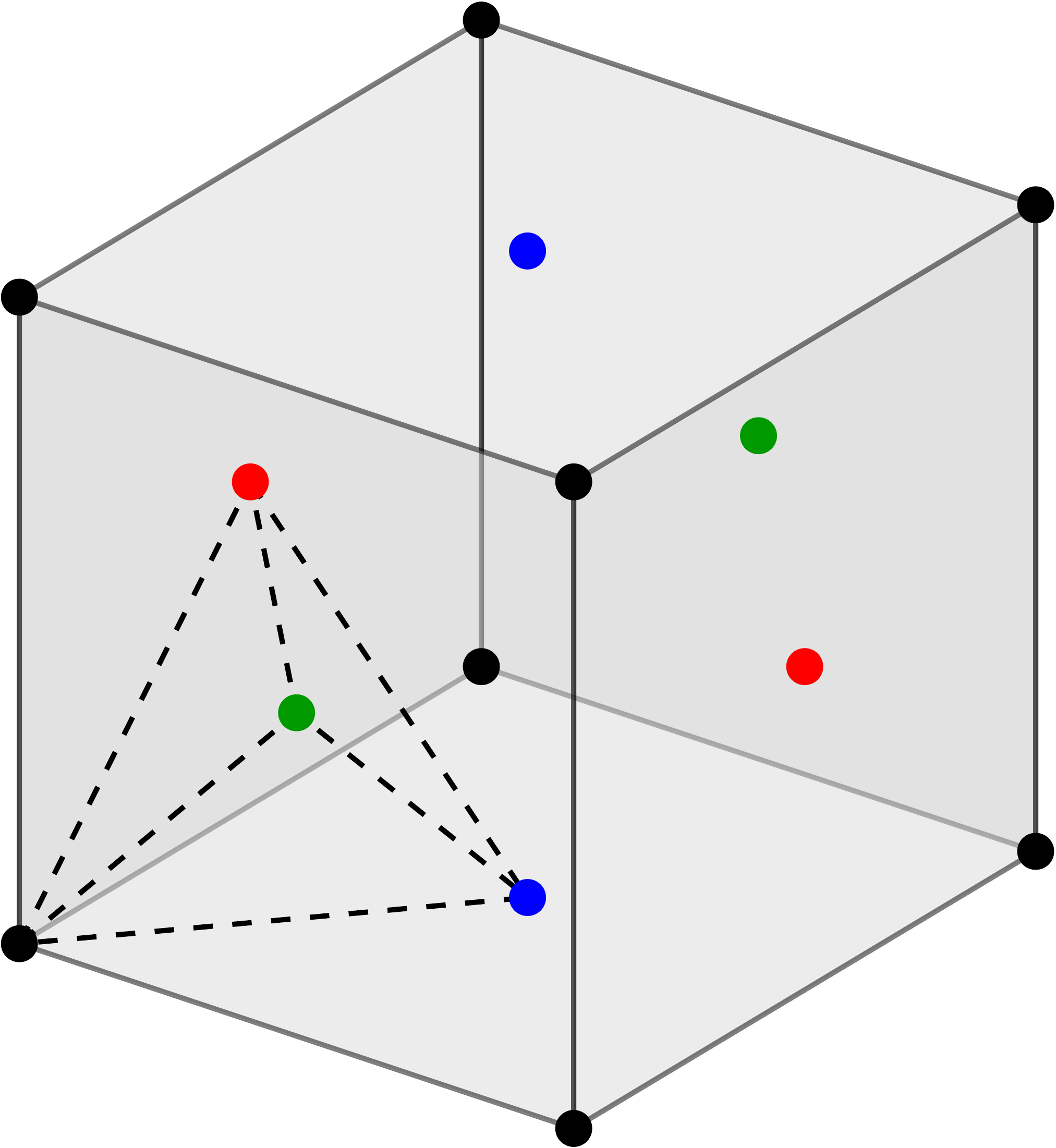}
\caption{}
\end{subfigure}
\begin{subfigure}{.5\textwidth}
\centering
\includegraphics[scale=0.3]{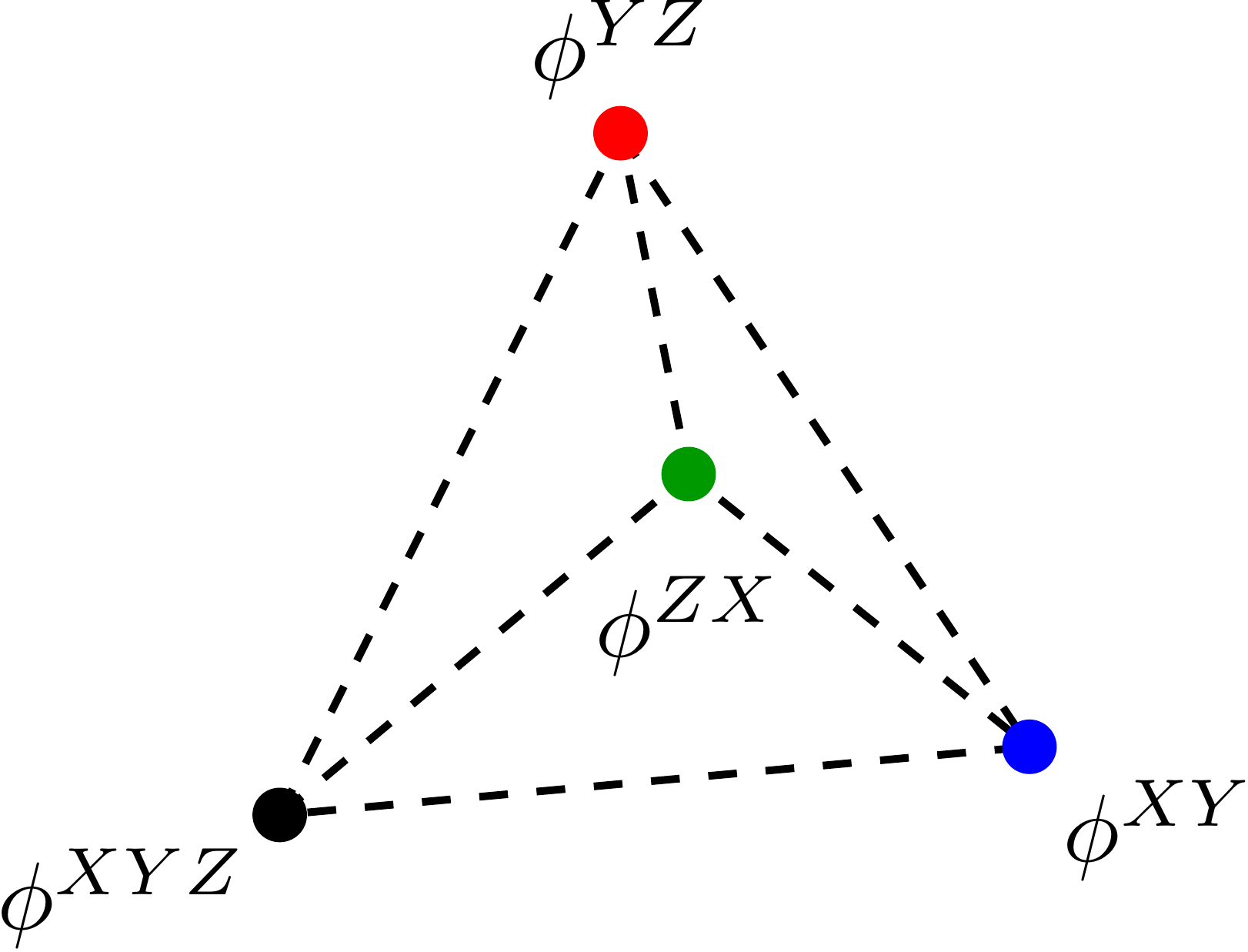}
\caption{}
\end{subfigure}
\caption{(a) The FCC lattice on which the XY tetrahedral model is defined. The black sites are the corners, while the red, green, and blue sites are at the face centers on the $yz$, $zx$, and $xy$ planes respectively. The distance between two nearest neighbor sites is $a/\sqrt{2}$. Each site participates in eight tetrahedra; or equivalently, in each cube, there are eight tetrahedra. (b) The tetrahedron of the interaction term in the second line of \eqref{H}. }\label{fig:XY-tetr}
\end{figure}

We will be working in the Hamiltonian formulation.
The spatial lattice is a three-dimensional face-centered cubic lattice, where the lattice site $s$ can either be at the corner of a cube or at the center of a face (see Figure \ref{fig:XY-tetr}).
We will label the lattice sites  at the corners by integers $(\hat x,\hat y,\hat z)$, while those on the faces by $(\hat x+\frac 12,\hat y+\frac12, \hat z )$, $(\hat x+\frac 12,\hat y, \hat z +\frac12)$, or $(\hat x,\hat y+\frac12, \hat z+\frac12 )$, with $\hat x, \hat y, \hat z\in \bZ$.

\begin{figure}[h!]
	\centering
	\begin{subfigure}{0.3\textwidth}
		\centering
		\includegraphics[width=0.8\linewidth]{XY-tetr-latt.pdf}
		\caption{}
	\end{subfigure}
	\begin{subfigure}{0.3\textwidth}
		\centering
		\includegraphics[width=0.8\linewidth]{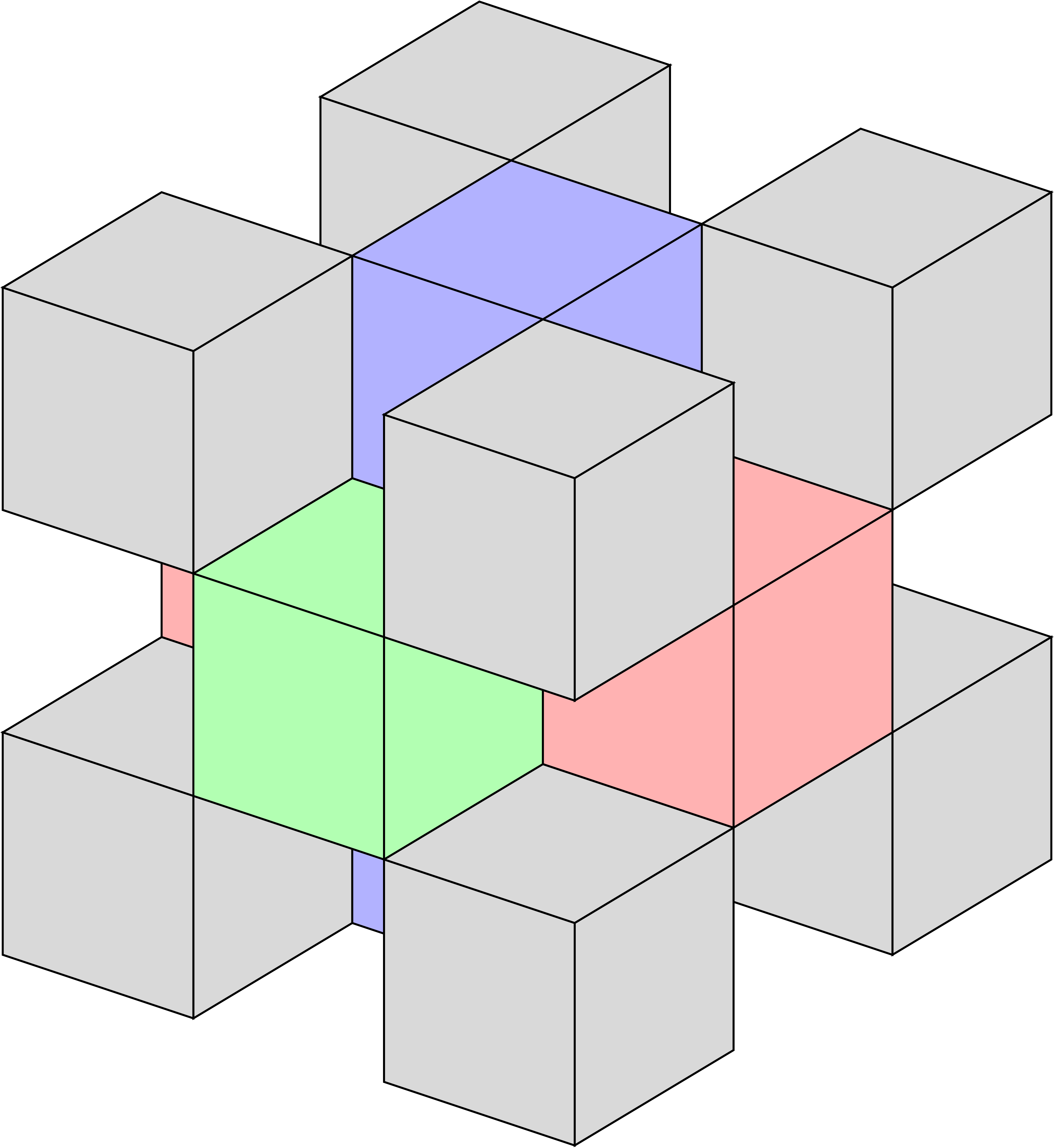}
		\caption{}
	\end{subfigure}
	\begin{subfigure}{0.3\textwidth}
		\centering
		\includegraphics[width=0.8\linewidth]{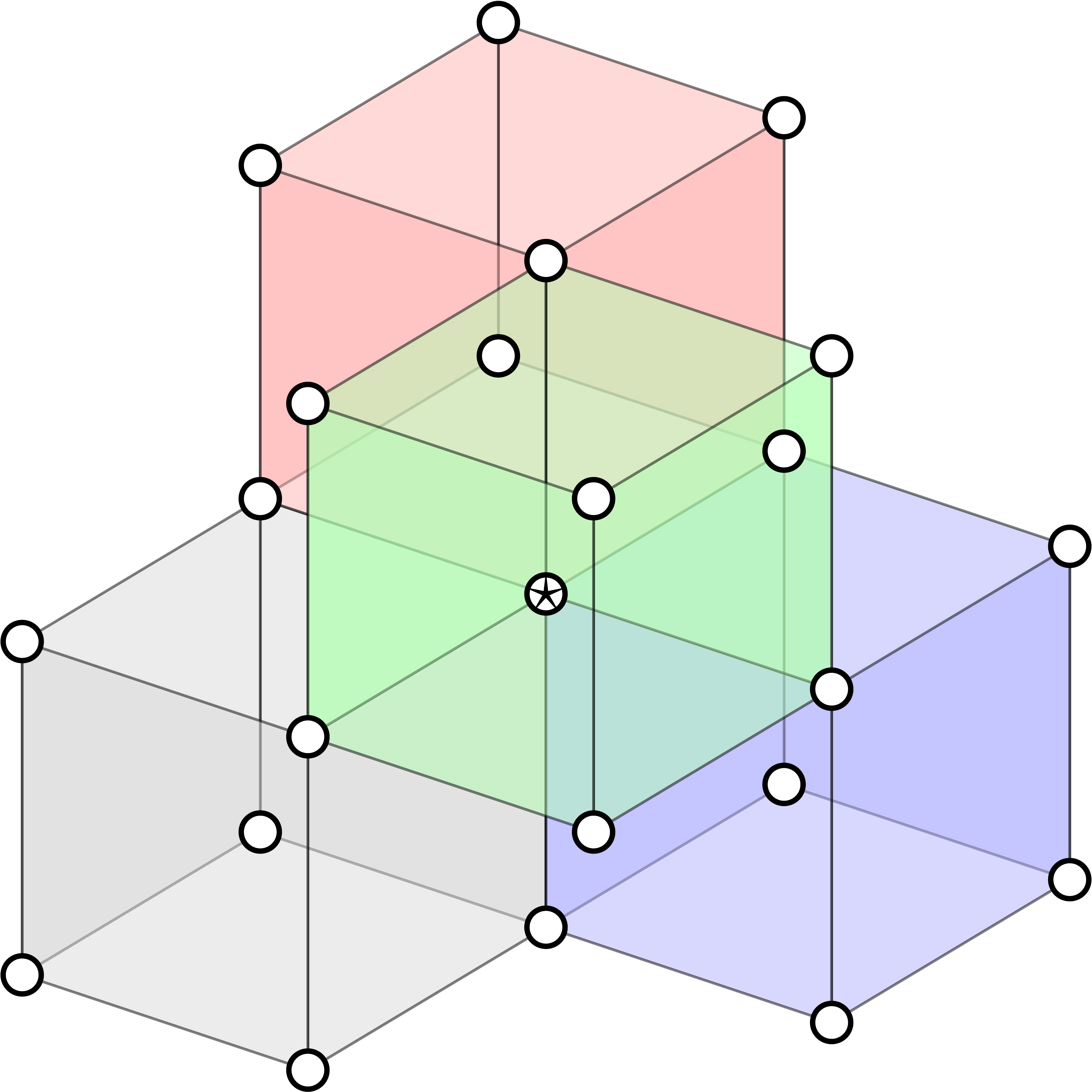}
		\caption{}
	\end{subfigure}
	\caption{The  lattice models in Section \ref{sec:XYtetra}, Section \ref{sec:Alattice}, and Section \ref{sec:ZNlattice} are formulated on an FCC lattice (see Figure (a)).  Equivalently, we can formulate it on the checkerboard  of a cubic lattice (see Figure (b)).  The sites  and the tetrahedra of the FCC lattice are mapped to the shaded cubes  and sites of the checkerboard, respectively.   The distance between two nearest neighbor sites in the FCC lattice is $a/\sqrt{2}$, while it is $a/2$ on the checkerboard.  For the XY-tetrahedral model of Section \ref{sec:XYtetra}, the phase variables and their conjugate momenta are placed on the sites of the FCC lattice, or equivalently the shaded cubes of the checkerboard.  Their interactions are associated with the tetrahedra of the FCC lattice, or equivalently, the four shaded cubes that share the same site in Figure (c) of the checkerboard.   For the $U(1)$ and $\bZ_N$ FCC gauge theories of Section \ref{sec:Alattice} and Section \ref{sec:ZNlattice}, both the gauge parameters and the magnetic interactions are associated with the sites of the FCC lattice, or equivalently, the shaded cubes of the checkerboard.  The gauge fields are placed on the tetrahedra, or equivalently, the sites of the checkerboard.}
	\label{fig:U1-CB}
\end{figure}

For every corner $s$, there is a $U(1)$ phase $e^{i \phi_s^{XYZ}}$ and its conjugate variable $\pi^{XYZ}_s$ at every lattice site, obeying the commutation relation $[\phi^{XYZ}_s, \pi^{XYZ}_s ] = i $ if they belong to the same site and commute otherwise.
For every face center $s$ on the $IJ$-plane,  there is a $U(1)$ phase $e^{i \phi^{IJ}_s}$ and its conjugate variable $\pi^{IJ}_s$ with the same commutation relations.

We emphasize that the indices with capital letters, such as $I,J,K$ and $X,Y,Z$, are neither $SO(3)$ indices nor $S_4$ indices. Throughout the paper, we will use indices with lowercase letters, such as $i,j,k$ and $x,y,z$, to denote $S_4$ indices.

For every site at the corner,  we define eight different kinds of four-spin interactions, each forming a tetrahedron:
\ie
\Delta_{\epsilon_x\epsilon_y\epsilon_z}  \phi \(\hat x,\hat y,\hat z\) = &\phi^{XYZ}\(\hat x,\hat y, \hat z \)+
\phi^{XY}\(\hat x+ { \epsilon_x \over 2} , \hat y+{ \epsilon_y \over2}, \hat z\)
\\
    &+\phi^{ZX} \(\hat x+ {\epsilon_x \over 2},\hat y, \hat z +{\epsilon_z\over 2}\)
        +\phi^{YZ} \(\hat x,\hat y+{\epsilon_y\over2}, \hat z +{\epsilon_z\over2}\)
        \,,
\fe
where $\epsilon_x, \epsilon_y,\epsilon_z =\pm1$.

The Hamiltonian is
\ie\label{H}
 H=& {u\over 2 }\left(
   \sum_{s\in \bZ^3} (  \pi^{XYZ}_s)^2
 +   \sum_{ s\in \( \bZ+\frac12 ,\bZ+\frac12, \bZ\)  }   (\pi_s^{XY})^2
  + \sum_{s\in ( \bZ ,\bZ+\frac12, \bZ+\frac12)}   (\pi_s^{YZ})^2
 +  \sum_{s\in ( \bZ+\frac12 ,\bZ, \bZ+\frac12)}   (\pi_s^{ZX})^2
 \right)\\
    &- K  \sum_{s\in \bZ^3}   \sum_{\epsilon_x,\epsilon_y,\epsilon_z =\pm1}  \cos\left (\Delta_{\epsilon_x\epsilon_y\epsilon_z}  \phi_s\right)~,
\fe
For simplicity, we have taken various lattice coupling constants to be equal.

For every $xy$-plane at a fixed \textit{integer} $\hat z_0\in \bZ$, there is a $U(1)$ subsystem symmetry shifting all the $\phi_s^{XYZ}$ at the corners and all the $\phi^{XY}_s$ at the faces by an opposite amount:
\ie\label{sub1}
&\phi _s^{XY} \to \phi_s^{XY}  + \varphi \,,~~~~~~~\text{if}~~s\in \(\bZ+\frac12, \bZ+{1\over 2} ,\hat z_0\)\,,\\
&\phi _s^{XYZ} \to \phi_s^{XYZ}  - \varphi \,,~~~~~~~\text{if}~~s\in (\bZ, \bZ ,\hat z_0)\,.
\fe
There are similar  subsystem symmetries associated with the $yz$ and $xz$ planes. For every $xy$-plane at a fixed \textit{half intger} $\hat z_0+\frac{1}{2}\in \bZ+1/2$, there is a $U(1)$ subsystem symmetry shifting all the $\phi_s^{YZ}$ and $\phi_s^{ZX}$ by an opposite amount:
\ie\label{sub2}
&\phi _s ^{ZX}\to \phi_s ^{ZX} + \varphi \,,~~~~~~~\text{if}~~s\in \(\bZ+\frac12, \bZ ,\hat z_0+\frac12\)\,,\\
&\phi _s^{YZ} \to \phi_s^{YZ}  - \varphi \,,~~~~~~~\text{if}~~s\in \(\bZ, \bZ +\frac12,\hat z_0+\frac12\)\,.
\fe

On the lattice, we have a  shift symmetry in the $x,y$ direction, which  acts on the fields as
\ie\label{Z2}
&\phi^{XYZ}_s \rightarrow \phi^{XY}_{s+(\frac 12, \frac 12, 0)} \,,\\
&\phi^{XY}_s \rightarrow \phi^{XYZ}_{s + (\frac 12, \frac 12,0)}\, ,\\
& \phi^{YZ}_s \rightarrow \phi^{ZX}_{s+ (\frac 12, \frac 12,0)}\, ,\\
&\phi^{ZX} _s  \rightarrow \phi^{YZ}_{s+ (\frac 12, \frac 12,0)}\, .
\fe
 The symmetry along $yz$ and $xz$ planes is similar.
  The composition of these three shifts is a lattice translation by $(a,a,a)$.
 In the continuum limit,  these three shifts reduce to a $\bZ_2\times \bZ_2$ internal symmetry.
Finally, there is an internal $\mathbb Z_2^C$ symmetry on the lattice which acts on the fields as
\ie\label{Z2_flip}
\phi^{XY} \rightarrow -\phi^{XY}\, ,\quad \phi^{YZ} \rightarrow -\phi^{YZ}\, ,\quad \phi^{ZX} \rightarrow -\phi^{ZX}  \,,\quad\phi^{XYZ} \rightarrow -\phi^{XYZ}\, .
\fe
Together, we have a $(\bZ_2)^3$ internal symmetry in the continuum.

The lattice model can also be formulated on a three-dimensional checkerboard. The sites  and the tetrahedra of the FCC lattice are mapped to the shaded cubes and the sites of the checkerboard, respectively. See Figure \ref{fig:U1-CB}.
On the checkerboard, the  phase variables $\phi$'s and their conjugate momenta $\pi$'s are placed at the shaded cubes.
 The tetrahedron interaction then becomes the interaction of four shaded cubes that share the same site (see Figure \ref{fig:U1-CB}c).
We emphasize that this reformulation from the FCC lattice to the checkerboard is not a duality transformation.
It is simply a reassignment of the same fields and the same interactions to different geometric objects on a different lattice.

\subsection{The continuum limit}

We now take the continuum limit $a\to0$ while holding
\ie
\mu \equiv 2Ka^2 =\text{fixed}\,.
\fe
To leading order in this limit, we obtain the constraint on the continuum fields
\ie\label{constraint}
\phi^{XYZ} (t,x,y,z)+ \phi^{XY}  (t,x,y,z)+ \phi^{YZ} (t,x,y,z) +\phi^{ZX} (t,x,y,z)= {\cal O}(a^2)\,.
\fe
Substituting $\phi^{XYZ}$ in terms of $\phi^{IJ}$ the Lagrangian in the continuum limit becomes
\ie
{\cal L}& = {\mu_0\over 2} \left[
(\partial_0\phi^{XY})^2+(\partial_0\phi^{YZ})^2  +(\partial_0\phi^{ZX})^2
+(\partial_0\phi^{XY}+\partial_0\phi^{YZ}+\partial_0\phi^{ZX})^2
\right]\\
&
- {\mu\over2}
\left[
  \partial_x( \phi^{XY}  + \phi^{ZX})\right]^2
  +\left[ \partial_y( \phi^{XY}  + \phi^{YZ})\right]^2
  +\left[ \partial_z( \phi^{ZX}  + \phi^{YZ})
\right]^2
+{\cal O}(a^2)\,.
\fe

Next we define
\ie\label{philowerupper}
&\phi^x = \phi^{XY} + \phi^{ZX}\,,\\
&\phi^y = \phi^{XY} + \phi^{YZ}\,,\\
&\phi^z = \phi^{ZX} + \phi^{YZ}\,.
\fe
Note that these are linear combinations of fields at different lattice points.  However, since these different lattice points correspond to the same point in the continuum limit, such a definition keeps these fields local. Conversely,
\ie\label{phiupperlower}
&\phi^{XY } = \frac 12\left(  \phi^x+\phi^y-\phi^z\right)\,,\\
&\phi^{ZX } = \frac 12\left(  \phi^z+\phi^x-\phi^y\right)\,,\\
&\phi^{YZ } = \frac 12\left(  \phi^y+\phi^z-\phi^x\right)\,.
\fe
The transformations \eqref{philowerupper} and \eqref{phiupperlower} relate different bases of the $\bf 3$ representation of $S_4$.\footnote{Recall that the uppercase indices $X$, $Y$, and $Z$ do not correspond to standard three vectors of $S_4$.} The Lagrangian can then be written as
\ie\label{L}
{\cal L} =
{\mu_0\over 2} \left[  \, (\partial_0\phi^x)^2+(\partial_0\phi^y)^2+(\partial_0\phi^z)^2\,\right]
-{ \mu\over 2} \left[ (\partial_x\phi^x )^2+(\partial_y\phi^y )^2 +(\partial_z\phi^z )^2 \right]  +{\cal O}(a^2)\,.
\fe
This continuum Lagrangian has an $S_4$ rotation symmetry that acts on $\phi^i$ as in $\mathbf{3}$.\footnote{Alternatively, we can redefine the action of the rotation group $S_4$ by composing the group elements of odd permutation in $S_4$ with the $\bZ_2^C$ symmetry $\phi^i \to -\phi^i$ \eqref{Z2_flip}.  The fields $\phi^i$ are then in the $\mathbf{3}'$ of the new $S_4$.  In the following we will make a choice of the $S_4$ symmetry such that $\phi^i$ is in the $\mathbf{3}$. }

The subsystem symmetries \eqref{sub1} and \eqref{sub2} from the lattice (along with those from the other directions) act as
\ie\label{latticesub}
&\phi^x \to \phi^x + f^x_y(y) + f^x_z(z)\,,\\
&\phi^y\to \phi^y+f^y_x(x) + f^y_z(z)\,,\\
&\phi^z \to \phi^z  + f^z_x(x) + f^z_y(y)\,.
\fe
In addition we have an  internal $(\bZ_2)^3 = \bZ_2^{(x)}\times \bZ_2^{(y)}\times \bZ_2^{(z)}$ symmetry from \eqref{Z2} and \eqref{Z2_flip}.
 $\bZ_2^{(i)}$  acts on the fields in the continuum as
\ie\label{Z2_cont}
\bZ_2^{(i)}:~~\phi^j(t,x,y,z)  \to (-1)^{\delta_{ij}} \, \phi^j (t,x,y,z)\,.
\fe
Note that this internal $(\bZ_2)^3$ symmetry   does not commute with the $S_4$ rotation symmetry.
It is a discrete vector symmetry.

\bigskip\bigskip\centerline{\it  Higher order corrections}\bigskip

The continuum Lagrangian \eqref{L} has a much larger, accidental subsystem symmetry
\ie\label{accidental}
&\phi^x \to \phi^x + f^x(y,z)\,,\\
&\phi^y\to \phi^y+f^y(z,x)\,,\\
&\phi^z \to \phi^z  + f^z(x,y)\,.
\fe
We can break the accidental symmetries \eqref{accidental} to the subsystem symmetries from the lattice \eqref{latticesub} by expanding the continuum Lagrangian to higher order in $a$.

The minimum of the potential is shifted at order $a^2$. This modifies the constraint \eqref{constraint} to
\ie
&\phi^{XYZ} + \phi^{XY}  + \phi^{YZ} +\phi^{ZX} \\
&+\frac{a^2}{4} \partial_x^2(\phi^{XY}+\phi^{ZX})
+\frac{a^2}{4} \partial_y^2(\phi^{XY}+\phi^{YZ})
+\frac{a^2}{4} \partial_z^2(\phi^{ZX}+\phi^{YZ})
={\cal O}(a^4)\,.
\fe
Next we define
\ie
&\phi^x = \phi^{XY} + \phi^{ZX}-\frac{a^2}{8} \left[\partial_x^2(\phi^{XY} + \phi^{ZX})+\partial_y^2(\phi^{YZ} + \phi^{ZX})+\partial_z^2(\phi^{ZX} + \phi^{YZ})\right]\,,\\
&\phi^y = \phi^{YZ} + \phi^{ZX}-\frac{a^2}{8} \left[\partial_x^2(\phi^{XY} + \phi^{ZX})+\partial_y^2(\phi^{YZ} + \phi^{ZX})+\partial_z^2(\phi^{ZX} + \phi^{YZ})\right]\,,\\
&\phi^z = \phi^{ZX} + \phi^{YZ}-\frac{a^2}{8} \left[\partial_x^2(\phi^{XY} + \phi^{ZX})+\partial_y^2(\phi^{YZ} + \phi^{ZX})+\partial_z^2(\phi^{ZX} + \phi^{YZ})\right]\,.
\fe
The terms that are quadratic in $\phi^i$ at order $a^2$ are
\begin{equation}
\begin{aligned}
\mathcal{L}\,\supset\,
&- \frac{\mu a^2}{8}\left(\frac{1}{2}\sum_{i\neq j\neq k \atop \text{cyclic}}(\partial_i\partial_j\phi^k)^2-\frac{1}{3}\sum_i(\partial_i^2\phi^i)^2-\sum_{j\neq i}(\partial_i\partial_j\phi^i)^2\right)~,
\end{aligned}
\end{equation}
where we discard the total derivative term.

\section{The continuum $\phi$-Theory} \label{sec:tetrphitheory}

In this section, we analyze the continuum field theory that we have just derived.

\subsection{Lagrangian}\label{sectpo}

Instead of focusing on a particular short-distance lattice model, we now look for the most general continuum field theory respecting  the global symmetry of our lattice system, including the subsystem symmetry \eqref{latticesub}.
At two-derivative order, we can add
\ie
(\partial_x \phi^x) (\partial_y \phi^y) \,,~~~(\partial_x \phi^x) (\partial_z \phi^z) \,,~~~(\partial_y \phi^y) (\partial_z \phi^z) \,.
\fe
These additional terms also respect the accidental symmetry \eqref{accidental}. However, they are not allowed by the $(\mathbb Z_2)^3$ symmetry \eqref{Z2_cont}.

At four-derivative order, we can have any $S_4$ symmetric quadratic functions of
\ie
&(\partial_i \partial_j \phi^k )\, ~~i\neq j ~~(\text{but $k$ can be the same as $i$ or $j$})~,\\
&(\partial_i^2 \phi^i)\,.
\fe
For example, we can add $(\partial_y \partial_z \phi^x)^2 \,,~(\partial_z \partial_x\phi^y)^2\,,~(\partial_x \partial_y\phi^z)^2$.
Cross-terms involving different fields are not allowed by the $(\mathbb Z_2)^3$ symmetry \eqref{Z2_cont}.  More generally, if we limit ourselves to terms that are quadratic in the fields, it appears that our continuum theory is made out of three decoupled theories, one for each of the three fields $\phi^x,\ \phi^y,\ \phi^z$.
We will soon see that this is not the case due to global issues.  (See also Section \ref{sec:anisotropic} for more details.)

Let us assume our continuum Lagrangian is a sum of the above symmetry-preserving two-derivative and four-derivative terms, with finite coefficients
\ie\label{eq:Lag_generic}
{\cal L} = {\mu_0\over 2} \sum_i (\partial_0\phi^i)^2
-{ \mu\over 2} \sum_i (\partial_i\phi^i )^2
- {\mu_1 \over 2} \sum_{\substack{i\ne j\ne k \\ \text{cyclic}}} (\partial_j \partial_k \phi^i)^2- {K_1 \over 2} \sum_{i} (\partial_i^2 \phi^i)^2- {K_2 \over 2} \sum_{i\ne j} (\partial_i\partial_j \phi^i)^2 \,.
\fe
We have imposed time-reversal symmetry to forbid cross terms between $\partial_0\phi^i$ and spatial derivatives of $\phi$'s.

For a generic mode of $\phi^i$ that depends on all the spacetime coordinates,  the four-derivative terms are less relevant than the two-derivative. The energy of this configuration is of order one.

Next, consider a  special configuration where, for example, $\phi^x$ is independent of $x$, i.e., $\phi^x = f^x(t,y,z)$.  For such a configuration, the terms with two spatial derivatives vanish.
The only nontrivial four-derivative term respecting the subsystem symmetry \eqref{latticesub} for this configuration is $(\partial_y\partial_z\phi^x)^2$. The energy of this configuration is again of order one.

Note that while the four-derivative terms are negligible for a generic mode in the momentum expansion, they are the only potential terms for the special modes such as $\phi^x = f^x(t,y,z)$.
If we had not included these four-derivative terms, such modes would not have had a potential term.  We can state that roughly, these four-derivative terms are ``dangerously irrelevant'' terms.  Although they are irrelevant relative to the two-derivative terms, they become relevant in some regions of field space.

Finally,  consider an even more special mode where $\phi^x$ depends only on $y$, i.e., $\phi^x = f^x_y(t,y)$.
To all derivative order, there is no nontrivial symmetry-preserving term for this configuration.
Therefore, these modes do not have  a potential term, and their classical energy is negligible.

From the discussion above, it is clear that the $K_1,K_2$ terms always give subleading corrections to either a generic mode or a non-generic mode. By contrast, this is not the case for the $(\partial_i\partial_j\phi^k)^2$ term, which is important for a non-generic mode  $\phi^k = f^k(t,x^i,x^j)$. Therefore we   will set $K_1=K_2=0$ in the following.

Let us discuss the periodicities of the $\phi^i$ field.
We impose the following gauge symmetry:
\ie
&\phi^{XY}(t,x,y,z) \sim \phi^{XY} (t,x,y,z) + 2\pi w^z(z)\,,\\
&\phi^{ZX}(t,x,y,z)\sim \phi^{ZX}(t,x,y,z)\,,\\
&\phi^{YZ}(t,x,y,z)\sim \phi^{YZ}(t,x,y,z)\,,
\fe
where $w^z(z)\in \bZ$.  Similarly for the other directions.
In addition, we impose  another gauge symmetry
\ie
&\phi^{XY}(t,x,y,z) \sim \phi^{XY} (t,x,y,z) \,,\\
&\phi^{ZX}(t,x,y,z)\sim \phi^{ZX}(t,x,y,z)+ 2\pi m^z(z)\,,\\
&\phi^{YZ}(t,x,y,z)\sim \phi^{YZ}(t,x,y,z)-2\pi m^z(z)\,,
\fe
where $m^z(z)\in \bZ$.  Similarly for the other directions.

In terms of the $\phi^i$, the periodicities are
\ie\label{eq:period_phi}
&\phi^x(t,x,y,z)  \sim \phi^x(t,x,y,z) + 2\pi (w^y(y) -m^y(y)) + 2\pi  ( w^z(z) +m^z(z)) \, ,\\
&\phi^y(t,x,y,z) \sim \phi^y(t,x,y,z)  + 2\pi (w^z(z) -m^z(z)) + 2\pi  ( w^x(x) +m^x(x)) \, ,\\
&\phi^z(t,x,y,z) \sim \phi^z(t,x,y,z)  + 2\pi (w^x(x) -m^x(x)) + 2\pi  ( w^y(y) +m^y(y)) \, ,
\fe
where $w^i(x^i), m^i(x^i)\in \bZ$.
Since the periodicities of $\phi^i$ are correlated as above, our Lagrangian \eqref{eq:Lag_generic} is not a decoupled Lagrangian for these three fields. See  Section \ref{sec:anisotropic} for more details.

As discussed in Section \ref{sec:intro}, we can describe the target space in the following way.
  Our target space is parameterized by the three fields $\phi^i$ in $\bR^3$ modded out the FCC lattice.  This means that we identify
\ie\label{eq:2.7}
\phi^i \sim \phi^i +2\pi n^i \qquad {\rm with}  \qquad   n^i \in\bZ~, \qquad n^x+n^y+n^z \in 2\bZ\,.
\fe
In a standard theory, these $n^i$ are independent of positions.  But in our case, they can depend on the spatial coordinates.  We can contemplate a general $n^i(x,y,z)$, but this should be such that the Lagrangian \eqref{eq:Lag_generic} is invariant under the identification.  (In other words, the identification should be a subgroup of the momentum symmetry discussed in \eqref{eq:2.20}).  This constrains the dependence to
\ie\label{eq:2.8}
\partial_xn^x=\partial_y\partial_zn^x=0~,
\fe
and similarly for the other directions.  This is the same as the identification \eqref{eq:period_phi}.

\subsection{Global symmetry}

\subsubsection{Discrete symmetry}\label{sec:phidiscrete}

The (orientation-preserving) spatial rotation group is  $S_4$.
In addition, we also have a $(\mathbb{Z}_2)^3$ symmetry in \eqref{Z2_cont} that does not commute with the $S_4$.
Together, the discrete global symmetry is $\mathbb{Z}_2^C \times G$ with
\begin{equation}
G=(\mathbb{Z}_2\times \mathbb{Z}_2)\rtimes S_4~.
\end{equation}
The order 96 group $G$ is the orientation-preserving symmetry group of the FCC lattice modulo translations of the minimal cubic sublattice.
See Appendices \ref{app:fcc} and  \ref{app:G} for more detail of this group.
Here $\bZ_2^C$ is the charge conjugation symmetry that flips the signs of all $\phi^i$:
\ie
\bZ_2^C:~ \phi^x \to -\phi^x\,,~~~~\phi^y \to -\phi^y \,,~~~~\phi^z \rightarrow -\phi^z\,.
\fe

The discrete global symmetry $\bZ_2^C \times G$ guarantees that  locally $\phi^i$  remain decoupled   even when higher derivative terms that are quadratic in the fields are included.
The three $\phi^i$ interact with each other only through the identification  \eqref{eq:period_phi}.

In the above we have assigned  particular $S_4$ representations to our fields and we used an $S_4$ covariant notation.
Since the  global  symmetry  $\bZ_2^C\times G$ of this continuum field theory is larger, it allows us to choose different ways to label our fields in terms of $S_4$ representations related by outer automorphisms of $\bZ_2^C\times G$.
It is important to clarify that the theory is not modified by this.  Even the Lagrangian is not modified.  The only thing that changes is the way we label the group elements of $\bZ_2^C\times G$.

We choose the coordinates $x,y,z$ to be in the $(+,  \mathbf{3}_1)$ of $\bZ_2^C\times G$ (see Appendix \ref{app:G} for the representation theory of this group).
Here we use $+$ and $-$ to denote the trivial representation and the sign representation of $\bZ_2^C$, respectively.
With respect to this choice, there is a $\bZ_2\times \bZ_2$ normal subgroup of $G$ that acts trivially on the coordinates.
The quotient group
\ie\label{rotqu}
{G\over \bZ_2\times \bZ_2} = S_4
\fe
acts on the coordinates and is generated by 90 degree rotations.

Next, we discuss the representations of $\phi^i$ under $\bZ_2^C\times G$.
We can choose $\phi^i$ to be in one of the following four representations  of $\bZ_2^C\times G$:
\ie\label{fourreps}
(-, \mathbf{3}_2),~~(-, \mathbf{3}_2'),~~(-, \mathbf{3}_3),~~\text{or }~ (-, \mathbf{3}_3')\,.
\fe
The four choices are equivalent as they are related by those outer automorphisms of $\bZ_2^C\times G$ that leave $(-,\mathbf{3}_1)$ invariant.\footnote{Note that we cannot take $\phi^i$ to be in $(-,\mathbf{3}_1)$ or $(-,\mathbf{3}_1')$ or $(+,\mathbf{R})$ for any $\mathbf{R}$. For these choices  the global symmetry $\bZ_2^C\times G$ does not act faithfully.}

Instead of working in a  $\bZ_2^C\times G$ covariant way, we will label our fields by  their representations of an $S_4$ subgroup of $G$.\footnote{This $S_4$ subgroup should not be confused with the $S_4$ quotient group that acts on the coordinates \eqref{rotqu}.}
Having fixed such a choice of the $S_4$ subgroup, the fields $\phi$, as well as their derivatives appearing in the Lagrangian \eqref{eq:Lag_generic},
are labeled by the following $S_4$ representations:
\\
\renewcommand{\arraystretch}{1.3}
\ie\label{phirep}
\begin{tabular}{|c|c|c|c|}
\hline
$\bZ_2^C\times G$ & $\phi$, $\partial_0\phi$ & $\partial \phi$ & $\partial \partial \phi$ \\
\hline
$(-,\mathbf{3}_2)$ & $\mathbf{3}$ & $\mathbf{1}\oplus\mathbf{2}$ & $\mathbf{1}'\oplus\mathbf{2}$ \\
\hline
$(-,\mathbf{3}_2')$ & $\mathbf{3}'$ & $\mathbf{1}'\oplus\mathbf{2}$ & $\mathbf{1}\oplus\mathbf{2}$ \\
\hline
$(-,\mathbf{3}_3)$ & $\mathbf{1}'\oplus\mathbf{2}$ & $\mathbf{3}'$ & $\mathbf{3}$ \\
\hline
$(-,\mathbf{3}_3')$ & $\mathbf{1}\oplus\mathbf{2}$ & $\mathbf{3}$ & $\mathbf{3}'$ \\
\hline
\end{tabular}
\fe
\renewcommand{\arraystretch}{1}
\\
where we have suppressed the indices.
In the previous subsections, we have chosen to label our fields as in the first row of \eqref{phirep}.
These different presentations of the same theory will be useful in the following sections.

Below we will  encounter  fields in the  four different $S_4$ representations above. We will use the following tensors to denote these fields:
\ie
\mathbf{3} :& ~~ T^{i}\,\\
\mathbf{3}':& ~~ T^{i'}\,\\
\mathbf{1}\oplus \mathbf{2} :& ~~ T^{ii}\,\\
\mathbf{1}'\oplus \mathbf{2} :& ~\, T^{ii'}\,
\fe
See Appendix \ref{sec:app_reps_S4} for more details.

\subsubsection{Momentum symmetry}

Many of  the expressions below  contain several equations related by cyclically permuting $x,y,z$. In some of these expressions, we will write only one of the  equations to avoid cluttering.

The equations of motion are
\ie\label{eq:eom_phi}
&\mu_0 \partial_0^2 \phi^x = \mu \partial_x^2 \phi^x -\mu_1 \partial_y^2 \partial_z^2 \phi^x \, ,
\fe
and similarly for the other two equations with $x,y,z$ cyclically permuted.
This leads to a current conservation equation
\ie\label{eq:momentum_conserv}
&\partial_0 J_0^x = \partial_x J^{xx} - \partial_y \partial_z \mathbf J^{xx'}~,
\fe
where we define the currents\footnote{The spatial currents satisfy an additional differential condition,
\ie\label{eq:diffcondition}
\mu_1\partial^y \partial^z J^{xx} + \mu \partial^x \mathbf J^{xx'} = 0\, .
\fe
This differential condition is not satisfied if we include the $K_1,K_2$ terms in \eqref{eq:Lag_generic}.}
\ie\label{eq:momentum_current}
\mathbf{3}:~&J_0^x = \mu_0 \partial_0 \phi^x\,,\\
\mathbf{1}\oplus\mathbf{2}:~&J^{xx} = \mu \partial_x \phi^x  \,,\\
\mathbf{1}'\oplus\mathbf{2}:~&\mathbf J^{xx'} = \mu_1 \partial_y \partial_z \phi^x\, .
\fe

The charges,
are defined on the planes
\ie \label{eq:mom_charge_pm}
Q^\pm_x (x ) =\oint dy dz~ (J^y_0 \pm J^z_0)\,.
\fe
There are six charges in total, two along each plane.
As we will soon see,  the linear combination above is chosen  such that each $Q^\pm_i$ is a sum of delta functions with integer coefficients.
These charges satisfy three constraints
\ie
&\oint dy~ \left(Q^+_y(y) - Q^-_y(y)\right) = \oint dz~ \left(Q^+_z(z) + Q^-_z(z)\right)\,,\\
&\oint dz~ \left(Q^+_z(z) - Q^-_z(z)\right) = \oint dx~ \left(Q^+_x(x) + Q^-_x(x)\right)\,,\\
&\oint dx~ \left(Q^+_x(x) - Q^-_x(x)\right) = \oint dy~ \left(Q^+_y(y) + Q^-_y(y)\right)\,.
\fe
These charges generate the \emph{momentum symmetry}\footnote{Here, by \emph{momentum}, we mean the conjugate momentum of the field $\phi^i$ in the target space as opposed to the momentum in coordinate space. Indeed, the temporal current $J_0^i$ of the momentum symmetry is the conjugate momentum of $\phi^i$.} that act on the fields as \eqref{latticesub}
\ie\label{eq:2.20}
&\phi^x(t,x,y,z) \rightarrow \phi^x(t,x,y,z) + (f^+_y(y) - f^-_y(y)) + (f^+_z(z) + f^-_z(z))\, ,
\\
&\phi^y(t,x,y,z) \rightarrow \phi^y(t,x,y,z) + (f^+_z(z) - f^-_z(z)) + (f^+_x(x) + f^-_x(x))\, ,
\\
&\phi^z(t,x,y,z) \rightarrow \phi^z(t,x,y,z) + (f^+_x(x) - f^-_x(x)) + (f^+_y(y) + f^-_y(y))\, .
\fe
The point-wise periodicities in \eqref{eq:period_phi} gauge the $\mathbb Z$ part of this symmetry. Hence, the momentum symmetry generated by the six charges $Q^\pm_i(x^i)$ is $U(1)$ instead of $\mathbb R$.

The above linear combinations of charges are motivated by the following observation. Since the target space of the field $\phi^i$ is the quotient of $\mathbb R^3$ by an FCC lattice (see \eqref{eq:2.7} and \eqref{eq:2.8}), the target space of its conjugate momentum $J_0^i$ is a body-centered cubic (BCC) lattice (the reciprocal lattice of FCC lattice). On a BCC lattice, while $J_0^i$ could be half-integers, the six combinations $J_0^i \pm J_0^j$ should all be integers. These are exactly the combinations defined above.

\subsubsection{Winding symmetry}

The current of the winding symmetry is
\ie\label{conswindc}
\mathbf{1}\oplus \mathbf{2} :~&J_0^{xx} = {1\over2\pi}\partial_x\phi^x \,,\\
\mathbf{1}'\oplus \mathbf{2} :~&\mathbf J_0^{xx'} = {1\over4\pi}\partial_y \partial_z \phi^x \,,\\
\mathbf{3}:~&J^x ={1\over4\pi} \partial_0\phi^x\, .
\fe
The current conservation equations are
\ie
&\partial_0 J^{xx}_0 =2 \partial_x J^x\,,\\
&\partial_0 \mathbf J^{xx'}_0 = \partial_y \partial_z J^x\, .
\fe
There is  a differential condition:
\ie\label{diff}
\partial_y \partial_z J_0^{xx} =2 \partial_x \mathbf J_0^{xx'}~.
\fe

The conserved charges are
\ie
&Q^{xx}(y,z) = \oint dx~ J_0^{xx}\,,\\
\fe
Using the differential  condition $Q^{xx}(y,z)$ is  a sum of a function $y$ and a function of $z$:
\ie
&Q^{xx}(y,z)  = Q^{xx}_y(y) +Q^{xx}_z(z)\,.
\fe
The two charges $Q^{xx}_y(y), Q^{xx}_z(z)$ share a common zero mode.  Equivalently, there is a gauge ambiguity
\ie
Q^{xx}_y(y) \sim Q^{xx}_y(y)+   n^{xx}\,,~~~~ Q^{xx}_z(z) \sim Q^{xx}_z(z) -   n^{xx}\,.
\fe
We will soon see that $Q^{ii}_j(x^j)$ are integer-valued functions.

In addition, we also have the following conserved charges
\ie
\mathbf{Q}^{xx'}_y(y) =\oint dz~ \mathbf J_0^{xx'}\,,~~~~ \mathbf{Q}^{xx'}_z(z) = \oint dy~ \mathbf J_0^{xx'}\,.
\fe
Note that they are independent of $x$ because of \eqref{diff}.
We will soon see that the properly quantized charges are actually the following combination:
\ie\label{eq:Q-def}
Q^{xx'}_{y\,-}  (y) & = \mathbf{Q}^{xx'}_y (y) - \frac 12 \partial_y Q^{zz}\\
&=  \oint dz~ \mathbf J_0^{xx'} - \frac12\partial_y\oint dz~J_0^{zz}\,,\\
Q^{xx'}_{z\,-}  (z) & = \mathbf{Q}^{xx'}_z (z) - \frac 12 \partial_z Q^{yy}\\
&=  \oint dy~ \mathbf J_0^{xx'} - \frac12\partial_z\oint dy~J_0^{yy}\,.
\fe
They obey the constraint
\ie
\oint dy~ Q^{xx'}_{y\,-}(y) = \oint dz~ Q^{xx'}_{z\,-}(z)\,.
\fe
In summary, the conserved charges are $Q^{ii}_j(x^j)$ and $Q^{ii'}_{j\, - } (x^j ) $.  On a lattice, there are $2L^x+2L^y+2L^z-3$ conserved charges of the form $Q^{ii}_j(x^j)$ , and  $2L^x+2L^y+2L^z-3$ conserved charges of the form $Q^{ii'}_{j\, - } (x^j ) $.

We will sometimes also consider the following composite conserved charge:
\ie\label{eq:Q+def}
&Q^{xx'}_{y\,+}(y)   =Q^{xx'}_{y\,-}   (y)+ \partial_y Q^{zz} =  \oint dz~ \mathbf J_0^{xx'} +\frac12\partial_y \oint dz~J_0^{zz}\,,\\
&Q^{xx'}_{z\,+} (z) =Q^{xx'}_{z\,-}   (z)+ \partial_z Q^{yy}=  \oint dy~ \mathbf J_0^{xx'} +\frac12\partial_z \oint dy~J_0^{yy}  \,.
\fe
The identification \eqref{eq:period_phi} implies that the charges $Q^{ii'}_{j\,\pm}(x^j)$ are sum of delta functions with integer coefficients:
\ie\label{peculiarrelation}
Q^{xx'}_{y\,\pm}(y) =\frac{1}{4\pi}\partial_y \oint dz \,\partial_z (\phi^x\pm\phi^z)=\sum_\alpha N^{xx'}_{y\,\pm\,,\alpha}  \, \delta(y - y_\alpha)~,
\fe
where $N^{xx'}_{y\,\pm\,,\alpha}$'s are integers.

\subsection{Momentum modes}\label{sec:momentum}

Let's discuss the modes that are charged under the momentum symmetry. We will consider plane wave solutions in $\mathbb R^{3,1}$ of the form
\ie
\phi^x = C^x e^{i\omega t+ik_x x  + i k_y y+ik_zz}\,,~~~
\phi^y = C^y e^{i\omega t+ik_x x  + i k_y y+ik_zz}\,,~~~
\phi^z = C^z e^{i\omega t+ik_x x  + i k_y y+ik_zz}\,.
\fe
We can consider a basis of them with only one nonzero $C^i$.
The mode with $C^x\neq0$ and $C^y=C^z=0$  obeys the dispersion relation
\ie
\omega_{(x)}^2 = \frac{1}{\mu_0}\left( \mu k_x^2 + \mu_1 k_y^2 k_z^2  \right)\,.
\fe
The classical zero energy solutions correspond to $k_x=0$  and $k_y k _z =0$.
Similarly there are such modes in the other directions.
We will refer to  these modes as the \emph{momentum modes}\footnote{By momentum modes here  we mean  the target space momenta. They should be contrasted with the momenta $k_i$ of the coordinate space.   These momentum modes arise when the coordinate space momenta $k_i$ are non-generic, i.e., when $k_x=0$  and $k_y  k _z =0$.} because the momentum symmetry maps one such solution to another.

Let us quantize the momentum modes now.  They are of the form
\ie
&\phi^x(t,x,y,z) = f^x_y(t,y) + f^x_z(t,z)\,.
\fe
It will be more convenient to write these momentum modes in a different basis where each mode has a simple periodicity:
\ie\label{momentummphii}
&\phi^x(t,x,y,z) = \left[\,  f^+_y(t,y) - f^-_y(t,y)\, \right] + \left[ \, f^+_z(t,z) + f^-_z(t,z)\,\right]\,,\\
&\phi^y(t,x,y,z) = \left[\,  f^+_z(t,z) - f^-_z(t,z)\, \right] + \left[ \, f^+_x(t,x) + f^-_x(t,x)\,\right]\,,\\
&\phi^z(t,x,y,z) = \left[\,  f^+_x(t,x) - f^-_x(t,x)\, \right] + \left[ \, f^+_y(t,y) + f^-_y(t,y)\,\right]\,.
\fe
From \eqref{eq:period_phi}, we find that each $f^\pm_i(t,x^i)$ is pointwise $2\pi$-periodic.
They share three common zero modes $c_i(t)$, which correspond to the following gauge symmetry
\ie\label{eq:phi_mom_mode_gauge}
&f^+_x(t,x) \sim f^+_x(t,x) + c_y(t) - c_z(t)\,,~~~f^-_x(t,x) \sim f^-_x(t,x) + c_y(t) + c_z(t)\,,\\
&f^+_y(t,y) \sim f^+_y(t,y) + c_z(t) - c_x(t)\,,~~~f^-_y(t,y) \sim f^-_y(t,y) + c_z(t) + c_x(t)\,,\\
&f^+_z(t,z) \sim f^+_z(t,z) + c_x(t) - c_y(t)\,,~~~f^-_z(t,z) \sim f^-_z(t,z) + c_x(t) + c_y(t)\,.
\fe
The Lagrangian of these modes is
\ie
L = \mu_0 &\left[  \ell^y\ell^z \oint dx~\left((\dot f^+_x)^2 + (\dot f^-_x)^2 \right) +  \ell^x \oint dy dz~\left(\dot f^+_y - \dot f^-_y \right) \left( \dot f^+_z + \dot f^-_z \right) \right.
\\
& +  \ell^x\ell^z \oint dy~\left((\dot f^+_y)^2 + (\dot f^-_y)^2 \right) +  \ell^y \oint dx dz~\left(\dot f^+_z - \dot f^-_z \right) \left( \dot f^+_x + \dot f^-_x \right)
\\
& \left. +  \ell^x\ell^y \oint dz~\left((\dot f^+_z)^2 + (\dot f^-_z)^2 \right) +  \ell^z \oint dx dy~\left(\dot f^+_x - \dot f^-_x \right) \left( \dot f^+_y + \dot f^-_y \right) \right]\,.
\fe
The conjugate momenta are
\ie
&\pi^+_x(t,x) = \mu_0 \left(2 \ell^y\ell^z \dot f^+_x +\ell^z \oint dy~(\dot f^+_y + \dot f^-_y) + \ell^y \oint dz~(\dot f^+_z - \dot f^-_z) \right) \,,
\\
&\pi^-_x(t,x) = \mu_0 \left(2 \ell^y\ell^z \dot f^-_x -\ell^z \oint dy~(\dot f^-_y + \dot f^+_y) + \ell^y \oint dz~( \dot f^+_z - \dot f^-_z) \right) \,,
\fe
They are the charges of the momentum symmetry
\ie
&Q^\pm_x(x) = \mu_0 \oint dydz~ (\partial_0 \phi^y \pm \partial_0 \phi^z) = \pi^\pm_x(x)\,,\\
&Q^\pm_y(y) = \mu_0 \oint dzdx~ (\partial_0 \phi^z \pm \partial_0 \phi^x) = \pi^\pm_y(y)\,,\\
&Q^\pm_z(z) = \mu_0 \oint dxdy~ (\partial_0 \phi^x \pm \partial_0 \phi^y) = \pi^\pm_z(z)\,.
\fe
The gauge symmetry \eqref{eq:phi_mom_mode_gauge} implies the three constraints
\ie\label{eq:mom_const}
&\oint dx \left(\pi^+_x(x ) - \pi^-_x(x)\right) = \oint dy \left(\pi^+_y(y) + \pi^-_y(y)\right)\,,\\
&\oint dy \left(\pi^+_y(y) - \pi^-_y(y)\right) = \oint dz\left(\pi^+_z(z) + \pi^-_z(z)\right)\,,\\
&\oint dz \left(\pi^+_z(z) - \pi^-_z(z)\right) = \oint dx \left(\pi^+_x(x) + \pi^-_x(x)\right)\,.
\fe

The Hamiltonian of these modes is
\ie
H = \frac{1}{2\mu_0 \ell^x \ell^y \ell^z} \left[ \frac12 \sum_{i} \ell^i \oint dx^i~\left((\pi^+_i)^2 + (\pi^-_i)^2 \right) - \frac14 \sum_{i} \left( \oint dx^i \left(\pi^+_i - \pi^-_i\right) \right)^2  \right] \,.
\fe

\bigskip\bigskip\centerline{\it Minimally charged states}\bigskip

The point-wise periodicity $f_i^+(t,x^i)\sim f_i^+(t,x^i)+2\pi w^i(x^i)$,  $f_i^-(t,x^i)\sim f_i^-(t,x^i)+2\pi m^i(x^i)$ implies that $\pi^i_\pm$ is a linear combination of delta function with integer coefficients. In addition, they satisfy the constraint \eqref{eq:mom_const}. There are two kinds of lowest energy charged states.
The first kind includes
\ie\label{eq:mom_charge_min}
&\pi^+_x = \delta(x-x_0)~,\qquad &&\pi^+_y = \delta(y-y_0)~,\qquad &&\pi^+_z = \delta(z-z_0)~,
\\
&\pi^-_x = 0~,\qquad &&\pi^-_y = 0~,\qquad &&\pi^-_z = 0~,
\fe
and its charge conjugate.
The second kind includes
\ie\label{eq:mom_charge_min_genBCC}
&\pi^+_x = \delta(x-x_0)~,\qquad &&\pi^+_y = 0~,\qquad &&\pi^+_z = 0~,
\\
&\pi^-_x = 0~,\qquad &&\pi^-_y = \delta(y-y_0)~,\qquad &&\pi^-_z = -\delta(z-z_0)~.
\fe
with two other similar configurations in the other directions and their
their three charge conjugate configurations.
Together, we have eight different momentum modes with the same minimal energy.\footnote{The three charges given by \eqref{eq:mom_charge_min_genBCC} and its permutations generate a BCC lattice of all the momentum charges with support only at $x=x_0,y=y_0,z=z_0$. The eight charges are the body centers of the eight cubes around the origin of the BCC lattice. They are the ones with minimal energy.}
\ie
H = \frac{1}{2\mu_0 \ell^x \ell^y \ell^z} \left[ \frac12 (\ell^x + \ell^y + \ell^z) \delta(0) - \frac34 \right]~.
\fe

\bigskip\bigskip\centerline{\it General charged states}\bigskip

More general charged states are written as
\ie
&\pi^+_x(x)=\sum_\alpha N^+_{x\,\alpha} \delta(x-x_\alpha)~,\qquad &&\pi^-_x(x)=\sum_{\alpha'} N^-_{x\,\alpha'} \delta(x-x_{\alpha'})~,
\\
&\pi^+_y(y)=\sum_\beta N^+_{y\,\beta} \delta(y-y_\beta)~,\qquad &&\pi^-_y(y)=\sum_{ \beta'} N^-_{y\,\beta'} \delta(y-y_{\beta'})~,
\\
&\pi^+_z(z)=\sum_\gamma N^+_{z\,\gamma} \delta(z-z_\gamma)~,\qquad &&\pi^-_z(z)=\sum_{ \gamma'} N^-_{z\,\gamma'} \delta(z-z_{\gamma'})~,
\fe
where $N^+_{i\,\alpha},N^-_{i\,\alpha'}\in\mathbb{Z}$ satisfy the constraints
\ie
&N^{z} \equiv \sum_\alpha N^+_{x\,\alpha} - \sum_{\alpha'} N^-_{x\,\alpha'} = \sum_\beta N^+_{y\,\beta} + \sum_{ \beta'} N^-_{y\,\beta'}~,
\\
&N^{x} \equiv \sum_\beta N^+_{y\,\beta} - \sum_{ \beta'} N^-_{y\,\beta'} = \sum_\gamma N^+_{z\,\gamma} + \sum_{ \gamma'} N^-_{z\,\gamma'}~,
\\
&N^{y} \equiv \sum_\gamma N^+_{z\,\gamma} - \sum_{ \gamma'} N^-_{z\,\gamma'} = \sum_\alpha N^+_{x\,\alpha} + \sum_{\alpha'} N^-_{x\,\alpha'}~.
\fe
The minimal energy with these charges is
\ie
H = \frac{1}{2\mu_0 \ell^x \ell^y \ell^z} \left[ \frac12 \sum_{i} \ell^i \left(\sum_{\alpha}(N^+_{i\,\alpha})^2 + \sum_{\alpha'}(N^-_{i\,\alpha'})^2 \right)\delta(0) - \frac14 \sum_{i} (N^i)^2  \right]~.
\fe

\subsection{Winding modes}\label{sec:winding}

As we said above, our target space has three fields $\phi^i$ and it is modded out by the FCC lattice \eqref{eq:2.7}.  Because of this quotient, we have winding states.

We start by analyzing the winding states that do not carry any subsystem winding symmetry.  For simplicity, we focus on winding in the $x$ direction
\ie\label{globalwinx}
&\phi^{x} =
2\pi {x\over \ell^x} N^{x}_x  \,,\\
&\phi^{y} =
2\pi {x\over \ell^x} N^{y}_x  \,,\\
&\phi^{z} =
2\pi {x\over \ell^x} N^{z}_x \,,\\
&N^x_x+N^y_x+N^z_x \in 2\mathbb{Z}~,
\fe
(with similar expressions for winding in the other directions).  This winding is consistent with the FCC lattice identification \eqref{eq:2.7}.
They are parameterized by three integers $N^{i}_x$ and have energy of order one.  The integers $N^y_x$ and $N^z_x$ can be shifted by momentum modes \eqref{momentummphii} and do not contribute to the winding charges.

Discarding the momentum modes, and adding winding around the other directions we are left with three winding modes.  Up to shifts by momentum modes, they can be taken to be:
\ie\label{eq:windingzeromode}
&\phi^{x} =2\pi {x\over \ell^x} N^x+2\pi {z\over \ell^z} N^z\,,\\
&\phi^{y} =2\pi {y\over \ell^y} N^y+2\pi {x\over \ell^x} N^x\,,\\
&\phi^{z}=2\pi {z\over \ell^z} N^z +2\pi {y\over \ell^y} N^y\,.
\fe
The shift of $\phi^x$ by $2\pi {z\over \ell^z} N^z$ (and similarly in the other directions) is needed in order to preserve the selection rule on the winding to be in the FCC lattice.

Their charges are
\ie
&Q^{xx}= Q^{xx}_y(y ) +Q^{xx}_z(z)= N^x~,\\
&Q^{yy}= Q^{yy}_z(z ) +Q^{yy}_x(x)= N^y~,\\
&Q^{zz}= Q^{zz}_x(x ) +Q^{zz}_y(y)= N^z~,\\
&\mathbf{Q}^{ii'}_j(x^j) =0\,.
\fe
Clearly, these winding charges do not correspond to a subsystem symmetry.

$Q^{xx}$ measures the winding of $\phi^x$ in the $x$ direction.  It is not sensitive to the winding of $\phi^y$ and $\phi^z$ in the $x$ direction.  This is consistent with the comment following \eqref{globalwinx}.  Stated differently, the winding of $\phi^y$ around the $x$ direction could have been measured by $\oint dx \partial_x \phi^y$, but this expression is not invariant under our gauge symmetry \eqref{eq:period_phi}.

We would like to make one more comment about these winding modes.  We can wind around three directions ($x$, $y$, or $z$) and for each of them we have a shift by a vector in the FCC lattice \eqref{eq:2.7}.  So naively, we can expect nine such winding charges.  Instead, we have only three such charges.  The point is that the rotations in space are not independent of rotations in the target space.  As a result, these would-be nine charges are in a reducible representation of our global symmetry.  Here we focus only on charges in ${\bf 2}\oplus {\bf 1}$ and discard the other charges.  In fact, the charges in the other rotation representations are not even well defined.  Equivalently, the same statement applies to the conserved winding currents \eqref{conswindc}.

Next, we discuss modes charged under the subsystem winding symmetry.  The simplest states are
\ie\label{eq:orderawinding}
&\phi^x(t,x,y,z) = 4\pi {x\over \ell^x}  N^x_y(y) + 4\pi {x\over \ell^x}  N^x_z(z) \,,\\
&\phi^y(t,x,y,z) = 4\pi {y\over \ell^y}  N^y_z(z) + 4\pi {y\over \ell^y}  N^y_x(x) \,,\\
&\phi^z(t,x,y,z) = 4\pi {z\over \ell^z}  N^z_x(x) + 4\pi {z\over \ell^z}  N^z_y(y) \,,\\
&N^i_j (x^j )\in \bZ\,.
\fe
Their winding charges are
\ie\label{evenchar}
Q_j^{ii}(x^j) = 2 N_j^i(x^j)\,,~~~~~\mathbf{Q}^{ii'}_{j}(x^j)=0\,.
\fe
The energy of such winding modes is of order $a$ if we have only order one nonzero $N$'s. On the other hand, if the number of nonzero $N$'s is order $1/a$ (\emph{e.g.}\,the functions $N$ have supports on segments with non-zero size), then the energy of such winding mode is of order one.
The periodicity of the $\phi^i$ field is simpler in the $\phi^{IJ}$ basis \eqref{philowerupper} and \eqref{phiupperlower}.
In the $\phi^{IJ}$ basis, the modes with nontrivial $N^x_y (y ),\,N^x_z (z)$ are (for simplicity the other $N$'s are set to zero)
\ie\label{eq:phixywinding}
&\phi^{XY}(t,x,y,z) = 2\pi {x\over \ell^x}  N^x_y(y) + 2\pi {x\over \ell^x}  N^x_z(z) \,,\\
&\phi^{YZ}(t,x,y,z) = -2\pi {x\over \ell^x}  N^x_y(y) - 2\pi {x\over \ell^x}  N^x_z(z) \,,\\
&\phi^{ZX}(t,x,y,z) = 2\pi {x\over \ell^x}  N^x_y(y) + 2\pi {x\over \ell^x}  N^x_z(z) \,,\\
&N^x_y (y ),\,N^x_z (z)\in \bZ\,.
\fe

As we explained, the modes \eqref{eq:windingzeromode} wind according to an FCC lattice \eqref{eq:2.7}.  However, the winding charges are not sensitive to the winding of all the fields.  For a given direction, say $x$, the winding charge $Q^{xx}$ is not sensitive to the winding of $\phi^y$ and $\phi^z$ around the $x$ cycle. This is to be contrasted with the situation here \eqref{eq:orderawinding}, where the local winding charge of the subsystem symmetry is sensitive to the winding of all the fields.  However, here the allowed windings are constrained to be on the cubic sublattice of our FCC lattice, hence the even charges in \eqref{evenchar}.

We stress that both the winding states \eqref{eq:windingzeromode} and \eqref{eq:orderawinding} do not realize the winding symmetry faithfully.  In the modes  \eqref{eq:windingzeromode}, the charges are independent of position and in the modes \eqref{eq:orderawinding}, the twists are only in the cubic sublattice. Next, we will discuss more generic modes, which realize the charges $Q^{ii}$ faithfully.

We now study the most general modes carrying the subsystem winding symmetry.   For simplicity, we first limit ourselves to modes that depend only  on $y$ and $z$ and present it in the $\phi^{IJ}$ basis
\ie\label{generalwinding}
&\phi^{XY} =  2\pi \left[ {y\over \ell^y} \sum_{\gamma'} W^z_{\gamma'} \Theta(z-z_{\gamma'})+ {z\over \ell^z}    \sum_\beta M^y_\beta \Theta(y-y_\beta)  - W^{yz} {yz\over \ell^y\ell^z} \right] \,,\\
&\phi^{ZX} =  2\pi \left[ {y\over \ell^y} \sum_\gamma M^z_\gamma \Theta(z-z_\gamma)+ {z\over \ell^z}    \sum_{\beta'} W^y_{\beta'} \Theta(y-y_{\beta'})  - W^{yz} {yz\over  \ell^y\ell^z} \right]\,,\\
&\phi^{YZ} =  -  2\pi \left[ {y\over \ell^y} \sum_\gamma M^z_\gamma \Theta(z-z_\gamma)+ {z\over \ell^z}    \sum_\beta  M^y_\beta \Theta(y-y_\beta)  - W^{yz} {yz\over \ell^y\ell^z} \right] \,,\\
&W^{yz} \equiv  \sum_{\beta'} W^y_{\beta'}  = \sum_\beta M^y_\beta
 = \sum_{\gamma'} W^z_{\gamma'}  = \sum_\gamma M^z_\gamma \,,~~~~~W^i_\alpha,M^i_\alpha\in \bZ~,
\fe
whose nonzero charges are\footnote{Written in terms of $Q^{ii'}_{j\,\pm}(x^j)$ defined in \eqref{eq:Q-def}, \eqref{eq:Q+def}, the nonzero charges are
\ie
& Q^{xx'}_{y\,+}(y) = \sum_{\beta'} W^y_{\beta'} \delta(y-y_{\beta'})\,,\\
&  Q^{xx'}_{z\,+}(z) = \sum_{\gamma'} W^z_{\gamma'} \delta(z-z_{\gamma'})\,,\\
& Q^{xx'}_{y\,-}(y) = \sum_{\beta'} M^y_{\beta'} \delta(y-y_{\beta'})\,,\\
& Q^{xx'}_{z\,-}(z) = \sum_{\gamma'} M^z_{\gamma'} \delta(z-z_{\gamma'})\,.
\fe
}
\ie\label{eq:windingchargeforgeneral}
&Q^{yy}_z(z) =  \sum_{\gamma'} W^z_{\gamma'} \Theta(z-z_{\gamma'}) - \sum_\gamma M^z_\gamma \Theta(z-z_\gamma) \,,\\
&Q^{zz}_y(y) =  \sum_{\beta'} W^y_{\beta'} \Theta(y-y_{\beta'})  - \sum_\beta M^y_\beta \Theta(y-y_\beta) \,,\\
&\mathbf{Q}^{xx'}_y(y) =  \frac12 \left( \sum_{\beta'} W^y_{\beta'} \delta(y-y_{\beta'})  + \sum_\beta M^y_\beta \delta(y-y_\beta) \right)\,,\\
&\mathbf{Q}^{xx'}_z(z) =  \frac12 \left( \sum_{\gamma'} W^z_{\gamma'} \delta(z-z_{\gamma'}) + \sum_\gamma M^z_\gamma \delta(z-z_\gamma) \right)\,.
\fe
These modes reduce to the modes in \eqref{eq:orderawinding}, \eqref{eq:phixywinding} with energy of  order $a$ or order one if $\mathbf{Q}^{xx'}_y(y)=\mathbf{Q}^{xx'}_z(z)=0$. As long as $\mathbf{Q}^{xx'}_y(y)\neq0$ or $\mathbf{Q}^{xx'}_z(z)\neq0$, these modes have energy of order $1/a$.

In the two previous cases \eqref{eq:windingzeromode} and \eqref{eq:orderawinding}, the charges $\bf Q$ vanished.  Here they do not vanish.  In fact, as we can see in \eqref{eq:windingchargeforgeneral}, they can be half-integer.  The charges $Q_\pm$ differ from $\bf Q$ by the charges $Q$ and they are all integral.  As we now see, the fact that in the previous cases $\bf Q$ vanished is correlated with the fact that the charges $Q$ were not realized faithfully.

If we discretize space, the modes in \eqref{generalwinding} are labeled by $2L^y+2L^z-3$ integers.
Combining with the other sectors, the modes are labeled by $4L^x+4L^y+4L^z- 9$ integers.
In total,  \eqref{generalwinding} (as well as the analogs in other directions) and \eqref{eq:windingzeromode}  give the most general winding modes (modulo momentum modes),  labeled by $4L^x+4L^y+4L^z-6$ integers.

We see that most of the winding modes have energy of order $1\over a$, but some of the winding modes have low energy.  This could raise a question about the robustness of the model under perturbations violating the winding symmetry (see \cite{paper1}).  However, these states are created by line operators, rather than local, point operators and therefore they cannot ruin the system.

\section{(3+1)d FCC lattice gauge theory}\label{sec:Alattice}

In this section we discuss a pure gauge theory associated with the global symmetry of the theory in the previous sections.  We will start with its lattice presentation.

\subsection{The lattice model}
\label{sec:FCClatticemodel}

Unlike the previous sections, here we will use a Lagrangian formulation on a Euclidean lattice, i.e., the theory will be placed on a four-dimensional spacetime lattice.

The spatial lattice is a three-dimensional face-centered cubic lattice, where the lattice site $s$ can either be at the corner of a cube or at the center of a face (see Figure \ref{fig:U1-CB}).
Let the shortest distance between two sites be $a/\sqrt{2}$ and $x^i  = a\hat x^i$.
We will label the lattice sites  at the corners by integers $(\hat x,\hat y,\hat z)$, while those on the faces by $(\hat x+\frac 12,\hat y+\frac12, \hat z )$, $(\hat x+\frac 12,\hat y, \hat z +\frac12)$, or $(\hat x,\hat y+\frac12, \hat z+\frac12 )$, with $\hat x, \hat y, \hat z\in \bZ$.

There is a $U(1)$ phase at the center $c$ of every tetrahedron.  We will label them as
\ie\label{eq:lattice_gaugefield}
&e^{i {\cal A}_c} ~~ &&\text{ if } c\in\(\mathbb{Z}- \tfrac14 ,\mathbb{Z}- \tfrac14,\mathbb{Z}-\tfrac14\)~,\\
&e^{ i {\cal A}^X_c}~~&&\text{ if } c\in\(\mathbb{Z}- \tfrac14,\mathbb{Z}+\tfrac14,\mathbb{Z}+\tfrac14\)~,\\
&e^{i {\cal A}^Y_c}~~&&\text{ if } c\in\(\mathbb{Z}+\tfrac14,\mathbb{Z}-\tfrac14,\mathbb{Z}+\tfrac14\)~,\\
&e^{ i{\cal A}^Z_c} ~~&&\text{ if } c\in\(\mathbb{Z}+\tfrac14,\mathbb{Z}+\tfrac14,\mathbb{Z}-\tfrac14\)~,\\
&e^{i{\cal A}^{XY}_c}~~&&\text{ if } c\in\(\mathbb{Z}-\tfrac14,\mathbb{Z}-\tfrac14,\mathbb{Z}+\tfrac14\)~,\\
&e^{i{\cal A}^{YZ}_c}~~&&\text{ if } c\in\(\mathbb{Z}+\tfrac14,\mathbb{Z}-\tfrac14,\mathbb{Z}-\tfrac14\)~,\\
&e^{i{\cal A}^{ZX}_c}~~&&\text{ if } c\in\(\mathbb{Z}-\tfrac14,\mathbb{Z}+\tfrac14,\mathbb{Z}-\tfrac14\)~,\\
&e^{i{\cal A}^{XYZ}_c}~~&&\text{ if } c\in\(\mathbb{Z}+\tfrac14,\mathbb{Z}+\tfrac14,\mathbb{Z}+\tfrac14\)~.
\fe
Each site $s$ of the FCC lattice at time $\hat\tau$ is connected with a site $s$ at time $\hat\tau+1$ by a temporal link where there is a $U(1)$ phase.
We will label them as
\ie\label{eq:temp_gauge_loc}
&e^{i\mathcal{A}^{XYZ}_{\tau,s}}~~&&\text{ if } s\in (\bZ,\bZ,\bZ)\,,\\
&e^{i\mathcal{A}_{\tau,s}^{XY}}~~&&\text{ if }s\in (\bZ+\tfrac 12,\bZ+\tfrac 12 , \bZ)\,,\\
&e^{i\mathcal{A}_{\tau,s}^{YZ}}~~&&\text{ if }s\in (\bZ,\bZ+\tfrac 12 , \bZ+\tfrac 12)\,,\\
&e^{i\mathcal{A}_{\tau,s}^{ZX}}~~&&\text{ if }s\in (\bZ+\tfrac 12,\bZ , \bZ+\tfrac 12)\,.
\fe
The four gauge parameters $\eta^{XYZ}_s=e^{i\alpha^{XYZ}_s}$, $\eta^{IJ}_s=e^{i\alpha^{IJ}_s}$ are also placed at the same locations as in \eqref{eq:temp_gauge_loc}.

The gauge transformations act on the spatial gauge fields as
\ie\label{eq:spatial_gauge_trans}
&{\cal A}_c \rightarrow {\cal A}_c - \alpha^{XYZ}_s - \alpha^{XY}_{s+(-\frac12,-\frac12,0)} - \alpha^{YZ}_{s+(0,-\frac12,-\frac12)} - \alpha^{ZX}_{s+(-\frac12,0,-\frac12)}~,
\\
&{\cal A}^X_{c+(0,\frac12,\frac12)} \rightarrow {\cal A}^X_{c+(0,\frac12,\frac12)} - \alpha^{XYZ}_s - \alpha^{XY}_{s+(-\frac12,\frac12,0)} - \alpha^{YZ}_{s+(0,\frac12,\frac12)} - \alpha^{ZX}_{s+(-\frac12,0,\frac12)}~,
\\
&{\cal A}^Y_{c+(\frac12,0,\frac12)} \rightarrow {\cal A}^Y_{c+(\frac12,0,\frac12)} - \alpha^{XYZ}_s - \alpha^{XY}_{s+(\frac12,-\frac12,0)} - \alpha^{YZ}_{s+(0,-\frac12,\frac12)} - \alpha^{ZX}_{s+(\frac12,0,\frac12)}~,
\\
&{\cal A}^Z_{c+(\frac12,\frac12,0)} \rightarrow {\cal A}^Z_{c+(\frac12,\frac12,0)} - \alpha^{XYZ}_s - \alpha^{XY}_{s+(\frac12,\frac12,0)} - \alpha^{YZ}_{s+(0,\frac12,-\frac12)} - \alpha^{ZX}_{s+(\frac12,0,-\frac12)}~,
\\
&{\cal A}^{XY}_{c+(0,0,\frac12)} \rightarrow {\cal A}^{XY}_{c+(0,0,\frac12)} - \alpha^{XYZ}_s - \alpha^{XY}_{s+(-\frac12,-\frac12,0)} - \alpha^{YZ}_{s+(0,-\frac12,\frac12)} - \alpha^{ZX}_{s+(-\frac12,0,\frac12)}~,
\\
&{\cal A}^{YZ}_{c+(\frac12,0,0)} \rightarrow {\cal A}^{YZ}_{c+(\frac12,0,0)} - \alpha^{XYZ}_s - \alpha^{XY}_{s+(\frac12,-\frac12,0)} - \alpha^{YZ}_{s+(0,-\frac12,-\frac12)} - \alpha^{ZX}_{s+(\frac12,0,-\frac12)}~,
\\
&{\cal A}^{ZX}_{c+(0,\frac12,0)} \rightarrow {\cal A}^{ZX}_{c+(0,\frac12,0)} - \alpha^{XYZ}_s - \alpha^{XY}_{s+(-\frac12,\frac12,0)} - \alpha^{YZ}_{s+(0,\frac12,-\frac12)} - \alpha^{ZX}_{s+(-\frac12,0,-\frac12)}~,
\\
&{\cal A}^{XYZ}_{c+(\frac12,\frac12,\frac12)} \rightarrow {\cal A}^{XYZ}_{c+(\frac12,\frac12,\frac12)} - \alpha^{XYZ}_s - \alpha^{XY}_{s+(\frac12,\frac12,0)} - \alpha^{YZ}_{s+(0,\frac12,\frac12)} - \alpha^{ZX}_{s+(\frac12,0,\frac12)}~,
\fe
where $s$ is a corner on the lattice (i.e., $s\in(\bZ,\bZ,\bZ)$), and $c$ is the center of one of the eight tetrahedra around $s$. They are related by $s=c+\left(\frac14,\frac14,\frac14\right)$. The gauge transformation acts on the temporal gauge fields as
\ie\label{eq:temporal_gauge_trans}
&\mathcal{A}^{XYZ}_{\tau,s}\left(\hat \tau+\tfrac{1}{2}\right)\rightarrow\mathcal{A}^{XYZ}_{\tau,s}\left(\hat \tau+\tfrac{1}{2}\right)+ \alpha^{XYZ}_s(\hat \tau+1)-\alpha^{XYZ}_s(\hat\tau) \,,\\
&\mathcal{A}_{\tau,s}^{XY}\left(\hat \tau+\tfrac{1}{2}\right)\rightarrow\mathcal{A}_{\tau,s}^{XY}\left(\hat \tau+\tfrac{1}{2}\right)+\alpha_s^{XY}(\hat \tau+1)-\alpha_s^{XY}(\hat \tau) \,,\\
&\mathcal{A}_{\tau,s}^{YZ}\left(\hat \tau+\tfrac{1}{2}\right)\rightarrow\mathcal{A}_{\tau,s}^{YZ}\left(\hat \tau+\tfrac{1}{2}\right)+\alpha_s^{YZ}(\hat \tau+1)-\alpha_s^{YZ}(\hat \tau) \,,\\
&\mathcal{A}_{\tau,s}^{ZX}\left(\hat \tau+\tfrac{1}{2}\right)\rightarrow\mathcal{A}_{\tau,s}^{ZX}\left(\hat \tau+\tfrac{1}{2}\right)+\alpha_s^{ZX}(\hat \tau+1)-\alpha_s^{ZX}(\hat \tau) \,,\\
\fe

The lattice model can also be formulated on a four-dimensional spacetime lattice where the spatial lattice is replaced by a three-dimensional checkerboard. The sites of the original FCC lattice and the tetrahedra are mapped to the shaded cubes and the sites of the checkerboard in Figure \ref{fig:U1-CB}(b).
On the checkerboard, the  gauge parameters $\alpha$ are placed at the shaded cubes and the spatial gauge fields $\mathcal{A}$ are at the sites.
We emphasize that this reformulation from the FCC lattice to the checkerboard is not a duality transformation.
It is simply a reassignment of the same fields and the same interactions to different geometric objects on a different lattice.

\subsection{The continuum limit}

\bigskip\bigskip\centerline{\it $B^2$ terms} \bigskip

Let us first discuss the gauge invariant interaction involving the spatial gauge fields.
At every site $s$, there is a  gauge-invariant interaction
\ie\label{eq:magnetic_term}
&- \cos\left({\cal A}_c+{\cal A}^X_{c+(0,\frac12,\frac12)}+{\cal A}^Y_{c+(\frac12,0,\frac12)}+{\cal A}^Z_{c+(\frac12,\frac12,0)} \right.
\\
&\qquad \qquad\left.-{\cal A}^{XY}_{c+(0,0,\frac12)}-{\cal A}^{YZ}_{c+(\frac12,0,0)}-{\cal A}^{ZX}_{c+(0,\frac12,0)}-{\cal A}^{XYZ}_{c+(\frac12,\frac12,\frac12)}\right)~,
\fe
which is an oriented product of the eight gauge fields surrounding the site.
Here the center $c$ of a tetrahedron is related to the site $s$ by $c=  s  - (\frac 14, \frac 14, \frac14)$.
On the checkerboard, it is an oriented product of the eight sites around a shaded cube.

In the Lagrangian, we sum over the corner sites $s\in(\mathbb{Z},\bZ,\bZ)$ of the FCC lattice and for each corner site there are four such gauge-invariant interactions.
More specifically, each spatial gauge field ${\cal A}_c$ participates in four different such interactions in the Lagrangian:
\ie\label{magnetic}
L_{mag}=-K&\Big[ \cos\left({\cal A}_c+{\cal A}^X_{c+(0,\frac12,\frac12)}+{\cal A}^Y_{c+(\frac12,0,\frac12)}+{\cal A}^Z_{c+(\frac12,\frac12,0)}\right.
\\
&\qquad \qquad\left.-{\cal A}^{XY}_{c+(0,0,\frac12)}-{\cal A}^{YZ}_{c+(\frac12,0,0)}-{\cal A}^{ZX}_{c+(0,\frac12,0)}-{\cal A}^{XYZ}_{c+(\frac12,\frac12,\frac12)}\right)
  \\
& + \cos\left({\cal A}_c+{\cal A}^X_{c+(0,-\frac12,\frac12)}+{\cal A}^Y_{c+(-\frac12,0,\frac12)}+{\cal A}^Z_{c+(-\frac12,-\frac12,0)}\right.
\\
&\qquad \qquad \left.-{\cal A}^{XY}_{c+(0,0,\frac12)}-{\cal A}^{YZ}_{c+(-\frac12,0,0)}-{\cal A}^{ZX}_{c+(0,-\frac12,0)}-{\cal A}^{XYZ}_{c+(-\frac12,-\frac12,\frac12)}\right)
  \\
 &+\cos\left({\cal A}_c+{\cal A}^X_{c+(0,-\frac12,-\frac12)}+{\cal A}^Y_{c+(\frac12,0,-\frac12)}+{\cal A}^Z_{c+(\frac12,-\frac12,0)}\right.
 \\
 &\qquad \qquad \left.-{\cal A}^{XY}_{c+(0,0,-\frac12)}-{\cal A}^{YZ}_{c+(\frac12,0,0)}-{\cal A}^{ZX}_{c+(0,-\frac12,0)}-{\cal A}^{XYZ}_{c+(\frac12,-\frac12,-\frac12)}\right)
  \\
 &+\cos\left({\cal A}_c+{\cal A}^X_{c+(0,\frac12,-\frac12)}+{\cal A}^Y_{c+(-\frac12,0,-\frac12)}+{\cal A}^Z_{c+(-\frac12,\frac12,0)}\right.
 \\
 &\qquad \qquad \left.-{\cal A}^{XY}_{c+(0,0,-\frac12)}-{\cal A}^{YZ}_{c+(-\frac12,0,0)}-{\cal A}^{ZX}_{c+(0,\frac12,0)}-{\cal A}^{XYZ}_{c+(-\frac12,\frac12,-\frac12)}\right)
 \Big]~.
\fe
On the checkerboard,  these four interactions correspond to the four shaded cubes in Figure \ref{fig:U1-CB}(c).

Next, we take the continuum limit by expanding in $a$ for large $K$.  We expand the gauge field around $c$ and from this point on,  all the fields in a given expression will reside at the same spacetime point.
At leading order we find a constraint, which sets
\ie\label{solveA}
&{\cal A}= - {\cal A}^X - {\cal A}^Y - {\cal A}^Z + {\cal A}^{XY} + {\cal A}^{YZ} + {\cal A}^{ZX} + {\cal A}^{XYZ}   + {\cal O}(a^2)\,.
\fe

We now use the gauge freedom associated with $\eta_s$ to set ${\cal A}^{XYZ}=0$ and we are left with six gauge fields.
We parameterize them in terms of $A^{ii}, \mathbf A^{ii'}$ as
\ie\label{eq:field_redef}
&\mathcal{A}^X=a A^{xx}+\frac{a^2}{4}\left(
C^x-\mathbf A^{yy'}-\mathbf A^{zz'}\right)+\mathcal{O}(a^3)~,
\\
&\mathcal{A}^{XY}=a(A^{xx}+A^{yy})+\frac{a^2}{4}\left(
C^x+C^y
-\mathbf A^{xx'}-\mathbf A^{yy'}\right)+\mathcal{O}(a^3)~,
\fe
where $C^i$ will be determined shortly.
Next, we substitute these values of ${\cal A}^I ,{\cal A}^{IJ}$ in \eqref{magnetic}  and let $A^{ii}$ and $\mathbf A^{ii'}$  be of order one.  Expanding \eqref{magnetic} to order $\mathcal{O}(a^6)$, it becomes
\ie\label{lattBs}
L_{mag} =  {K\over 8}  a^6 \left[
(\mathbf{B}^{x'})^2+(\mathbf{B}^{y'})^2+(\mathbf{B}^{z'})^2
\right]
+{\cal O}(a^8)\,.
\fe
where we have defined
\ie
\mathbf{B}^{x'} = \partial_x \mathbf{A}^{xx'} - \partial_y\partial_z A^{xx}\,.
\fe
Note that this expression is independent of $C^i$, which is an order $a$ correction to $A^{ii}$.
In the continuum limit, we take  $a\to 0$ and $K\to \infty$ with fixed $g_m^2 \equiv{8\over Ka^2}$. The remaining factor of $a^4$ in \eqref{lattBs} turns the sum over the lattice points into an integral of the continuum Lagrangian density.

In the continuum, the spatial gauge fields $A^{ii}$ and $\mathbf A^{ii'}$ are in the $S_4$ representation $\mathbf{1} \oplus \mathbf{2}$ and $\mathbf{1}' \oplus \mathbf{2}$ respectively, as indicated by their indices. Unlike the lower case indices, the upper cases indices, e.g., $X$ and $Y$ in \eqref{eq:field_redef}, are not $S_4$ vector indices.

\bigskip\bigskip\centerline{\it Gauge transformations}  \bigskip

Let us study the gauge transformations in the continuum limit.
It constrains the gauge parameter $\alpha^{XYZ}_s$ in terms of the others as
\ie\label{eq:gauge_constraint}
\alpha^{XYZ}_s + \alpha^{XY}_{s+(\frac12,\frac12,0)} + \alpha^{YZ}_{s+(0,\frac12,\frac12)} + \alpha^{ZX}_{s+(\frac12,0,\frac12)} = 0~.
\fe
We use this constraint to solve $\alpha^{XYZ}$ in terms of the remaining three $\alpha^{IJ}$'s. Substituting $\alpha^{XYZ}$ into \eqref{eq:spatial_gauge_trans} and expanding the gauge transformation to $\mathcal{O}(a^2)$, we find that
\ie
&\mathcal{A}^X \sim \mathcal{A}^X+a\partial_x\alpha^x+\frac{a^2}{4}\left(\partial_x^2\alpha^x+\partial_x\partial_y(\alpha^y-\alpha^z)+\partial_x\partial_z(\alpha^z-\alpha^y)\right)
\\
&\qquad\ \ +\frac{a^3}{16}\left(\frac{1}{2}(\partial_y^2+\partial_z^2-\partial_y\partial_z)\partial_x\alpha^x+\partial_x^2(\partial_y-\partial_z)\alpha^y+\partial_x^2(\partial_z-\partial_y)\alpha^z+\frac{7}{6}\partial_x^3\alpha^x\right)+\mathcal{O}(a^4)~,
\\
&\mathcal{A}^{XY}\sim\mathcal{A}^{XY}+a(\partial_x\alpha^x+\partial_y\alpha^y)
\\
&\qquad\quad+\frac{a^2}{4}\left(\partial_x^2\alpha^x+\partial_y^2\alpha^y+\partial_x\partial_z(\alpha^z-\alpha^y)+\partial_y\partial_z(\alpha^z-\alpha^x)+\partial_x\partial_y(\alpha^x+\alpha^y)  \right)+\mathcal{O}(a^3)~,
\fe
where the three continuum gauge parameters $\alpha^i$ are
\ie
\alpha^x= \alpha^{XY} + \alpha^{ZX}\,,~~\alpha^y= \alpha^{XY} + \alpha^{YZ}\,,~~\alpha^z= \alpha^{ZX} + \alpha^{YZ}  ~.
\fe
The gauge transformations of $A^{ii}, \mathbf A^{ii'}$ are
\ie
&A^{xx} \sim A^{xx} + \partial_x \alpha^x + {\cal O}(a)~,
\\
&\mathbf A^{xx'} \sim \mathbf A^{xx'} + \partial_y \partial_z \alpha^x +   {\cal O}(a)~.
\fe
This expression holds true to leading order in $a$ for any $C^i$. We can fix $C^i$ and the $O(a^2)$ terms in \eqref{eq:field_redef} by demanding that the gauge transformations of $A^{ii}, \mathbf A^{ii'}$ are
\ie
&A^{xx} \sim A^{xx} + \partial_x \alpha^x + {\cal O}(a^3)~,
\\
&\mathbf A^{xx'} \sim \mathbf A^{xx'} + \partial_y \partial_z\alpha^x +   {\cal O}(a^2)~.
\fe
This fixes $C^i=\partial_i(A^{xx}+A^{yy}+A^{zz})$ in \eqref{eq:field_redef}.
Note that the magnetic fields $\mathbf{B}^{i'}$ are indeed invariant under this gauge transformation.

In the continuum limit, the gauge parameters $\alpha^i$ are in the $S_4$ representation $\mathbf{3}$, and as we said above, $A^{ii}$ are in $\mathbf{1}\oplus \mathbf{2}$, and $\mathbf A^{ii'}$ are in $\mathbf{1}'\oplus \mathbf{2}$.

\bigskip\bigskip\centerline{\it $E^2$ terms}  \bigskip

We will now discuss the gauge invariant kinetic terms involving the temporal gauge fields.
At every corner $s$, there are eight gauge-invariant kinetic terms (one per tetrahedron at that site), corresponding to the eight spatial gauge fields. The kinetic term for, say, ${\cal A}_c$ is
\ie
&L_{elec} \supset
\\
&-{1\over g_{e1}^2} \cos\({\cal A}_c(\hat \tau+\tfrac12)-{\cal A}_c(\hat \tau-\tfrac12)+{\cal A}^{XYZ}_{\tau ,s}(\hat \tau)+{\cal A}^{XY}_{\tau ,s-(\frac12,\frac12,0)}(\hat \tau)+{\cal A}^{YZ}_{\tau,s-(0,\frac12,\frac12)}(\hat \tau)+{\cal A}^{ZX}_{\tau,s-(\frac12,0,\frac12)}(\hat \tau)\)~.
\fe
Here the center $c$ of a tetrahedron is related to the site $s$ by $c=  s  - (\frac 14, \frac 14, \frac14)$.
In the Lagrangian, we sum over the corners and at each corner there are eight such gauge-invariant kinetic terms.

In the continuum limit, we take $a\to 0$ with small but fixed coupling constant $g_{e1}^2$. At the leading order we find a  constraint, which sets
\ie\label{solveA0}
&{\cal A}^{XYZ}_\tau = - {\cal A}_\tau ^{XY} - {\cal A}_\tau ^{YZ} - {\cal A}_\tau ^{ZX} + {\cal O}(a^2)\,.
\fe

Next, we expand ${\cal A}^{IJ}_\tau $ as
\ie
&\mathcal{A}^{XY}_\tau   =  \frac a2 (A^x_\tau  +A^y_\tau  - A^z_\tau ) -{a^3 \over32} \partial_z E^{zz}+ {\cal O}(a^4)\,,\\
&\mathcal{A}^{ZX}_\tau   =  \frac a2 (A^z_\tau  +A^x_\tau  - A^y_\tau ) -{a^3 \over32} \partial_y E^{yy}+ {\cal O}(a^4)\,,\\
&\mathcal{A}^{YZ}_\tau   =  \frac a2 (A^y_\tau  +A^z_\tau  - A^x_\tau ) -{a^3 \over32} \partial_x E^{xx}+ {\cal O}(a^4)\,,
\fe
with $A_\tau ^i$ of order one.

The electric field is defined as
\ie
E^{ii}&=\partial_\tau  A^{ii}  -  \partial_i A_\tau ^i~.
\fe
 The gauge transformations acts on the temporal gauge field $A_\tau ^i$ as
\ie
A_\tau ^i \sim A_\tau ^i + \partial_\tau  \alpha^i~.
\fe
In the continuum limit, $A^i_\tau$ is in the $S_4$ representation $\mathbf{3}$.

We now substitute all the ${\cal A}_\tau ,{\cal A}$ gauge fields in terms of the order one continuum gauge fields $A_\tau ,A$ into $L_{elec}$.
Discarding the total derivative terms and higher order terms in the time derivative, the quadratic terms in  $L_{elec}$ are
\ie\label{latticeelet}
L_{elec}&=
{a^4\over g_{e1}^2}\left[(E^{xx})^2
+(E^{yy})^2
+(E^{zz})^2\right]
+{a^6\over 16 g_{e1}^2}\Big[ (\mathbf E^{xx'} )^2+(\mathbf E^{yy'})^2+(\mathbf E^{zz'} )^2
\\
&\quad
-(\partial_x E^{yy})^2 - (\partial_y E^{xx} )^2
-(\partial_x E^{zz})^2-  (\partial_z E^{xx} )^2
-(\partial_y E^{zz} )^2- (\partial_z E^{yy} )^2\\
&\quad
- \frac 43 ( \partial_xE^{xx})^2
-\frac 43 ( \partial_y E^{yy})^2
-\frac 43 ( \partial_zE^{zz})^2
\Big]
+\mathcal{O}(a^7)~,
\fe
where we have defined
\ie
&\mathbf E^{xx'}&=\partial_\tau  \mathbf A^{xx'}  -  \partial_y\partial_z A_\tau ^x~,\\
&\mathbf E^{yy'}&=\partial_\tau  \mathbf A^{yy'}  -  \partial_z\partial_x A_\tau ^y~,\\
&\mathbf E^{zz'}&=\partial_\tau  \mathbf A^{zz'}  -  \partial_x\partial_y A_\tau ^z~,
\fe

In the continuum limit, the factor of $a^4$ in the first term in the lattice Lagrangian \eqref{latticeelet} turns the sum over the lattice points into an integral of the continuum Lagrangian density.  The second term appears to be suppressed by a factor of $a^2$.  However, this higher order term is the leading order kinetic term for the gauge fields $\mathbf{A}^{ii'}$ and therefore cannot be ignored.  It is similar to the dangerously irrelevant terms mentioned in Section \ref{sectpo}.

The terms in the second and  third lines of \eqref{latticeelet} are subleading for a generic mode in the continuum, but they can affect the quantitative results of certain charged states.  (See \cite{paper1,paper2}, for a discussion of such terms and their effects.)

In Section \ref{sec:tetrgaugetheory}, we will consider the general continuum gauge theory with these gauge fields and symmetry, where we allow the coefficients of the higher derivative terms to be more general.  For simplicity, we will ignore higher derivative terms as in the second and third lines of \eqref{latticeelet}.

\section{The continuum gauge theory}\label{sec:tetrgaugetheory}

In this section, we analyze the continuum field theory derived from the FCC lattice gauge theory of Section \ref{sec:Alattice}.  Many of  the expressions below  contain several equations related by cyclically permuting $x,y,z$. In some of these expressions, we will write only one of the  equations to avoid cluttering.

\subsection{Lagrangian}

Our gauge theory is determined by the following gauge transformations
\ie\label{gaugetrans}
\mathbf{3}: ~&A_0^x \sim A_0^x + \partial_0 \alpha^x\, ,
\\
\mathbf{1} \oplus \mathbf{2}: ~&A^{xx} \sim A^{xx} + \partial_x \alpha^x\, ,
\\
\mathbf{1}' \oplus \mathbf{2}:  ~&\mathbf A^{xx'} \sim \mathbf A^{xx'} + \partial_y \partial_z \alpha^x\, .
\fe
Importantly, the gauge fields $(A_0^i , A^{ii} ,\mathbf A^{ii'})$ with $i=x,y,z$ are not independent gauge fields.
Rather, they are coupled through the periodicities of their gauge parameters.
More specifically, the gauge parameters $\alpha^i$ have the same periodicities as $\phi^i$ in \eqref{eq:period_phi}:
\ie\label{alphaperiodicity}
&\alpha^x(t,x,y,z)  \sim \alpha^x(t,x,y,z) + 2\pi (w^y(y) -m^y(y)) + 2\pi  ( w^z(z) +m^z(z)) \, ,\\
&\alpha^y(t,x,y,z) \sim \alpha^y(t,x,y,z)  + 2\pi (w^z(z) -m^z(z)) + 2\pi  ( w^x(x) +m^x(x)) \, ,\\
&\alpha^z(t,x,y,z) \sim \alpha^z(t,x,y,z)  + 2\pi (w^x(x) -m^x(x)) + 2\pi  ( w^y(y) +m^y(y)) \, ,
\fe
where $w^i(x^i), m^i(x^i)\in \bZ$.

These gauge fields $(A_0^i , A^{ii} ,\mathbf A^{ii'})$  are  related to the $\phi$-theory of Section \ref{sec:tetrphitheory} as follows.
They are the gauge fields for the momentum symmetry \eqref{eq:momentum_conserv}, \eqref{eq:momentum_current}  of the $\phi$-theory.
The coupling is given by
\ie
\sum_i( J_0^i A_0^i - J^{ii} A^{ii} + \mathbf J^{ii'} \mathbf A^{ii'})\, ,
\fe
which is invariant under the gauge transformation \eqref{gaugetrans}.

We can define the gauge invariant electric and magnetic fields
\ie
\mathbf{1} \oplus\mathbf{2}: ~&E^{xx} = \partial_0 A^{xx} - \partial_x A^x_0\, ,
\\
\mathbf{1}' \oplus\mathbf{2}:  ~&\mathbf E^{xx'} = \partial_0 \mathbf A^{xx'} - \partial_y \partial_z A^x_0\, ,
\\
\mathbf{3}': ~&\mathbf B^{x'} = \partial_x \mathbf A^{xx'} - \partial_y \partial_z A^{xx}\, .
\fe

In addition to the rotation symmetry $S_4$, we also have an internal $(\bZ_2)^3 = \bZ_2^{(x)} \times \bZ_2^{(y)} \times \bZ_2^{(z)}$ symmetry.
For example, $\bZ_2^{(x)}$ acts as
\ie
\bZ_2^{(x)} :~~A_0^x \to - A_0^x \,,~~~~A^{xx} \to -A^{xx}\,,~~~~\mathbf A^{xx'}\to -\mathbf A^{xx'}~,
\fe
with the other fields invariant.  The actions of $\bZ_2^{(y)}$ and $\bZ_2^{(z)}$ are defined similarly.
Together, the global symmetry is $\bZ_2^C \times G= \bZ_2^C \times ( (\bZ_2\times \bZ_2)\rtimes S_4)$.
See Section \ref{sec:Adiscrete} for more discussions.

The Lagrangian is
\ie \label{eq:A_Lag}
{\cal L} &=  {1\over g_{e1}^2}\sum_i   E^{ii} E_{ii}  + {1\over g_{e2}^2 }\sum_{i }   \mathbf E^{ii'}\mathbf E_{ii'} - {1\over g_m^2}\sum_{i} \mathbf B^{i'}\mathbf B_{i'}
\\
&\qquad - \frac{1}{\kappa_1} \sum_i   (\partial_i E^{ii})^2 - \frac{1}{\kappa_2}\sum_{i\neq j}  (\partial_j E^{ii})^2~.
\fe
Note that there is no cross term between $\partial_j E^{ii}$ and $\mathbf E^{ii'}$ that respects the $\bZ_2^C \times G$ symmetry.
In addition, we impose time-reversal symmetry to forbid cross terms between the electric and the magnetic fields.
Since the three gauge parameters $\alpha^i$ have correlated periodicities \eqref{alphaperiodicity}, our $U(1)$ Lagrangian \eqref{eq:A_Lag} is not a sum of decoupled Lagrangians for the three gauge fields associated with the $x,y,z$ directions.
See  Section \ref{sec:anisotropic} for more details.

The terms with $\kappa_1$ and $\kappa_2$ are higher order in the derivative expansion.  However, as mentioned in the introduction and discussed in more detail in \cite{paper1,paper2}, such terms can affect the spectrum of states associated with discontinuous modes.  This is particularly important for the term with $\kappa_2$.  For simplicity, below we will ignore these terms.

The equation of motion for $A_0^x$ is Gauss law
\ie\label{AGaussLaw}
{1\over g_{e1}^2 }  \partial_x  E^{xx}    = {1\over g_{e2}^2}  \partial_y \partial_z \mathbf E^{xx'}~.
\fe
The equations of motion for $A^{xx}$ and $\mathbf A^{xx'}$ are
\ie\label{Aeom}
 &{1\over g_{e1}^2 } \partial_0 E^{xx}   = {1\over g_m^2} \partial_y\partial_z \mathbf B^{x'}\,,\\
 &{1\over g_{e2}^2} \partial_0 \mathbf E^{xx'}  = {1\over g_m^2} \partial_x \mathbf B^{x'}~.
\fe
The field strengths also satisfy a Bianchi identity
\ie \label{eq:Bianchi}
\partial_0 \mathbf B^{x'} = \partial_x \mathbf E^{xx'} - \partial_y \partial_z E^{xx}\,.
\fe

\subsection{Fluxes}\label{sec:flux}
Let us put the theory on a Euclidean $4$-torus of lengths $\ell^\tau,\ell^x,\ell^y,\ell^z$. Consider the gauge field configuration with nontrivial transition function $g_{(\tau)}^{i}(x,y,z)$ of the form $\phi^i$ given by \eqref{generalwinding} at $\tau = \ell^\tau$.
 We have ($i\neq j\neq k$)
\ie
&A^{ii}(\tau+\ell^\tau,x,y,z) = A^{ii}(\tau,x,y,z) + \partial_i g_{(\tau)}^{i}(x,y,z)~,
\\
&\mathbf A^{ii'}(\tau+\ell^\tau,x,y,z) = \mathbf A^{ii'}(\tau,x,y,z) + \partial_j \partial_k g_{(\tau)}^{i}(x,y,z)~.
\fe
The gauge fields that have non-periodic boundary conditions are $A^{yy}$, $A^{zz}$ and $\mathbf A^{xx'}$.
For example, we can have
\ie
&A_0^i=0\,,~~~A^{xx}=0\,,\\
&A^{yy} = 2\pi {\tau\over \ell^\tau } \left[
 {1\over \ell^y}  \sum_{\gamma'}W^z_{\gamma'}\Theta(z-z_{\gamma'})    - {1\over\ell^y}\sum_\gamma M^z_\gamma \Theta(z-z_\gamma)
\right] \,,\\
&A^{zz} = 2\pi {\tau\over \ell^\tau}  \left[
{1\over \ell^z} \sum_{\beta'} W^y_{\beta'} \Theta(y-y_{\beta'})  - {1\over \ell^z} \sum_\beta M^y_\beta  \Theta(y-y_\beta)
\right]\,,\\
&\mathbf A^{xx'} = 2\pi {\tau\over \ell^\tau} \left[
{1\over \ell^y}   \sum_{\gamma'}W^z_{\gamma'}\delta(z-z_{\gamma'})    +{1\over\ell^y}\sum_\gamma M^z_\gamma \delta(z-z_\gamma)\right.\\
&\left.+
{1\over \ell^z} \sum_{\beta'} W^y_{\beta'} \delta(y-y_{\beta'})  + {1\over \ell^z} \sum_\beta M^y_\beta  \delta(y-y_\beta)- {2W^{yz}\over \ell^y\ell^z}
\right]\,,\\
&\mathbf A^{yy'}=\mathbf A^{zz'}=0\,,\\
&W^{yz} \equiv  \sum_{\beta'} W^y_{\beta'}  = \sum_\beta M^y_\beta
 = \sum_{\gamma'} W^z_{\gamma'}  = \sum_\gamma M^z_\gamma \,,~~~~~W^i_\alpha,M^i_\alpha\in \bZ\,.
\fe

More generally, every configuration of our gauge fields has quantized Euclidean electric fluxes
\ie
&e^{yy}(x,z) = \oint d\tau \oint dy~ E^{yy} \in 2\pi \mathbb Z~,
\\
&e^{zz}(x,y) = \oint d\tau \oint dz~ E^{zz} \in 2\pi \mathbb Z~,
\fe
The Bianchi identity \eqref{eq:Bianchi} implies
\ie
&\partial_x \partial_z e^{yy}(x,z) = 0~,\qquad &&\partial_x \partial_y e^{zz}(x,y) = 0~.
\fe
Therefore, $e^{yy}$ ($e^{zz}$) is a sum of functions of $x$ and $z$ ($x$ and $y$):
\ie
e^{yy}(x,z) = e^{yy}_x(x) + e^{yy}_z(z)~,\qquad e^{zz}(x,y) = e^{zz}_x(x) + e^{zz}_y(y)~.
\fe

Similarly, we define
\ie\label{eq:naive_elec_flux}
&\mathbf e^{xx'}_y(y_1,y_2) = \oint d\tau \int_{y_1}^{y_2} dy \oint dz~ \mathbf E^{xx'}~,
\\
&\mathbf e^{xx'}_z(z_1,z_2) = \oint d\tau \oint dy \int_{z_1}^{z_2} dz~ \mathbf E^{xx'}~.
\fe
These fluxes are independent of  $x$ because of the  Bianchi identity \eqref{eq:Bianchi}
\ie\label{eq:quant_elec_flux}
&\partial_x \mathbf e^{xx'}_y  = 0~,\qquad &&\partial_x \mathbf e^{xx'}_z = 0~,
\fe
The second set of quantized electric fluxes are
\ie
&e^{xx'}_{y\,-}(y_1,y_2) =\frac12 \left(\mathbf e^{xx'}_y(y_1,y_2) + e^{zz}_y(y_2) - e^{zz}_y(y_1) \right) \in 2\pi \mathbb Z~,
\\
&e^{xx'}_{z\,-}(z_1,z_2) = \frac12 \left(\mathbf e^{xx'}_z(z_1,z_2) + e^{yy}_z(z_2) - e^{yy}_z(z_1) \right) \in 2\pi \mathbb Z~.
\fe
We use a different font for $\mathbf e^{ii'}_j$ because not all the integer values of ${1\over 2\pi}\mathbf e^{ii'}_j$ are realized. For example, $\mathbf e^{ii'}_j(0,\ell^j)\in 4\pi \mathbb Z$.

Nontrivial magnetic fluxes are realized in a bundle whose transition function at $x=\ell^x$ is
\ie\label{eq:min_trans_function_1}
&g^x_{(x)}(y,z) = 2\pi \left[ \frac{y}{\ell^y} \Theta(z-z_0) + \frac{z}{\ell^z} \Theta(y-y_0) - \frac{yz}{\ell^y\ell^z} \right]~,
\\
&g^y_{(x)}(y,z) = 0~,
\\
&g^z_{(x)}(y,z) = -g^x_{(x)}(y,z)~,
\fe
and the transition function at $y=\ell^y$ is
\ie\label{eq:min_trans_function_2}
&g^x_{(y)}(x,z) = 0~,
\\
&g^y_{(y)}(x,z) = 2\pi \left[ \frac{x}{\ell^x} \Theta(z-z_0) + \frac{z}{\ell^z} \Theta(x-x_0) - \frac{xz}{\ell^x\ell^z} \right]~,
\\
&g^z_{(y)}(x,z) = -g^y_{(y)}(x,z)~.
\fe
These transition functions do not correspond to well-defined $\phi$ configurations in the $\phi$-theory. Nevertheless, we should consider them, since the configurations with these transition functions lead to the minimal energy state carrying the magnetic charge.  A similar situation was encountered in the $\hat A$-theory in \cite{paper2}.

An example of such a gauge field configuration is
\ie\label{eq:min_gauge_field}
&A^{xx} = A^{yy} = 0~,
\\
&A^{zz} = -2\pi \left[ \frac{y}{\ell^y\ell^z} \Theta(x-x_0) + \frac{x}{\ell^x\ell^z} \Theta(y-y_0) + \frac{xy}{\ell^x\ell^y} \delta(z-z_0) - 2 \frac{xy}{\ell^x\ell^y\ell^z} \right]~,
\\
&\mathbf A^{xx'} = \frac{2\pi x}{\ell^x} \left[ \frac{1}{\ell^y} \delta(z-z_0) + \frac{1}{\ell^z} \delta(y-y_0) - \frac{1}{\ell^y\ell^z} \right]~,
\\
&\mathbf A^{yy'} = \frac{2\pi y}{\ell^y} \left[ \frac{1}{\ell^x} \delta(z-z_0) + \frac{1}{\ell^z} \delta(x-x_0) - \frac{1}{\ell^x\ell^z} \right]~,
\\
&\mathbf A^{zz'} = -\frac{2\pi}{\ell^x\ell^y} \left[ \Theta(z-z_0) - \frac{z}{\ell^z} \right]~.
\fe
The transition functions \eqref{eq:min_trans_function_1} and \eqref{eq:min_trans_function_2} can be changed by a non-periodic gauge transformation. For example, the transformation
\ie
&\alpha^x = -2\pi \left[ {yz \over \ell^y \ell^z} \Theta(x-x_0) + {xz \over \ell^x \ell^z} \Theta(y-y_0) + {xy \over \ell^x \ell^y} \Theta(z-z_0) - 2{xyz \over \ell^x \ell^y \ell^z} \right]~,
\\
&\alpha^y = 0~,
\\
&\alpha^z = -\alpha^x~,
\fe
exchanges $x$ and $z$ in \eqref{eq:min_trans_function_1} and \eqref{eq:min_trans_function_2}. Although the transition functions are changed, the bundle remains the same.
Such gauge field configurations have nontrivial magnetic fields
\ie
&\mathbf B^{x'} = \frac{2\pi}{\ell^x} \left[ \frac{1}{\ell^y} \delta(z-z_0) + \frac{1}{\ell^z} \delta(y-y_0) - \frac{1}{\ell^y\ell^z} \right]~,
\\
&\mathbf B^{y'} = \frac{2\pi}{\ell^y} \left[ \frac{1}{\ell^x} \delta(z-z_0) + \frac{1}{\ell^z} \delta(x-x_0) - \frac{1}{\ell^x\ell^z} \right]~,
\\
&\mathbf B^{z'} = \frac{2\pi}{\ell^z} \left[ \frac{1}{\ell^x} \delta(y-y_0) + \frac{1}{\ell^y} \delta(x-x_0) - \frac{1}{\ell^x\ell^y} \right]~.
\fe
We define
\ie
&\mathbf{b}^{x'}_y(y_1,y_2) = \oint dx \int_{y_1}^{y_2} dy \oint dz~ \mathbf B^{x'} ~,
\\
&\mathbf{b}^{x'}_z(z_1,z_2) = \oint dx \oint dy \oint_{z_1}^{z_2} dz~ \mathbf B^{x'} ~,
\fe
and similarly in other directions. The Bianchi identity \eqref{eq:Bianchi} implies
\ie
\partial_\tau\mathbf b^{i'}_j = 0~.
\fe
The quantized magnetic fluxes are
\ie\label{eq:magneticflux_constraint}
b^{\pm}_x (x_1,x_2)=\frac12 \left( \mathbf b^{y'}_x(x_1,x_2) \pm\mathbf b^{z'}_x(x_1,x_2) \right) \in 2\pi \mathbb Z~.
\fe
and similarly in other directions. The magnetic fluxes are conserved.
We use a different font for $\mathbf b^{i'}_j$ because not all the integer values of ${1\over 2\pi}\mathbf b^{i'}_j$ are realized. For example, $\mathbf b^{i'}_j(0,\ell^j) \pm\mathbf b^{k'}_j(0,\ell^j) \in 4\pi \mathbb Z$.

\subsection{Global symmetry}

\subsubsection{Discrete symmetry}\label{sec:Adiscrete}

Similar to the discussion in Section \ref{sec:phidiscrete} for the $\phi^i$ theory, we can label the fields of the $A$ gauge theory with respect to different $S_4$ subgroups of the full global symmetry $\bZ_2^C\times G$.

In the $A$ gauge theory, we can take the gauge parameters $\alpha^i$ (and hence $A_0^i$) to be in one of the four representations of $\bZ_2^C\times G$ in \eqref{fourreps}, which are related by outer automorphisms.
Their $S_4$ representations are then
\\
\renewcommand{\arraystretch}{1.3}
\ie
\begin{tabular}{|c|c|c|c|c|}
\hline
$\bZ_2^C\times G$ & $\alpha,A_0$ & $A$, $E$ & $\mathbf A$, $\mathbf E$ & $\mathbf B$ \\
\hline
$(-,\mathbf{3}_2)$ & $\mathbf{3}$ & $\mathbf{1}\oplus\mathbf{2}$ & $\mathbf{1}'\oplus\mathbf{2}$ & $\mathbf{3}'$ \\
\hline
$(-,\mathbf{3}_2')$ & $\mathbf{3}'$ & $\mathbf{1}'\oplus\mathbf{2}$ & $\mathbf{1}\oplus\mathbf{2}$ & $\mathbf{3}$ \\
\hline
$(-,\mathbf{3}_3)$ & $\mathbf{1}'\oplus\mathbf{2}$ & $\mathbf{3}'$ & $\mathbf{3}$ & $\mathbf{1}\oplus\mathbf{2}$\\
\hline
$(-,\mathbf{3}_3')$ & $\mathbf{1}\oplus\mathbf{2}$ & $\mathbf{3}$ & $\mathbf{3}'$ & $\mathbf{1}'\oplus\mathbf{2}$ \\
\hline
\end{tabular}
\fe
\renewcommand{\arraystretch}{1}
\\
where we suppress the indices on the gauge fields and their field strengths.

\subsubsection{Electric symmetry}

Define the conserved currents for the electric symmetry as
\ie
\mathbf{1} \oplus\mathbf{2}: ~&\mathbf J_0^{xx}  =  {2\over g_{e1}^2} E^{xx}\,,\\
\mathbf{1}' \oplus\mathbf{2}: ~&J_0^{xx'} = {4\over g_{e2}^2} \mathbf E^{xx'} \,,\\
\mathbf{3}': ~&  J^{x'} = {2\over g_m^2}\mathbf B^{x'}~.
\fe
Here, we ignore the $\kappa_1$ and $\kappa_2$ terms of the Lagrangian \eqref{eq:A_Lag}.

The conservation equations follow from the equations of motion \eqref{Aeom}:
\ie\label{eq:curr_elec}
&  \partial_0 \mathbf J_0^{xx} =  \partial_y\partial_z J^{x'}\,,\\
 &\partial_0 J_0^{xx'}  = 2 \partial_x J^{x'}\,.
\fe
Gauss law \eqref{AGaussLaw} becomes a differential condition
\ie\label{eq:diff_cond_elec}
2\partial_x \mathbf J_0^{xx} =  \partial_y\partial_z J_0^{x'}\,.
\fe

Let us discuss the conserved charges.
From the second current conservation equation of \eqref{eq:curr_elec}, we have
\ie
Q^{xx'}(y,z) = \oint dx~J_0^{xx'}\,.
\fe
They can be written as a sum of functions of $y$ and $z$ using the differential condition \eqref{eq:diff_cond_elec}:
\ie
Q^{xx'}(y,z) = Q^{xx'}_y(y) + Q^{xx'}_z(z)\,.
\fe
We will soon see that $Q^{ii'}_{j}$ are  integer-valued functions.
There is a gauge ambiguity in these charges:
\ie
(Q^{xx'}_y(y), Q^{xx'}_z(z)) \sim (Q^{xx'}_y(y)+  n^{xx'}, Q^{xx'}_z(z)-  n^{xx'})
\fe
with constant $n^{xx'}$.
In addition, we also have the following conserved charges
\ie
\mathbf{Q}^{xx}_y(y) =\oint dz~\mathbf J_0^{xx}\,,~~~~~\mathbf{Q}^{xx}_z(z) = \oint dy~\mathbf J_0^{xx}\,,
\fe
which are independent of $x$ because of the differential condition \eqref{eq:diff_cond_elec}.
As we will soon see, the properly quantized $U(1)$ charges are actually the following combinations:
\ie\label{electricQm}
Q^{xx}_{y\, - } (y )  &= \mathbf{Q}^{xx}_y(y) -\frac12 \partial_yQ^{zz'}\\
& =\oint dz~\mathbf J_0^{xx}  -  \frac12   \partial_y \oint dz~J_0^{zz'} \,,\\
Q^{xx}_{z\, - } (z)  &= \mathbf{Q}^{xx}_z(z) -\frac12 \partial_z Q^{yy'}\\
& =\oint dy~\mathbf J_0^{xx}  -  \frac12   \partial_z \oint dy~J_0^{yy'} \,.
\fe
More precisely, we will show that $Q^{ii}_{j\,-}$ are sums of delta functions with integer coefficients.
They are subject to the constraint
\ie
\oint dy~Q^{xx}_{y\, -}(y) = \oint dz~Q^{xx}_{z\,-}(z)\,.
\fe

In summary, the conserved charges are $Q^{ii'}_j(x^j)$ and $Q^{ii}_{j\, - } (x^j ) $.  On a lattice, there are $2L^x+2L^y+2L^z-3$ conserved charges of the form $Q^{ii'}_j(x^j)$, and  $2L^x+2L^y+2L^z-3$ conserved charges of the form $Q^{ii}_{j\, - } (x^j ) $.

We will also sometimes use the conserved charges
\ie
&Q^{xx}_{y\, + } (y ) =  Q^{xx}_{y\, -  } (y )  +  \partial_y Q^{zz'}  =\oint dz~\mathbf J_0^{xx}  +  \frac12   \partial_y  \oint dz~J_0^{zz'} \,,\\
&Q^{xx}_{z\, + } (z )  =  Q^{xx}_{y\, -  } (z )  +  \partial_z Q^{yy'}  =\oint dy~\mathbf J_0^{xx}  +  \frac12   \partial_z  \oint dy~J_0^{yy'}\,.
\fe

The first symmetry operator is a line operator
\ie
\mathcal U^{zz'}(\beta;x,y) =\exp\left[-i\beta  \left(\, Q^{zz'}_x(x)+Q^{zz'}_y(y)\, \right) \right] =  \exp \left[ -i \frac{4\beta}{g_{e2}^2}\oint dz~ \mathbf E^{zz'} \right]~.
\fe
The charged operator is a strip operator
\ie
&W^{zz'}_{x\,-}(z,x_1,x_2) = \exp \left[ \frac i2 \int_{x_1}^{x_2} dx \oint dy~ \left( \mathbf A^{zz'} - \partial_x A^{yy} \right) \right]~,\\
&W^{zz'}_{y\,-}(z,y_1,y_2) = \exp \left[ \frac i2 \int_{y_1}^{y_2} dy \oint dx~ \left( \mathbf A^{zz'} - \partial_y A^{xx} \right) \right]~.
\fe
Only integers powers of $W^{zz'}_{x\,-}$ and $W^{zz'}_{y\,\pm}$ are invariant under the large gauge transformation \eqref{generalwinding}. The operators $\mathcal U^{zz'}$ and $W^{zz'}_{y\,-}$ satisfy the commutation relations
\ie
\mathcal U^{zz'}(\beta;x,y)  \, W^{zz'}_{y\,-}(z,y_1,y_2) = e^{-i\beta} \, W^{zz'}_{y\,-}(z,y_1,y_2)\,  \mathcal U^{zz'}(\beta;x,y)~,\quad\text{if }~y_1<y<y_2~,
\fe
and they commute otherwise.

The second symmetry operator is a strip operator
\ie
\mathcal U^{xx}_{y\,-}(\beta;x,y_1,y_2)
&= \exp \left[ -i\beta \int_{y_1}^{y_2} dy \, Q^{xx}_{y\,-}   \right]\\
&= \exp \left[ -i\beta \int_{y_1}^{y_2} dy \oint dz~ \left( \frac{2}{g_{e1}^2} E^{xx} - \frac{2}{g_{e2}^2} \partial_y \mathbf E^{zz'} \right) \right]~.
\fe
The charged operator is a line operator
\ie
W^{xx}(y,z) = \exp \left[ i \oint dx~ A^{xx} \right]~.
\fe
Only integers powers of $W^{xx}$ are invariant under the large gauge transformation \eqref{eq:windingzeromode}. The operators $\mathcal U^{xx}_{y\,-}$ and $W^{xx}$ satisfy the commutation relations
\ie
\mathcal U^{xx}_{y\,-}(\beta;y_1,y_2)  \, W^{xx}(y,z) = e^{-i\beta}  \,  W^{xx}(y,z)  \, \mathcal U^{xx}_{y\,-}(\beta;y_1,y_2)~,\quad\text{if }~y_1<y<y_2~,
\fe
and they commute otherwise.

We can also define
\ie
\mathcal U^{xx}_{y\,+}(\beta;x,y_1,y_2) = \mathcal U^{zz'}(\beta;x,y_1)^{-1}\,  \mathcal U^{xx}_{y\,-}(\beta;x,y_1,y_2) \,\mathcal U^{zz'}(\beta;x,y_2)~,
\fe
and
\ie
W^{zz'}_{y\,+}(z,y_1,y_2) = W^{xx}(z,y_1)^{-1}  \, W^{zz'}_{y\,-}(z,y_1,y_2)  \,  W^{xx}(z,y_2)^{}~.
\fe
The operators $\mathcal U^{xx}_{y\,\pm}$ and $W^{zz'}_{y\mp}$ satisfy the commutation relations
\ie
&\mathcal U^{xx}_{y\,\pm}(\beta;x,y_1,y_2)  \, W^{zz'}_{y\,\mp}(z,y_3,y_4)\\
&=
\begin{cases}
e^{\mp i\beta} \, W^{zz'}_{y\,\mp}(z,y_3,y_4) \, \mathcal U^{xx}_{y\,\pm}(\beta;x,y_1,y_2)~,\qquad &\text{if }~y_1<y_3<y_2<y_4~,
\\
e^{\pm i\beta} \, W^{zz'}_{y\,\mp}(z,y_3,y_4) \, \mathcal U^{xx}_{y\,\pm}(\beta;x,y_1,y_2)~,\qquad &\text{if }~y_3<y_1<y_4<y_2~,
\end{cases}
\fe
and they commute otherwise.

\subsubsection{Magnetic symmetry}
Define the conserved currents for the magnetic symmetry as
\ie
\mathbf{3}': ~&  J_0^{x'} = \frac 1{4\pi} \mathbf B^{x'}\,,\\
\mathbf{1} \oplus\mathbf{2}: ~&\mathbf J^{xx}  = \frac 1{4\pi} E^{xx}\,,\\
\mathbf{1}' \oplus\mathbf{2}: ~&J^{xx'} = \frac 1{4\pi} \mathbf E^{xx'} \,.
\fe
The conservation equation follows from the Bianchi identity \eqref{eq:Bianchi}:
\ie
\partial_0 J_0^{x'} = \partial_x J^{xx'} - \partial_y \partial_z \mathbf J^{xx}\,.
\fe

The conserved charge operator for the magnetic symmetry is
\ie
Q_{x'}^\pm(x) = \oint dy dz (J_0^{y'} \pm J_0^{z'})\,.
\fe
These charges satisfy the constraint
\ie
&\oint dy~ \left(Q^+_{y'}(y) - Q^-_{y'}(y)\right) = \oint dz~ \left(Q^+_{z'}(z) + Q^-_{z'}(z)\right)\,,\\
&\oint dz~ \left(Q^+_{z'}(z) - Q^-_{z'}(z)\right) = \oint dx~ \left(Q^+_{x'}(x) + Q^-_{x'}(x)\right)\,,\\
&\oint dx~ \left(Q^+_{x'}(x) - Q^-_{x'}(x)\right) = \oint dy~ \left(Q^+_{y'}(y) + Q^-_{y'}(y)\right)\,.
\fe

The magnetically charged objects are local point operators, which can be thought of as monopole operators. They can be written as $e^{i\tilde \phi^{i'}}$ in terms of the dual field $\tilde \phi^{i'}$ (see Section \ref{sec:duality}).

\subsection{Electric modes}\label{sec:elec_modes}

The electric modes arise from field configurations with vanishing classical energy.  Therefore, their magnetic field vanishes.

We analyze these modes by first picking the temporal gauge $A_0^i =0$.
Let us focus on $\mathbf A^{xx'}$ and $A^{xx}$; fields in the other directions work in a similar way.
The vanishing of $\mathbf B^{x'}  =\partial_x \mathbf A^{xx'}   - \partial_y\partial_z A^{xx}$ implies that  $\mathbf A^{xx'} = \mathbf f(y,z)$ and $A^{xx} =f_1(x,y)+f_2(x,z)$. Using gauge transformations, we can bring them to the form:
\ie
&A^{xx}  =  \frac{1}{\ell^x} f^{xx}_y(y) + \frac{1}{\ell^x} f^{xx}_z(z)\,,
\\
&\mathbf A^{xx'}  = \frac{2}{\ell^z} \mathbf f^{xx'}_y(y) + \frac{2}{\ell^y} \mathbf f^{xx'}_z(z)\,.
\fe
And similarly for the other directions.

We now quantize these electric modes. There is a gauge ambiguity due to zero modes,
\ie\label{eq:elec_gauge_amb}
&f^{xx}_y(t,y) \sim f^{xx}_y(t,y) + \ell^x c^{xx}(t)\,,\qquad &&f^{xx}_z(t,z) \sim f^{xx}_z(t,z) - \ell^x c^{xx}(t)\,,
\\
&\mathbf f^{xx'}_y(t,y) \sim \mathbf f^{xx'}_y(t,y) + \ell^z \mathbf c^{xx'}(t)\,,\qquad &&\mathbf f^{xx'}_z(t,z) \sim \mathbf f^{xx'}_z(t,z) - \ell^y \mathbf c^{xx'}(t)\,,
\fe
and similarly in other directions. Let us define gauge invariant variables
\ie
&\bar {\mathbf f}^{xx'}_y(t,y) = \mathbf f^{xx'}_y(t,y) + \frac{1}{\ell^y} \oint dz~ \mathbf f^{xx'}_z(t,z)\,,
\\
&\bar {\mathbf f}^{xx'}_z(t,z) = \mathbf f^{xx'}_z(t,z) + \frac{1}{\ell^z} \oint dy~ \mathbf f^{xx'}_y(t,y)\,.
\fe
They satisfy a constraint
\ie\label{eq:elec_constraint}
\oint dy~ \bar {\mathbf f}^{xx'}_y(t,y) = \oint dz~ \bar {\mathbf f}^{xx'}_z(t,z)\,.
\fe

The Lagrangian for these modes is
\ie
L &= \frac{1}{g_{e1}^2\ell^x} \left[ \ell^z \oint dy~ (\dot f^{xx}_y)^2 + \ell^y \oint dz~ (\dot f^{xx}_z)^2 + 2\oint dy dz~ \dot f^{xx}_y \dot f^{xx}_z  \right]\,
\\
& + \frac{4\ell^x}{g_{e2}^2\ell^y\ell^z} \left[ \ell^y \oint dy~ (\dot{\bar {\mathbf f}}^{xx'}_y)^2 + \ell^z \oint dz~ (\dot{\bar {\mathbf f}}^{xx'}_z)^2 -\oint dy dz~ \dot{\bar {\mathbf f}}^{xx'}_y \dot{\bar {\mathbf f}}^{xx'}_z \right]\,.
\fe
The conjugate momenta are
\ie
&\pi^{xx}_y = \frac{2}{g_{e1}^2\ell^x} \left( \ell^z \dot f^{xx}_y + \oint dz~ \dot f^{xx}_z \right)\,,\quad &&\pi^{xx}_z = \frac{2}{g_{e1}^2\ell^x} \left( \ell^y \dot f^{xx}_z + \oint dy~ \dot f^{xx}_y \right)\,,
\\
&\bar \Pi^{xx'}_y = \frac{4\ell^x}{g_{e2}^2\ell^y\ell^z} \left( 2\ell^y \dot{\bar {\mathbf f}}^{xx'}_y - \oint dz~ \dot{\bar {\mathbf f}}^{xx'}_z \right)\,,\quad &&\bar \Pi^{xx'}_z = \frac{4\ell^x}{g_{e2}^2\ell^y\ell^z} \left( 2\ell^z \dot{\bar {\mathbf f}}^{xx'}_z - \oint dy~ \dot{\bar {\mathbf f}}^{xx'}_y \right)\,.
\fe
The gauge ambiguity \eqref{eq:elec_gauge_amb} in $f$ implies the constraint
\ie
\oint dy~\pi^{xx}_y(y) = \oint dz~\pi^{xx}_z(z)\,.
\fe
On the other hand, the constraint \eqref{eq:elec_constraint} on $\bar g$ implies the gauge ambiguity
\ie\label{eq:momentumgaugeambiguity}
\bar \Pi^{xx'}_y(y) \sim \bar \Pi^{xx'}_y(y) +   n^{yz}\,, \qquad \bar \Pi^{xx'}_z(z) \sim \bar \Pi^{xx'}_z(z) -   n^{yz}\,,
\fe
We shall show in a moment that the conjugate momenta are quantized to be half-integers.

The charges of the electric symmetry in terms of the conjugate momenta are
\ie
&\qquad Q^{xx'}(y,z) = \frac{4}{g_{e2}^2} \oint dx~ \mathbf E^{xx'} = \bar \Pi^{xx'}_y(y) + \bar \Pi^{xx'}_z(z)\,,\\
\mathbf Q^{xx}_y(y) &= \frac{2}{g_{e1}^2} \oint dz~ E^{xx} = \pi^{xx}_y(y)\,, \qquad \mathbf Q^{xx}_z(z) = \frac{2}{g_{e1}^2} \oint dy~ E^{xx} = \pi^{xx}_z(z)\,,
\fe
where in the first line we used the constraint \eqref{eq:elec_constraint}.

The Hamiltonian for these modes is
\ie
H &= \frac{g_{e1}^2 \ell^x}{4\ell^y \ell^z} \left[ \ell^y \oint dy~ (\pi^{xx}_y)^2 + \ell^z \oint dz~ (\pi^{xx}_z)^2 - \oint dy dz~ \pi^{xx}_y \pi^{xx}_z \right]
\\
& + \frac{g_{e2}^2}{16\ell^x} \left[ \ell^z \oint dy~ (\bar \Pi^{xx'}_y)^2 + \ell^y \oint dz~ (\bar \Pi^{xx'}_z)^2  + 2 \oint dy dz~ \bar \Pi^{xx'}_y \bar \Pi^{xx'}_z \right]\,.
\fe

Consider a large gauge transformation of the form \eqref{generalwinding}. It shifts the gauge fields by
\ie \label{eq:large_gauge}
A^{xx} &\sim A^{xx}\,,\\
A^{yy} &\sim A^{yy}  +{2\pi\over \ell^y} \sum_{\gamma'} W^z_{\gamma'}\Theta(z-z_{\gamma'}) - {2\pi\over \ell^y}\sum_\gamma M^z_\gamma \Theta(z-z_\gamma)\,,\\
A^{zz} &\sim A^{zz}  +{2\pi\over \ell^z} \sum_{\beta'} W^y_{\beta'}\Theta(y-y_{\beta'})-{2\pi\over \ell^z} \sum_\beta M^y_\beta\Theta(y-y_\beta)\,,\\
\mathbf A^{xx'} &\sim \mathbf A^{xx'}  +{2\pi\over \ell^y} \sum_{\gamma'} W^z_{\gamma'}\delta(z-z_{\gamma'}) +{2\pi\over \ell^y}\sum_\gamma M^z_\gamma \delta(z-z_\gamma)\\
&
\qquad +{2\pi\over \ell^z} \sum_{\beta'} W^y_{\beta'}\delta(y-y_{\beta'}) + {2\pi\over \ell^z} \sum_\beta M^y_\beta\delta(y-y_\beta)
-  {4\pi \over \ell^y\ell^z}W^{yz}\,,\\
\mathbf A^{yy'}&\sim \mathbf A^{yy'}\,,~~~~~~\\
\mathbf A^{zz'}&\sim \mathbf A^{zz'}\,.
\fe
The corresponding shifts in the electric modes are
\ie
&f^{zz}_y(t,y) \sim f^{zz}_y(t,y) + 2\pi \sum_{\beta'} W^y_{\beta'}\Theta(y-y_{\beta'})- 2\pi \sum_\beta M^y_\beta\Theta(y-y_\beta)\,,
\\
&f^{yy}_z(t,z) \sim f^{yy}_z(t,z) + 2\pi \sum_{\gamma'} W^z_{\gamma'}\Theta(z-z_{\gamma'}) - 2\pi\sum_\gamma M^z_\gamma \Theta(z-z_\gamma)\,,
\\
&\bar {\mathbf f}^{xx'}_y(t,y) \sim \bar {\mathbf f}^{xx'}_y(t,y) + \pi \sum_{\beta'} W^y_{\beta'}\delta(y-y_{\beta'}) + \pi \sum_\beta M^y_\beta\delta(y-y_\beta)\,,
\\
&\bar {\mathbf f}^{xx'}_z(t,z) \sim \bar {\mathbf f}^{xx'}_z(t,z) + \pi \sum_{\gamma'} W^z_{\gamma'}\delta(z-z_{\gamma'}) +\pi \sum_\gamma M^z_\gamma \delta(z-z_\gamma)\,.
\fe
Consider a particular large gauge transformation with $W^y_{\beta'}=M^y_\beta=1$ at the same point $y_0$ and $W^z_{\gamma'}=M^z_\gamma=1$ at the same point $z_0$.  It shifts the electric modes by
\ie
&f^{zz}_y(t,y) \sim f^{zz}_y(t,y) \,,
\\
&f^{yy}_z(t,z) \sim f^{yy}_z(t,z) \,,
\\
&\bar {\mathbf f}^{xx'}_y(t,y) \sim \bar {\mathbf f}^{xx'}_y(t,y) + 2\pi \delta(y-y_0)\,,
\\
&\bar {\mathbf f}^{xx'}_z(t,z) \sim \bar {\mathbf f}^{xx'}_z(t,z) + 2\pi \delta(z-z_0)\,,
\fe
We obtain the following identification on $\bar{{\mathbf f}}^{ii'}_j$
\ie
&\bar {\mathbf f}^{xx'}_y(t,y) \sim \bar {\mathbf f}^{xx'}_y(t,y) + 2\pi \delta(y-y_0)\,,
\\
&\bar {\mathbf f}^{xx'}_z(t,z) \sim \bar {\mathbf f}^{xx'}_z(t,z) + 2\pi \delta(z)\,,
\fe
for each $y_0$, and
\ie
&\bar {\mathbf f}^{xx'}_y(t,y) \sim \bar {\mathbf f}^{xx'}_y(t,y)\,,
\\
&\bar {\mathbf f}^{xx'}_z(t,z) \sim \bar {\mathbf f}^{xx'}_z(t,z) + 2\pi \delta(z-z_0) - 2\pi \delta(z)\,,
\fe
for each $z_0$. This implies that $\bar{\Pi}^{xx'}_y(y)$ is a half-integer function up to the gauge ambiguity \eqref{eq:momentumgaugeambiguity}. We can consider a more general large gauge transformation with $W^y_{\beta'}=M^y_\beta=1$ at    $y_{\beta'}$ and $y_\beta$, respectively:
\ie
&f^{zz}_y(t,y) \sim f^{zz}_y(t,y)+2\pi [\Theta(y-y_{\beta'})-\Theta(y-y_\beta)] \,,
\\
&\bar {\mathbf f}^{xx'}_y(t,y) \sim \bar {\mathbf f}^{xx'}_y(t,y) +\pi [\delta(y-y_{\beta'})+\delta(y-y_\beta)]\,,
\fe
This implies that the following combination of conjugate momenta is an integer function
\begin{equation}\label{eq:momenta_relation}
\int_{y_{\beta'}}^{y_\beta} dy~\pi^{zz}_y(y)+\frac12 \bar{\Pi}^{xx'}_y(y_\beta)+\frac12\bar{\Pi}^{xx'}_y(y_{\beta'})~.
\end{equation}
Taking the differential of the above combination with respect to $y_\beta$ or $y_{\beta'}$ implies that
\begin{equation}
Q^{zz}_{y\,\pm}(y) =   \pi^{zz}_y(y)\pm \frac12\partial_y\bar{\Pi}^{xx'}_y(y)~,
\end{equation}
is a sum of delta functions with integer coefficients.
This explains the quantization of the charges defined in \eqref{electricQm}.

\bigskip\bigskip\centerline{\it General charged states}\bigskip

It is convenient to parametrize the conjugate momenta as
\ie\label{generalelectric}
&Q^{zz'}_y(y)= \bar{\Pi}^{zz'}_y(y)=  \sum_{\beta'} \widetilde{W}^y_{\beta'} \Theta(y-y_{\beta'})  - \sum_\beta \widetilde{M}^y_\beta \Theta(y-y_\beta)  ~,
\\
&Q^{yy'}_z(z)= \bar{\Pi}^{yy'}_z(z)= \sum_{\gamma'} \widetilde{W}^z_{\gamma'} \Theta(z-z_{\gamma'})  - \sum_\gamma \widetilde{M}^z_\gamma \Theta(z-z_\gamma) ~,
\\
&\mathbf{Q}^{xx}_y(y) = \pi^{xx}_y(y)=\frac{1}{2}\left(\sum_{\beta'} \widetilde{W}^y_{\beta'} \delta(y-y_{\beta'})  + \sum_\beta \widetilde{M}^y_\beta \delta(y-y_\beta)\right)~,
\\
&\mathbf{Q}^{xx}_z(z) =  \pi^{xx}_z(z)=\frac{1}{2}\left(\sum_{\gamma'} \widetilde{W}^z_{\gamma'} \delta(z-z_{\gamma'})  + \sum_\gamma \widetilde{M}^z_\gamma \delta(z-z_\gamma)\right)~,
\fe
where, the integers $\widetilde W^i_\alpha,\widetilde M^i_\alpha\in \bZ$ satisfy the constraint
\ie
\widetilde W^{yz} \equiv  \sum_{\beta'}\widetilde W^y_{\beta'}  = \sum_\beta\widetilde M^y_\beta = \sum_{\gamma'}\widetilde W^z_{\gamma'}  = \sum_\gamma\widetilde M^z_\gamma \,.
\fe
The $Q^{ii}_{j\,-}$ charges are
\ie
&Q^{xx}_{y\,-}(y)=\sum_\beta \widetilde{M}^y_\beta \delta(y-y_\beta)~,
\\
&Q^{xx}_{z\,-}(z)=\sum_\gamma \widetilde{M}^z_\gamma \delta(z-z_\gamma)~.
\fe
The momenta in the other directions are parametrized similarly.
On a lattice, we have $4L^x+4L^y+4L^z-9$ such electric modes.

In addition to the above electric modes, we also have three zero modes  parametrized as
\ie
Q^{xx'}(y,z) &= Q^{xx'}_y(y) + Q^{xx'}_z(z)\\
&= \bar \Pi^{xx'}_y(y) + \bar \Pi^{xx'}_z(z) =\widetilde N^{xx'} ~,\\
Q^{yy'}(z,x)& = Q^{yy'}_z(z) + Q^{yy'}_x(x)\\
& = \bar\Pi^{yy'}_z(z) + \bar\Pi^{yy'}_x(x)=\widetilde N^{yy'} ~,\\
Q^{zz'}(x,y) &= Q^{zz'}_x(x) + Q^{zz'}_y(y)\\
&= \bar \Pi^{zz'}_x(x) + \bar\Pi^{zz'}_y(y)=\widetilde N^{zz'} ~,\\
 Q^{ii}_{j\,-}(x^j)& = 0~,
\fe
with $\widetilde N^{kk'} \in \mathbb Z$.
Combining with the modes \eqref{generalelectric} (and their counterparts along the other directions), we end up with $4L^x+4L^y+4L^z -6$ electric modes.

\subsection{Magnetic modes}\label{sec:magnetic}

\bigskip\bigskip\centerline{\it Minimally charged states}\bigskip

A minimal magnetic charge is realized in the bundle with the transition functions \eqref{eq:min_trans_function_1} at $x=\ell^x$ and \eqref{eq:min_trans_function_2} at $y=\ell^y$.
The gauge field configuration \eqref{eq:min_gauge_field} that realizes these transitions functions have magnetic charges
\ie
&Q^+_{x'}(x)=\delta(x-x_0)~,\quad &&Q^+_{y'}(y)=\delta(y-y_0)~,\quad &&Q^+_{z'}(z)=\delta(z-z_0)~,
\\
&Q^-_{x'}(x)=0~,\quad &&Q^-_{y'}(y)=0~,\quad &&Q^-_{z'}(z)=0~.
\fe

Other minimal magnetic charge configurations are realized in the bundles whose transition function at $x=\ell^x$ is
\ie
&g^x_{(x)}(y,z) = 2\pi \epsilon^x \left[ \frac{y}{\ell^y} \Theta(z-z_0) + \frac{z}{\ell^z} \Theta(y-y_0) - \frac{yz}{\ell^y\ell^z} \right]~,
\\
&g^y_{(x)}(y,z) = 0~,
\\
&g^z_{(x)}(y,z) = -2\pi \epsilon^z \left[ \frac{y}{\ell^y} \Theta(z-z_0) + \frac{z}{\ell^z} \Theta(y-y_0) - \frac{yz}{\ell^y\ell^z} \right]~,
\fe
and the transition function at $y=\ell^y$ is
\ie
&g^x_{(y)}(x,z) = 0~,
\\
&g^y_{(y)}(x,z) = 2\pi \epsilon^y \left[ \frac{x}{\ell^x} \Theta(z-z_0) + \frac{z}{\ell^z} \Theta(x-x_0) - \frac{xz}{\ell^x\ell^z} \right]~,
\\
&g^z_{(y)}(x,z) = -2\pi \epsilon^z \left[ \frac{x}{\ell^x} \Theta(z-z_0) + \frac{z}{\ell^z} \Theta(x-x_0) - \frac{xz}{\ell^x\ell^z} \right]~,
\fe
where $\epsilon^i = \pm 1$. This is realized by a gauge field configuration given by
\ie
&A^{xx} = A^{yy} = 0~,
\\
&A^{zz} =- 2\pi \epsilon^z \left[ \frac{y}{\ell^y\ell^z} \Theta(x-x_0) + \frac{x}{\ell^x\ell^z} \Theta(y-y_0) + \frac{xy}{\ell^x\ell^y} \delta(z-z_0) - 2 \frac{xy}{\ell^x\ell^y\ell^z} \right]~,
\\
&\mathbf A^{xx'} = \frac{2\pi \epsilon^x x}{\ell^x} \left[ \frac{1}{\ell^y} \delta(z-z_0) + \frac{1}{\ell^z} \delta(y-y_0) - \frac{1}{\ell^y\ell^z} \right]~,
\\
&\mathbf A^{yy'} = \frac{2\pi \epsilon^y y}{\ell^y} \left[ \frac{1}{\ell^x} \delta(z-z_0) + \frac{1}{\ell^z} \delta(x-x_0) - \frac{1}{\ell^x\ell^z} \right]~,
\\
&\mathbf A^{zz'} =- \frac{2\pi \epsilon^z}{\ell^x\ell^y} \left[ \Theta(z-z_0) - \frac{z}{\ell^z} \right]~.
\fe
Its magnetic fields are
\ie
&\mathbf B^{x'} = \frac{2\pi\epsilon^x}{\ell^x} \left[ \frac{1}{\ell^y} \delta(z-z_0) + \frac{1}{\ell^z} \delta(y-y_0) - \frac{1}{\ell^y\ell^z} \right]~,
\\
&\mathbf B^{y'} = \frac{2\pi\epsilon^y}{\ell^y} \left[ \frac{1}{\ell^x} \delta(z-z_0) + \frac{1}{\ell^z} \delta(x-x_0) - \frac{1}{\ell^x\ell^z} \right]~,
\\
&\mathbf B^{z'} = \frac{2\pi\epsilon^z}{\ell^z} \left[ \frac{1}{\ell^x} \delta(y-y_0) + \frac{1}{\ell^y} \delta(x-x_0) - \frac{1}{\ell^x\ell^y} \right]~.
\fe
Its magnetic charge is
\ie
&Q^\pm_{x'}(x)=\frac12(\epsilon^y \pm \epsilon^z)\delta(x-x_0)~,
\\
&Q^\pm_{y'}(y)=\frac12(\epsilon^z \pm \epsilon^x)\delta(y-y_0)~,
\\
&Q^\pm_{z'}(z)=\frac12(\epsilon^x \pm \epsilon^y)\delta(z-z_0)~.
\fe

The minimal energy with any of the above minimal magnetic charges is
\ie
H &= {1\over g_m^2}\oint dx \oint dy \oint dz~\sum_{i } \mathbf B^{i'}\mathbf B_{i'}
\\
&={4\pi^2 \over g_m^2 \ell^x \ell^y \ell^z} \left[ 2(\ell^x+\ell^y+\ell^z)\delta(0) - 3 \right]~.
\fe

\bigskip\bigskip\centerline{\it General charged states}\bigskip

A more general gauge field configuration carrying magnetic charges is
\ie
A^{xx} &= A^{yy} = 0~,
\\
A^{zz} &= -2\pi \left[ \frac{y}{\ell^y\ell^z} \sum_\alpha N^+_{x\,\alpha}\Theta(x-x_\alpha) - \frac{y}{\ell^y\ell^z} \sum_{\alpha'} N^-_{x\,\alpha'}\Theta(x-x_{\alpha'}) + \frac{x}{\ell^x\ell^z} \sum_\beta N^+_{y\,\beta}\Theta(y-y_\beta) \right.
\\
&\qquad \qquad \left. + \frac{x}{\ell^x\ell^z} \sum_{\beta'} N^-_{y\,\beta'}\Theta(y-y_{\beta'}) + \frac{xy}{\ell^x\ell^y} \sum_\gamma N^+_{z\,\gamma}\delta(z-z_\gamma) + \frac{xy}{\ell^x\ell^y} \sum_{\gamma'} N^-_{z\,\gamma'} \delta(z-z_{\gamma'}) \right.
\\
&\qquad \qquad \left. + \frac{xy}{\ell^x\ell^y} (N^z-N^x)\delta(z-z_0) - 2N^z \frac{xy}{\ell^x\ell^y\ell^z} \right]~,
\\
\mathbf A^{xx'} &= \frac{2\pi x}{\ell^x} \left[\frac{1}{\ell^z} \sum_\beta N^+_{y\,\beta} \delta(y-y_\beta) - \frac{1}{\ell^z} \sum_{ \beta'} N^-_{y\,\beta'} \delta(y-y_{\beta'}) \right.
\\
& \qquad \qquad \left. + \frac{1}{\ell^y}  \sum_\gamma N^+_{z\,\gamma} \delta(z-z_\gamma) + \frac{1}{\ell^y} \sum_{ \gamma'} N^-_{z\,\gamma'} \delta(z-z_{\gamma'}) - \frac{N^x}{\ell^y\ell^z} \right]~,
\\
\mathbf A^{yy'} &= \frac{2\pi y}{\ell^y} \left[\frac{1}{\ell^x} \sum_\gamma N^+_{z\,\gamma} \delta(z-z_\gamma) - \frac{1}{\ell^x} \sum_{ \gamma'} N^-_{z\,\gamma'} \delta(z-z_{\gamma'}) \right.
\\
& \qquad \qquad \left. + \frac{1}{\ell^z}  \sum_\alpha N^+_{x\,\alpha} \delta(x-x_\alpha) + \frac{1}{\ell^z} \sum_{ \alpha'} N^-_{x\,\alpha'} \delta(x-x_{\alpha'}) - \frac{N^y}{\ell^x\ell^z} \right]~,
\\
\mathbf A^{zz'} &= -\frac{2\pi}{\ell^x\ell^y} \left[ \sum_\gamma N^+_{z\,\gamma} \Theta(z-z_\gamma) + \sum_{\gamma'} N^-_{z\,\gamma'} \Theta(z-z_{\gamma'}) + (N^z-N^x)\Theta(z-z_0) - N^z \frac{z}{\ell^z} \right]~,
\fe
where $N^+_{i\,\alpha},N^-_{i\,\alpha'}\in\mathbb{Z}$ satisfy the constraints
\ie
&N^{x} = \sum_\beta N^+_{y\,\beta} - \sum_{ \beta'} N^-_{y\,\beta'} = \sum_\gamma N^+_{z\,\gamma} + \sum_{ \gamma'} N^-_{z\,\gamma'}~,
\\
&N^{y} = \sum_\gamma N^+_{z\,\gamma} - \sum_{ \gamma'} N^-_{z\,\gamma'} = \sum_\alpha N^+_{x\,\alpha} + \sum_{\alpha'} N^-_{x\,\alpha'}~,
\\
&N^{z} = \sum_\alpha N^+_{x\,\alpha} - \sum_{\alpha'} N^-_{x\,\alpha'} = \sum_\beta N^+_{y\,\beta} + \sum_{ \beta'} N^-_{y\,\beta'}~.
\fe

Its bundle is characterized by the transition function at $x=\ell^x$
\ie
g^x_{(x)}(y,z) &= 2\pi \left[\frac{z}{\ell^z} \sum_\beta N^+_{y\,\beta} \Theta(y-y_\beta) - \frac{z}{\ell^z} \sum_{ \beta'} N^-_{y\,\beta'} \Theta(y-y_{\beta'}) \right.
\\
& \qquad \qquad \left. + \frac{y}{\ell^y}  \sum_\gamma N^+_{z\,\gamma} \Theta(z-z_\gamma) + \frac{y}{\ell^y} \sum_{ \gamma'} N^-_{z\,\gamma'} \Theta(z-z_{\gamma'}) - N^x \frac{yz}{\ell^y\ell^z} \right]~,
\\
g^y_{(x)}(y,z) &= 0~,
\\
g^z_{(x)}(y,z) &= -2\pi \left[\frac{z}{\ell^z} \sum_\beta N^+_{y\,\beta} \Theta(y-y_\beta) + \frac{z}{\ell^z} \sum_{ \beta'} N^-_{y\,\beta'} \Theta(y-y_{\beta'}) + \frac{y}{\ell^y}  \sum_\gamma N^+_{z\,\gamma} \Theta(z-z_\gamma) \right.
\\
&\qquad \qquad \left. + \frac{y}{\ell^y}  \sum_{\gamma'} N^-_{z\,\gamma'} \Theta(z-z_{\gamma'}) + \frac{y}{\ell^y}  (N^z-N^x)\Theta(z-z_0) - N^z \frac{yz}{\ell^y\ell^z} \right]~,
\fe
and the transition function at $y=\ell^y$
\ie
g^x_{(y)}(x,z) &= 0~,
\\
g^y_{(y)}(x,z) &= 2\pi \left[\frac{x}{\ell^x} \sum_\gamma N^+_{z\,\gamma} \Theta(z-z_\gamma) - \frac{x}{\ell^x} \sum_{ \gamma'} N^-_{z\,\gamma'} \Theta(z-z_{\gamma'}) \right.
\\
& \qquad \qquad \left. + \frac{z}{\ell^z}  \sum_\alpha N^+_{x\,\alpha} \Theta(x-x_\alpha) + \frac{z}{\ell^z} \sum_{ \alpha'} N^-_{x\,\alpha'} \Theta(x-x_{\alpha'}) - N^y \frac{xz}{\ell^x\ell^z} \right]~,
\\
g^z_{(y)}(x,z) &= -2\pi \left[ \frac{z}{\ell^z}  \sum_\alpha N^+_{x\,\alpha} \Theta(x-x_\alpha) - \frac{z}{\ell^z} \sum_{ \alpha'} N^-_{x\,\alpha'} \Theta(x-x_{\alpha'}) + \frac{x}{\ell^x} \sum_\gamma N^+_{z\,\gamma} \Theta(z-z_\gamma) \right.
\\
&\qquad \qquad \left. + \frac{x}{\ell^x}  \sum_{\gamma'} N^-_{z\,\gamma'} \Theta(z-z_{\gamma'}) + \frac{x}{\ell^x}  (N^z-N^x)\Theta(z-z_0) - N^z \frac{xz}{\ell^x\ell^z} \right]~.
\fe
The transition functions are chosen such that their transitions functions satisfy a more general periodicity than the one stated in \eqref{eq:period_phi}, but still are part of an FCC lattice.
Its magnetic fields are
\ie
\mathbf B^{x'} &= \frac{2\pi}{\ell^x} \left[ \frac{1}{\ell^z} \sum_\beta N^+_{y\,\beta} \delta(y-y_\beta) - \frac{1}{\ell^z} \sum_{ \beta'} N^-_{y\,\beta'} \delta(y-y_{\beta'}) \right.
\\
&\qquad \qquad \left. + \frac{1}{\ell^y}  \sum_\gamma N^+_{z\,\gamma} \delta(z-z_\gamma) + \frac{1}{\ell^y} \sum_{ \gamma'} N^-_{z\,\gamma'} \delta(z-z_{\gamma'}) - \frac{N^x}{\ell^y\ell^z} \right]~,
\fe
and similarly in the other directions. Its magnetic charge is
\ie
&Q^+_{x'}(x)=\sum_\alpha N^+_{x\,\alpha} \delta(x-x_\alpha)~,\qquad &&Q^-_{x'}(x)=\sum_{\alpha'} N^-_{x\,\alpha'} \delta(x-x_{\alpha'})~,
\\
&Q^+_{y'}(y)=\sum_\beta N^+_{y\,\beta} \delta(y-y_\beta)~,\qquad &&Q^-_{y'}(y)=\sum_{ \beta'} N^-_{y\,\beta'} \delta(y-y_{\beta'})~,
\\
&Q^+_{z'}(z)=\sum_\gamma N^+_{z\,\gamma} \delta(z-z_\gamma)~,\qquad &&Q^-_{z'}(z)=\sum_{ \gamma'} N^-_{z\,\gamma'} \delta(z-z_{\gamma'})~,
\fe

The minimal energy with this magnetic charge is
\ie
H ={4\pi^2 \over g_m^2 \ell^x \ell^y \ell^z} \left[ 2 \sum_{i} \ell^i \left(\sum_{\alpha}(N^+_{i\,\alpha})^2 + \sum_{\alpha'}(N^-_{i\,\alpha'})^2 \right)\delta(0) - \sum_{i} (N^i)^2  \right]~,
\fe
which is infinite in the continuum limit.

\subsection{Fractons and lineons as defects}

We first define $A_0^{IJ}$  in the same way as $\phi^{IJ}$ \eqref{philowerupper} and \eqref{phiupperlower}:
\ie\label{Aupperlower}
&A_0^{XY} =\frac12( A_0^x + A_0^y -A_0^z)\,,\\
&A_0^{ZX} =\frac12( A_0^z + A_0^x -A_0^y)\,,\\
&A_0^{YZ} =\frac12( A_0^y + A_0^z -A_0^x)\,,
\fe
or conversely,
\ie\label{Alowerupper}
&A_0^x  = A_0^{XY} +A_0^{ZX}\,,\\
&A_0^y  = A_0^{YZ} +A_0^{XY}\,,\\
&A_0^z  = A_0^{YZ} +A_0^{ZX}\,.
\fe
And similarly for $\alpha^{IJ}$.  Their gauge transformations are
\ie
&A_0^{XY} \sim A_0^{XY} +\partial_0 \alpha^{XY}\,,\\
&A^{xx} \sim A^{xx} +\partial_x(\alpha^{XY} +\alpha^{ZX} )\,,\\
&\mathbf A^{xx'}\sim \mathbf A^{xx'}  +\partial_y\partial_z  (\alpha^{XY} +\alpha^{ZX})\,.
\fe

Let us start with gauge-invariant defects at a fixed point in space and extend in time.
We have three such defects:
\ie
&\exp \left[  i \int_{-\infty}^\infty dt ~A_0^{XY} \right]\,,\\
&\exp \left[  i  \int_{-\infty}^\infty dt ~A_0^{ZX} \right]\,,\\
&\exp \left[  i  \int_{-\infty}^\infty dt ~A_0^{YZ} \right]\,.
\fe
The exponent of, say,  the first defect is quantized by a large gauge transformation $\alpha^{IJ}$ that winds around the Euclidean time direction
\ie
\alpha^{XY} = 2\pi {\tau\over\ell^\tau}  \,,~~~\alpha^{YZ} =\alpha^{ZX}=0\,.
\fe

The defect $\exp \left[  i \int_{-\infty}^\infty dt ~A_0^{IJ} \right]$ cannot move by itself -- it  is a fracton.

The composite defects
\ie
&\exp \left[  i \int_{-\infty}^\infty dt~  (A_0^{XY} +A_0^{ZX}) \right] \,,\\
&\exp \left[  i \int_{-\infty}^\infty dt~  (A_0^{XY} +A_0^{YZ}) \right] \,,\\
&\exp \left[  i \int_{-\infty}^\infty dt~  (A_0^{YZ} +A_0^{ZX}) \right] \,,
\fe
by contrast can move in the $x,y,z$ directions, respectively.
These are lineons.
The motion of these lineons is captured by the defects
\ie
&\exp \left[  i \int_{{\cal C}^x} \left(  dt~  (A_0^{XY} +A_0^{ZX})    +dx~ A^{xx}\right) \right] \,,\\
&\exp \left[  i \int_{{\cal C}^y} \left(  dt~  (A_0^{XY} +A_0^{YZ})    +dy~ A^{yy}\right) \right] \,,\\
&\exp \left[  i \int_{{\cal C}^z} \left(  dt~  (A_0^{YZ} +A_0^{ZX})    +dz~ A^{zz}\right) \right] \,,\\
\fe
where ${\cal C}^i$ is a spacetime curve in $(t,x^i)$.

\section{Duality}\label{sec:duality}

In this section, we will show that the $\phi$ theory of Section \ref{sec:tetrphitheory} is dual to the $A$ theory of Section \ref{sec:tetrgaugetheory}.  We will work in  Euclidean signature.

 Many of  the expressions below  contain several equations related by cyclically permuting $x,y,z$. In some of these expressions, we will write only one of the  equations to avoid cluttering.

We will start with the $A$ gauge theory in the following $S_4$ presentations
 \ie
&\text{gauge fields}: ~~A_0^i :~\mathbf{3} ,  ~~ A^{ii}:~\mathbf{1}\oplus\mathbf{2} ,~~ \mathbf A^{ii'}:~\mathbf{1}'\oplus\mathbf{2}\\
&\text{field strengths}: ~~ \mathbf B^{i'}:~\mathbf{3}',~~E^{ii}:~\mathbf{1}\oplus \mathbf{2} ,~~\mathbf E^{ii'}:~\mathbf{1}'\oplus \mathbf{2}
\fe
The Lagrangian is
\ie
\mathcal L_E &= {1\over g_{e1}^2}\sum_i   E^{ii} E_{ii}  + {1\over g_{e2}^2 }\sum_{i}   \mathbf E^{ii'}\mathbf E_{ii'} + {1\over g_m^2}\sum_{i} \mathbf B^{i'}\mathbf B_{i'}
\\
&+ {i \over 4\pi} \sum_i \widetilde {\mathbf B}_{ii} \left(\partial_\tau A^{ii} - \partial_i A_\tau^i - E^{ii} \right) \\
&+ {i \over 4\pi} \sum_{i \ne j \ne k} \widetilde B_{ii'} \left( \partial_\tau \mathbf A^{ii'} - \partial_j \partial_k A_\tau^i - \mathbf E^{ii'} \right)
\\
&+ {i \over 4\pi} \sum_{i \ne j \ne k} \widetilde E_{i'} \left(  \partial_j \partial_k A^{ii} - \partial_i \mathbf A^{ii'} + \mathbf B^{i'} \right)~,
\fe
where $E^{ii}$, $\mathbf E^{ii'}$, $\mathbf B^{i'}$, $\widetilde {\mathbf B}_{ii}$, $\widetilde B_{ii'}$ and $\widetilde E_{i'}$ are independent fields. If we integrate out the Lagrangian multipliers $\widetilde {\mathbf B}_{ii}$, $\widetilde B_{ii'}$ and $\widetilde E_{i'}$, we get back the original Lagrangian \eqref{eq:A_Lag}.

On the other hand, if we integrate out $E^{xx}$, $\mathbf E^{xx'}$ and $\mathbf B^{x'}$, we get
\ie
E^{xx} = {ig_{e1}^2 \over 8\pi}\widetilde {\mathbf B}_{xx}~, \qquad \mathbf E^{xx'} = {ig_{e2}^2 \over 8\pi}\widetilde B_{xx'}~,\qquad \mathbf B^{x'} = {ig_{m}^2 \over 8\pi}\widetilde E_{x'}~.
\fe
The Lagrangian then becomes
\ie
\mathcal L_E &= {g_{e1}^2 \over 64\pi^2}\sum_i   \widetilde {\mathbf B}^{ii} \widetilde {\mathbf B}_{ii}  + {g_{e2}^2 \over 64\pi^2}\sum_{i}   \widetilde B^{ii'} \widetilde B_{ii'} + {g_m^2 \over 64\pi^2} \sum_{i} \widetilde E^{i'} \widetilde E_{i'}
\\
&+ {i \over 4\pi} \sum_i \widetilde {\mathbf B}_{ii} \left(\partial_\tau A^{ii} - \partial_i A_\tau^i\right) \\
&+ {i \over 4\pi} \sum_{i \ne j \ne k} \widetilde B_{ii'} \left( \partial_\tau \mathbf A^{ii'} - \partial_j \partial_k A_\tau^i \right)
\\
&+ {i \over 4\pi} \sum_{i \ne j \ne k} \widetilde E_{i'} \left( \partial_j \partial_k A^{ii}-\partial_i \mathbf A^{ii'}  \right)~.
\fe
Next, integrating over the gauge fields gives the constraints
\ie
&\partial_x \widetilde {\mathbf B}_{xx} - \partial_y \partial_z\widetilde B_{xx'} = 0~,
\\
&\partial_\tau \widetilde {\mathbf B}_{xx} - \partial_y \partial_z \widetilde E_{x'} = 0~,
\\
&\partial_\tau \widetilde B_{xx'} - \partial_x \widetilde E_{x'} = 0~.
\fe
These constraints are locally solved by a field $\tilde \phi^{i'}$ in the $\mathbf{3}'$:
\ie
\mathbf{1}\oplus\mathbf{2}:~~~&\widetilde {\mathbf B}_{xx} = \partial_y\partial_z \tilde \phi^{x'}~,
\\
\mathbf{1}'\oplus\mathbf{2}:~~~&\widetilde B_{xx'} = \partial_x \tilde \phi^{x'}~,
\\
\mathbf{3}':~~~&\widetilde E_{x'} = \partial_\tau \tilde \phi^{x'}~.
\fe
The Euclidean Lagrangian can be written in terms of $\tilde \phi^{i'}$:
\ie
\mathcal L_E = {g_m^2 \over 64\pi^2} \sum_{i} (\partial_\tau \tilde \phi^{i'})^2 + {g_{e2}^2 \over 64\pi^2}\sum_{i} (\partial_i \tilde \phi^{i'})^2 + {g_{e1}^2 \over 64\pi^2}\sum_{i \ne j \ne k\atop \text{cyclic}}   (\partial_j \partial_k \tilde \phi^{i'})^2.
\fe
This is the theory  in Section \ref{sec:tetrphitheory}, presented using different $S_4$ representations related to the ones there by an outer automorphism.
Specifically,  these are the $S_4$ representations in the second row of \eqref{phirep}.

The nontrivial fluxes of $E^{ii},\mathbf E^{ii'},\mathbf B^{i'}$ (see Section \ref{sec:flux}) mean that the periods of  $\widetilde E_{i'} ,\widetilde {\mathbf B}_{ii},\widetilde B_{ii'}$ are quantized, corresponding to the periodicities of $\tilde\phi^i$ in \eqref{eq:period_phi}.

Going back to  Lorentzian signature, we have
\ie
E^{xx} = -{g_{e1}^2 \over 8\pi}\partial_y \partial_z\tilde \phi^{x'}~, \qquad \mathbf E^{xx'} = -{g_{e2}^2 \over 8\pi}\partial_x \tilde \phi^{x'}~,\qquad \mathbf B^{x'} = {g_{m}^2 \over 8\pi}\partial_0 \tilde \phi^{x'}~.
\fe
The  Lagrangian is
\ie
\mathcal L = {g_m^2 \over 64\pi^2} \sum_{i } (\partial_0 \tilde \phi^{i'})^2 - {g_{e2}^2 \over 64\pi^2}\sum_{i } (\partial_i \tilde \phi^{i'})^2 - {g_{e1}^2 \over 64\pi^2}\sum_{i \ne j \ne k \atop \text{cyclic}}   (\partial_j \partial_k \tilde \phi^{i'})^2.
\fe
Comparing with \eqref{eq:Lag_generic}, and ignoring the terms with $K_1$ and $K_2$, the duality maps
\ie\label{dualitycoupling}
\mu_0 = {g_m^2 \over 32\pi^2}~,\qquad \mu = {g_{e2}^2 \over 32\pi^2}~,\qquad \mu_1 = {g_{e1}^2 \over 32\pi^2}~.
\fe

Under the duality, the momentum charges of the $\tilde \phi$ theory are mapped to the magnetic charges of the $A$ theory as
\ie
\tilde Q^\pm_x(x)  =   \mu_0 \oint dy dz\, (  \partial_0 \tilde \phi^{y'} \pm  \partial_0 \tilde \phi^{z'})
~  \longleftrightarrow ~
 Q^\pm_x(x) = {1\over 4\pi}  \oint dydz\, (\mathbf B^{y'}\pm \mathbf B^{z'})~.
\fe
In particular, the quantization of the momentum charges $\tilde Q^\pm_x(x)$ (discussed in Section \ref{sec:momentum}) matches with the quantization of magnetic charges $ Q^\pm_x(x)$ (discussed in Section \ref{sec:magnetic}).

Similarly, the winding charges of the $\tilde \phi$ theory are mapped to the electric charges of the $A$ theory as
\ie
&\tilde Q^{xx'}(y,z) = \frac{1}{2\pi} \oint dx~ \partial_x \tilde \phi^{x'}
~\longleftrightarrow~
 Q^{xx'}(y,z) = \frac{4}{g_{e2}^2} \oint dz~ \mathbf E^{xx'}~,\\
&\widetilde{ \mathbf{Q}}^{xx}_y(y) = \frac{1}{4\pi} \oint dz~ \partial_y \partial_z \tilde \phi^{x'}
~\longleftrightarrow~
 \mathbf{Q}^{xx}_y(y) = \frac{2}{g_{e1}^2} \oint dz~ E^{xx}~.
\fe
In particular, the quantization of the winding charges  $\tilde Q^{xx'}(y,z),\widetilde{ \mathbf{Q}}^{xx}_y(y) $ (discussed in Section \ref{sec:winding}) matches with the quantization of electric  charges $ Q^{xx'}(y,z),    \mathbf{Q}^{xx}_y(y) $ (discussed in Section \ref{sec:elec_modes}).

\section{The $\bZ_N$ theory}\label{ZNgaugetheory}

\subsection{The $\bZ_N$ FCC lattice gauge theory and the $\bZ_N$ checkerboard model}\label{sec:ZNlattice}

\bigskip\bigskip\centerline{\it $\mathbb{Z}_N$ FCC lattice gauge theory}\bigskip

We now discuss a $\mathbb{Z}_N$ version of the $U(1)$ FCC lattice gauge theory of Section \ref{sec:Alattice}. But unlike the discussion there, here we will follow a Hamiltonian formalism with a spatial lattice.

We start with a three-dimensional face-centered cubic lattice. The shortest distance between two sites is $a/\sqrt{2}$ and $x^i=a\hat x^i$. The lattice sites at the corners are labeled by integers $(\hat x, \hat y, \hat z)$, while those on the faces are labeled by $(\hat x+\frac{1}{2}, \hat y+\frac{1}{2}, \hat z)$, $(\hat x+\frac{1}{2}, \hat y, \hat z+\frac{1}{2})$, $(\hat x, \hat y+\frac{1}{2}, \hat z+\frac{1}{2})$ with integer $\hat x,\hat y, \hat z$.

The gauge parameters are $\mathbb{Z}_N$ phases placed on the sites. We will label them as
\ie\label{eq:ZN_gauge_loc}
&\eta_{s}^{XYZ}\quad &&\text{ if } s\in (\bZ,\bZ,\bZ)\,,\\
&\eta_{s}^{XY}\quad &&\text{ if }s\in (\bZ+\tfrac 12,\bZ+\tfrac 12 , \bZ)\,,\\
&\eta_{s}^{YZ}\quad &&\text{ if }s\in (\bZ,\bZ+\tfrac 12 , \bZ+\tfrac 12)\,,\\
&\eta_{s}^{ZX}\quad &&\text{ if }s\in (\bZ+\tfrac 12,\bZ , \bZ+\tfrac 12)\,.
\fe
The spatial gauge fields are $\mathbb{Z}_N$ phases placed at the tetrahedra, which we label by their center coordinates $c$. These $\bZ_N$ phases are denoted by
\ie
&U_c\quad &&\text{ if } c\in\(\mathbb{Z}- \tfrac14 ,\mathbb{Z}- \tfrac14,\mathbb{Z}-\tfrac14\)~,\\
&U_c^X\quad &&\text{ if } c\in\(\mathbb{Z}- \tfrac14,\mathbb{Z}+\tfrac14,\mathbb{Z}+\tfrac14\)~,\\
&U_c^Y\quad &&\text{ if } c\in\(\mathbb{Z}+\tfrac14,\mathbb{Z}-\tfrac14,\mathbb{Z}+\tfrac14\)~,\\
&U_c^Z\quad &&\text{ if } c\in\(\mathbb{Z}+\tfrac14,\mathbb{Z}+\tfrac14,\mathbb{Z}-\tfrac14\)~,\\
&U_c^{XY}\quad &&\text{ if } c\in\(\mathbb{Z}-\tfrac14,\mathbb{Z}-\tfrac14,\mathbb{Z}+\tfrac14\)~,\\
&U_c^{YZ}\quad &&\text{ if } c\in\(\mathbb{Z}+\tfrac14,\mathbb{Z}-\tfrac14,\mathbb{Z}-\tfrac14\)~,\\
&U^{ZX}_c\quad &&\text{ if } c\in\(\mathbb{Z}-\tfrac14,\mathbb{Z}+\tfrac14,\mathbb{Z}-\tfrac14\)~,\\
&U^{XYZ}_c\quad &&\text{ if } c\in\(\mathbb{Z}+\tfrac14,\mathbb{Z}+\tfrac14,\mathbb{Z}+\tfrac14\)~.
\fe

The gauge transformations multiply the spatial gauge fields by the gauge parameters on the vertices of the tetrahedron
\ie\label{eq:ZN_spatial_gauge_trans}
&U_c \rightarrow U_c \left(\eta^{XYZ}_s \eta^{XY}_{s+\left(-\frac12,-\frac12,0\right)} \eta^{YZ}_{s+\left(0,-\frac12,-\frac12\right)} \eta^{ZX}_{s+\left(-\frac12,0,-\frac12\right)}\right)^{-1}~,
\fe
where $s$ is a corner on the lattice (i.e., $s\in(\bZ,\bZ,\bZ)$), and $c$ is the center of one of the eight tetrahedra around $s$. They are related by $s=c+\left(\frac14,\frac14,\frac14\right)$. The gauge transformations act on the other spatial gauge fields in a similar way.

Let $V$ be the conjugate momenta for $U$. They obey the $\mathbb{Z}_N$ clock and shift algebra $UV=e^{2\pi i/N}VU$. Similarly, we define the conjugate momenta $V^X$, $V^Y$, $V^Z$, $V^{XY}$, $V^{YZ}$, $V^{ZX}$ and $V^{XYZ}$. Since we use a Hamiltonian formalism, we pick the temporal gauge, and then Gauss law is imposed as an operator equation on the states. For every site $s$ with integer $(\hat x,\hat y,\hat z)$, there are four Gauss laws
\ie
&G_s^{++}=V_c V^X_{c+(0,\frac12,\frac12)} V^Y_{c+(\frac12,0,\frac12)} V^Z_{c+(\frac12,\frac12,0)}  V^{XY}_{c+(0,0,\frac12)}    V^{YZ}_{c+(\frac12,0,0)}    V^{ZX}_{c+(0,\frac12,0)}    V^{XYZ}_{c+(\frac12,\frac12,\frac12)}  =1
\\
&G_s^{-+}=V_c V^X_{c+(0,-\frac12,\frac12)} V^Y_{c+(-\frac12,0,\frac12)} V^Z_{c+(-\frac12,-\frac12,0)}
 V^{XY}_{c+(0,0,\frac12)}    V^{YZ}_{c+(-\frac12,0,0)}     V^{ZX}_{c+(0,-\frac12,0)}    V^{XYZ}_{c+(-\frac12,-\frac12,\frac12)}  =1
  \\
&G_s^{--}=V_cV^X_{c+(0,-\frac12,-\frac12)}V^Y_{c+(\frac12,0,-\frac12)}V^Z_{c+(\frac12,-\frac12,0)}  V^{XY}_{c+(0,0,-\frac12)}    V^{YZ}_{c+(\frac12,0,0)}    V^{ZX}_{c+(0,-\frac12,0)}     V^{XYZ}_{c+(\frac12,-\frac12,-\frac12)}  =1
  \\
&G_s^{+-}=V_c V^X_{c+(0,\frac12,-\frac12)} V^Y_{c+(-\frac12,0,-\frac12)} V^Z_{c+(-\frac12,\frac12,0)}  V^{XY}_{c+(0,0,-\frac12)}  V^{YZ}_{c+(-\frac12,0,0)}  V^{ZX}_{c+(0,\frac12,0)}  V^{XYZ}_{c+(-\frac12,\frac12,-\frac12)} =1~.
\fe
The site $s$ is the related to the center of the tetrahedron by $s=c+(\frac14,\frac14,\frac14)$.

Let us discuss the gauge invariant local terms in the Hamiltonian. For every site $s$ with integer $(\hat x,\hat y,\hat z)$, there are four terms
\ie
&L_s^{++}=U_c U^X_{c+(0,\frac12,\frac12)} U^Y_{c+(\frac12,0,\frac12)} U^Z_{c+(\frac12,\frac12,0)} \left(U^{XY}_{c+(0,0,\frac12)} U^{YZ}_{c+(\frac12,0,0)}U^{ZX}_{c+(0,\frac12,0)} U^{XYZ}_{c+(\frac12,\frac12,\frac12)}\right)^{-1}
\\
&L_s^{-+}=U_c U^X_{c+(0,-\frac12,\frac12)} U^Y_{c+(-\frac12,0,\frac12)} U^Z_{c+(-\frac12,-\frac12,0)}
\left(U^{XY}_{c+(0,0,\frac12)} U^{YZ}_{c+(-\frac12,0,0)}  U^{ZX}_{c+(0,-\frac12,0)} U^{XYZ}_{c+(-\frac12,-\frac12,\frac12)}\right)^{-1}
  \\
&L_s^{--}=U_cU^X_{c+(0,-\frac12,-\frac12)}U^Y_{c+(\frac12,0,-\frac12)}U^Z_{c+(\frac12,-\frac12,0)} \left(U^{XY}_{c+(0,0,-\frac12)} U^{YZ}_{c+(\frac12,0,0)} U^{ZX}_{c+(0,-\frac12,0)}  U^{XYZ}_{c+(\frac12,-\frac12,-\frac12)}\right)^{-1}
  \\
&L_s^{+-}=U_c U^X_{c+(0,\frac12,-\frac12)} U^Y_{c+(-\frac12,0,-\frac12)} U^Z_{c+(-\frac12,\frac12,0)} \left(U^{XY}_{c+(0,0,-\frac12)} U^{YZ}_{c+(-\frac12,0,0)} U^{ZX}_{c+(0,\frac12,0)} U^{XYZ}_{c+(-\frac12,\frac12,-\frac12)}\right)^{-1}~.
\fe
Each term is a product of eight tetrahedra whose centers form a cube of size $a/2$. The Hamiltonian is a sum of the above interactions over the integer sites. It is
\ie\label{ZNlatticef}
H=-\frac{1}{\hat g_e^2}\sum_{\text{tetrahedra}} V_c - \frac{1}{\hat{g}_m^2}\sum_{\text{integer }s} (L_s^{++}+L_s^{-+}+L_s^{--}+L_s^{+-})+c.c.~.
\fe
In the first sum the conjugate momenta should be decorated with appropriate superscripts according to the position of the tetrahedron. Here we suppress the superscripts.

Alternatively, we can impose Gauss law energetically by adding another term to the Hamiltonian:
\ie\label{nogauss}
H=&-\frac{1}{\hat g_e^2}\sum_{\text{tetrahedra}} V_c - \frac{1}{\hat{g}_m^2}\sum_{\text{integer $s$}} (L_s^{++}+L_s^{-+}+L_s^{--}+L_s^{+-})
\\
&- \frac{1}{\hat{g}^2}\sum_{\text{integer $s$}} (G_s^{++}+G_s^{-+}+G_s^{--}+G_s^{+-})+c.c.~.
\fe

\bigskip\bigskip\centerline{\it $\mathbb{Z}_N$ checkerboard model}\bigskip

Similar to the lattice models of Section \ref{sec:XYtetra} and Section \ref{sec:Alattice}, we can equivalently formulate the $\bZ_N$ FCC lattice gauge theory on  the checkerboard of a cubic lattice.
As illustrated in Figure \ref{fig:U1-CB}, the sites and the tetrahedra of the FCC lattice are mapped to the shaded cubes and the sites of the checkerboard, respectively.
As a result, the $\bZ_N$ phase variables $U$ and $V$ are placed on the sites of the checkerboard.
The gauge transformations and the gauge invariant interactions $L^{\pm\pm}_s$ are now associated with the shaded cubes of the checkerboard.
We emphasize that this reformulation from the FCC lattice to the checkerboard is not a duality transformation.
It is simply a reassignment of the same fields and the same interactions to different geometric objects on a different lattice.

Let us perform this reformulation more explicitly.
We start with a cubic lattice with lattice spacing $a/2$ and label the sites by integer or half-integers $(\hat x,\hat y,\hat z)$ (see Figure \ref{fig:U1-CB}(b)). At every site, there is a $\mathbb{Z}_N$ phase variable $U_s$ and its conjugate momentum $V_s$. These objects satisfy the clock and shift algebra at every site
\ie\label{ZNcomm}
&U_sV_s=e^{2\pi i/N}V_sU_s~,\\
&U_s^N=V_s^N=1~.
\fe
and commute at different sites.

\begin{figure}
\begin{subfigure}{.5\textwidth}
\centering
\includegraphics[scale=0.3]{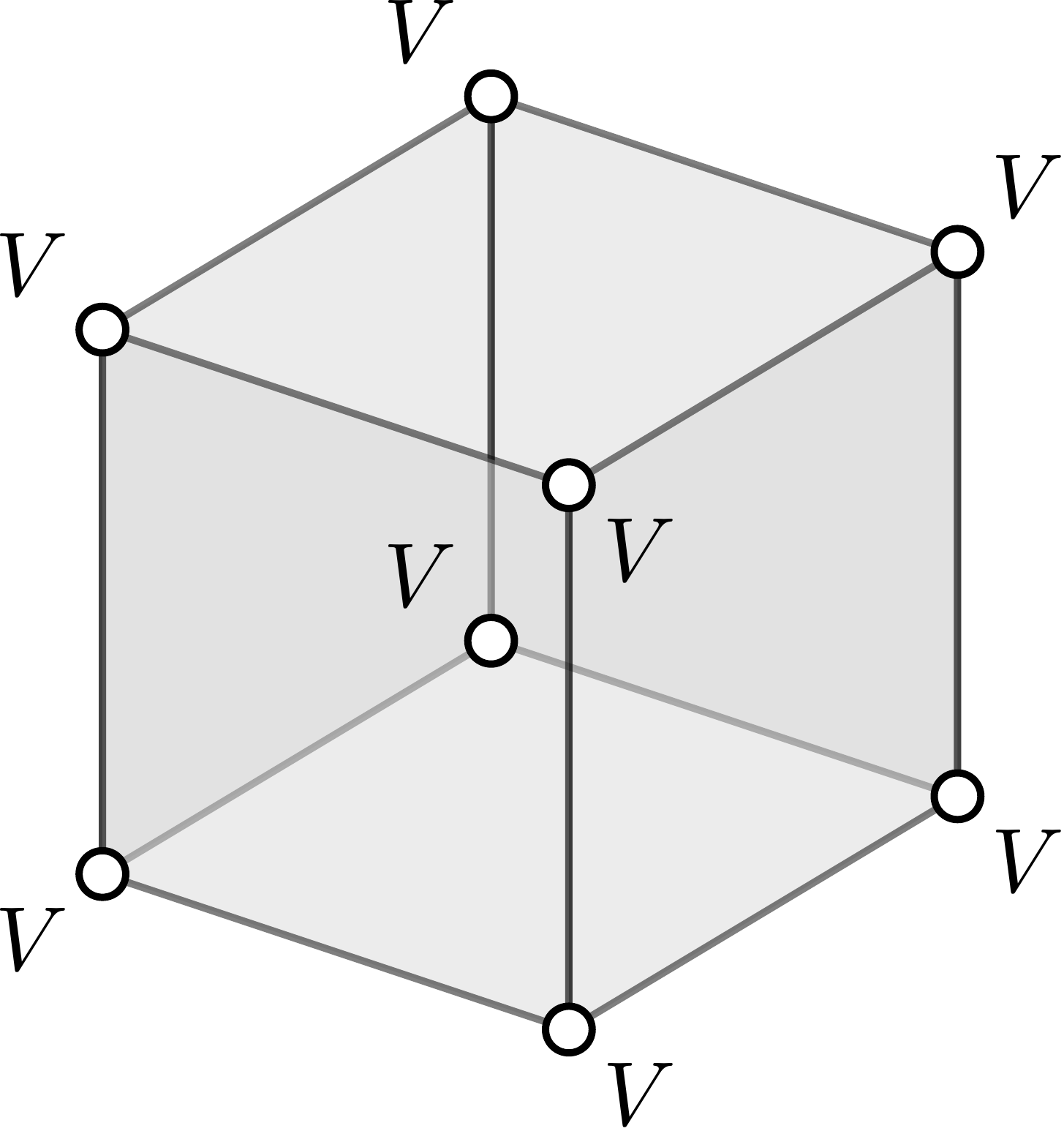}
\caption{}
\end{subfigure}
\begin{subfigure}{.5\textwidth}
\centering
\includegraphics[scale=0.3]{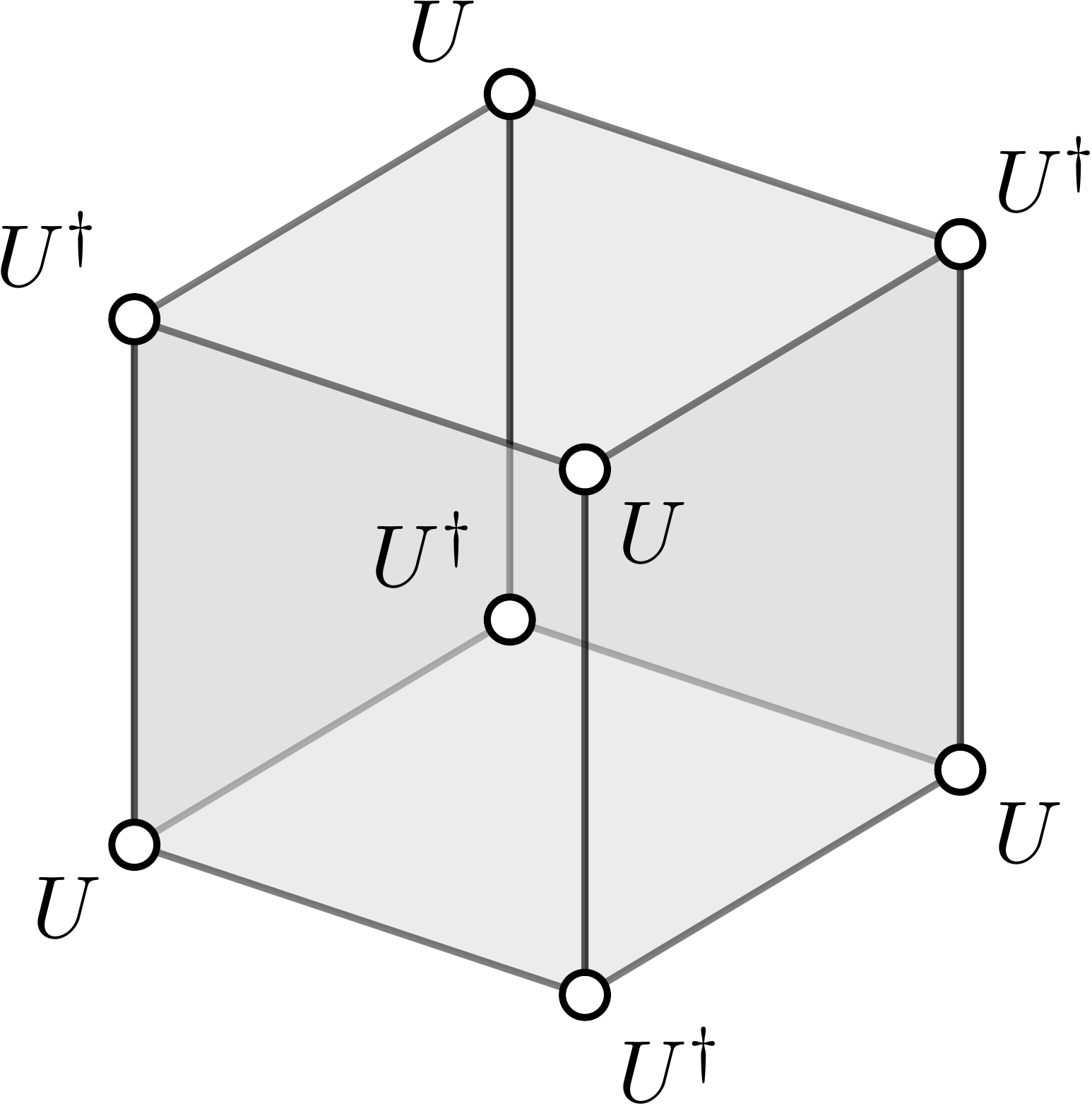}
\caption{}
\end{subfigure}
\caption{(a) The $G_c$ (Gauss law) term, and (b) the $L_c$ term in the Hamiltonian \eqref{checkerboardH} of the $\mathbb Z_N$ checkerboard model. The cube shown here is any shaded cube of the checkerboard in Figure \ref{fig:U1-CB}(b).}\label{fig:ZN-CB}
\end{figure}

The Hamiltonian consists of two kinds of interactions (see Figure \ref{fig:ZN-CB}). The first kind is a product of $V$ around a cube
\begin{equation}\label{latticeG}
G_c=V_s V_{s+(\frac{1}{2},0,0)} V_{s+(0,\frac{1}{2},0)} V_{s+(0,0,\frac{1}{2})} V_{s+(\frac{1}{2},\frac{1}{2},0)} V_{s+(\frac{1}{2},0,\frac{1}{2})} V_{s+(0,\frac{1}{2},\frac{1}{2})} V_{s+(\frac{1}{2},\frac{1}{2},\frac{1}{2})}~,
\end{equation}
where $c$ denotes the cube these eights sites surround.
The second kind is an oriented product of $U$ around a cube
\begin{equation}\label{latticeL}
L_c=U_s U_{s+(\frac{1}{2},0,0)}^{-1} U_{s+(0,\frac{1}{2},0)}^{-1} U_{s+(0,0,\frac{1}{2})}^{-1} U_{s+(\frac{1}{2},\frac{1}{2},0)} U_{s+(\frac{1}{2},0,\frac{1}{2})} U_{s+(0,\frac{1}{2},\frac{1}{2})} U_{s+(\frac{1}{2},\frac{1}{2},\frac{1}{2})}^{-1}~.
\end{equation}

The Hamiltonian is a sum over the interactions $G_c$ and $L_c$ only over the shaded cubes (gray, green, blue, red) of the checkerboard in Figure \ref{fig:U1-CB}(b):
\begin{equation}\label{checkerboardH}
H=- \frac{1}{\hat{g}_m^2}\sum_{c\in\text{shaded}\atop\text{cubes}} L_c- \frac{1}{\hat{g}^2}\sum_{c\in\text{shaded}\atop\text{cubes}} G_c+c.c.~.
\end{equation}
Note that all the terms in the Hamiltonian commute with each other.\footnote{We can replace some of the $U$ and $V$ in \eqref{latticeL} and \eqref{latticeG} by $U^{-1}$ and $V^{-1}$ in such a way that all the terms in the Hamiltonian still commute with each other.  The Hamiltonian \eqref{checkerboardH} is special as being $S_4$ invariant. }

This is a $\bZ_N$ generalization of the checkerboard model in \cite{Vijay:2016phm}.

Comparing with the $\bZ_N$ FCC lattice gauge theory Hamiltonian \eqref{ZNlatticef}, or its version \eqref{nogauss} where Gauss law is imposed energetically, we  see that this system lacks the first term in these Hamiltonians $-\frac{1}{\hat g_e^2}\sum_{\text{tetrahedra}} V_c$.  In other words, this checkerboard model is the $g_e\rightarrow\infty$ limit of  \eqref{nogauss}.  In fact, as we will see, this model is gapped and has no local operators.  Therefore, we can add to it this term without changing the long-distance behavior, provided its coefficient is small enough.

The system enjoys an interesting self-duality. The map
\ie\label{dualitytran}
&U_s\rightarrow V_s^{-1},\quad V_s\rightarrow U_s\quad \text{ for }\hat{x}+\hat{y}+\hat{z}\in \mathbb{Z}~,
\\
&U_s\rightarrow V_s,\quad V_s\rightarrow U_s^{-1}\quad \text{ for }\hat{x}+\hat{y}+\hat{z}\in \mathbb{Z}+\frac{1}{2}~,
\fe
preserves the commutation relation \eqref{ZNcomm} and maps the Hamiltonian \eqref{checkerboardH} with $(\hat g_m, \hat g)$ to itself with $(\hat g, \hat g_m)$.  The low-energy physics is obtained by taking $\hat g_m, \hat g\to 0$ and it is independent of the values of these coupling constants.  As a result, \eqref{dualitytran} is a  global symmetry of the low-energy physics of this system.
We will refer to it as $\bZ_4^D$.

\subsection{Continuum Lagrangian}

Let us Higgs the $U(1)$ gauge theory to a $\mathbb{Z}_N$ theory using $\phi^i$'s of charge $N$.
As in Section \ref{sec:tetrphitheory} and \ref{sec:tetrgaugetheory}, we  start by choosing $\phi^i$ to be in the $\mathbf{3}$ and $A_\tau^i, A^{ii}, \mathbf A^{ii'}$ to be in the $\mathbf{3}, \mathbf{1}\oplus\mathbf{2}, \mathbf{1}'\oplus\mathbf{2}$, respectively.
The Lagrangian is
\ie\label{eq:ZNLag_Higgs}
\mathcal L_E = - \frac i {4\pi} \sum_{i\neq j\neq k\atop \text{cyclic}} \tilde E^{ii'} \left(\partial_j \partial_k \phi^i - N \mathbf A^{ii'} \right)
 - \frac i {4\pi}\sum_i \tilde {\mathbf E}^{ii} \left( \partial_i \phi^i - N A^{ii} \right)
 + \frac i {4\pi} \sum_i \tilde {\mathbf B}^i \left( \partial_\tau \phi^i - N A_\tau^i \right)~,
\fe
where $\phi^i$ are Higgs fields for the gauge fields $(A_\tau^i,A^{ii},\mathbf A^{ii'})$, and the fields $\tilde E^{ii'}$, $\tilde {\mathbf E}^{ii}$ and $\tilde {\mathbf B}^i$ are Lagrangian multipliers. We rewrite the Lagrangian as
\ie
\mathcal L_E = \frac{iN}{4\pi}
\sum_i \left(\tilde E^{ii'} \mathbf A^{ii'} + \tilde {\mathbf E}^{ii} A^{ii} - \tilde {\mathbf B}^i A_\tau^i \right)
+ \frac i {4\pi}\sum_{i\neq j\neq k\atop \text{cyclic}} \phi^i \left( - \partial_\tau \tilde {\mathbf B}^i + \partial_i \tilde {\mathbf E}^{ii} - \partial_j \partial_k \tilde E^{ii'} \right)~,
\fe
Now, we treat $\phi^i$ as a Lagrangian multiplier which imposes the constraint
\ie
\partial_\tau \tilde {\mathbf B}^x= \partial_x \tilde {\mathbf E}^{xx} - \partial_y \partial_z \tilde E^{xx'}~.
\fe
Locally, we can solve this constraint using gauge fields $(\tilde A_\tau^{i'},\tilde A^{ii'}, \tilde {\mathbf A}^{ii})$:
\ie
&\tilde {\mathbf E}^{xx} = \partial_\tau \tilde {\mathbf A}^{xx} - \partial_y\partial_z \tilde A_\tau^{x'}~,
\\
&\tilde E^{xx'} = \partial_\tau \tilde A^{xx'} - \partial_x \tilde A_\tau^{x'}~,
\\
&\tilde {\mathbf B}^x = \partial_x\tilde {\mathbf A}^{xx} - \partial_y \partial_z \tilde A^{xx'}~.
\fe
We can then write the Lagrangian as
\ie\label{eq:ZNLag_BF_oldbasis}
\mathcal L_E =
\frac{iN}{4\pi} \left[ \sum_i \mathbf A^{ii'} ( \partial_\tau \tilde A^{ii'} - \partial_i \tilde A_\tau^{i'} )
+\sum_{i\neq j\neq k\atop \text{cyclic}}  A^{ii} ( \partial_\tau \tilde {\mathbf A}^{ii} - \partial_j \partial_k \tilde A_\tau^{i'} )
- \sum_{i\neq j\neq k\atop \text{cyclic}}A_\tau^i ( \partial_i \tilde {\mathbf A}^{ii} - \partial_j \partial_k \tilde A^{ii'} ) \right]~.
\fe

To summarize, this $BF$-type Lagrangian for the $\mathbb{Z}_N$ theory involves  two sets of gauge fields in the following $S_4$ representations
\ie\label{eq:ZN_gauge_trans_A}
\mathbf{3}: ~&A_0^x \sim A_0^x + \partial_0 \alpha^x\, ,
\\
\mathbf{1}\oplus\mathbf{2}: ~&A^{xx} \sim A^{xx} + \partial_x \alpha^x\, ,
\\
\mathbf{1}'\oplus\mathbf{2} :  ~&\mathbf A^{xx'} \sim \mathbf A^{xx'} + \partial_y \partial_z \alpha^x\, .
\fe
and
\ie\label{eq:ZN_gauge_trans_tildeA}
\mathbf{3}': ~&\tilde A_0^{x'} \sim\tilde A_0^{x'}+ \partial_0 \tilde\alpha^{x'}\, ,
\\
\mathbf{1}'\oplus\mathbf{2}: ~&\tilde A^{xx'} \sim \tilde A^{xx'} + \partial_x\tilde  \alpha^{x'}\, ,
\\
\mathbf{1}\oplus\mathbf{2} :  ~&\tilde {\mathbf A}^{xx} \sim \tilde {\mathbf A}^{xx} + \partial_y \partial_z\tilde\alpha^{x'}\, .
\fe
 Importantly, the gauge fields $(A_0^i , A^{ii} ,\mathbf A^{ii'})$ with $i=x,y,z$ are not independent gauge fields.
Their gauge parameters are coupled through their periodicities \eqref{alphaperiodicity}.
The same is true for the $\tilde A$ gauge fields.
Therefore, the $\bZ_N$ Lagrangian \eqref{eq:ZNLag_BF_oldbasis} is not a sum of three decoupled Lagrangians.
See  Section \ref{sec:anisotropic} for more details.

\bigskip\bigskip\centerline{\it An alternative presentation}\bigskip

As discussed in Section \ref{sec:Adiscrete}, we can assign the gauge fields to different $S_4$ representations related by outer automorphisms of the global symmetry $\bZ_2^C\times G$.
For example, instead of the choice above, we can assign the two sets of gauge fields to the following $S_4$ representations:
\ie\label{ZNArep}
\mathbf{1}\oplus\mathbf{2}: ~&A_0^{xx} \sim A_0^{xx}+ \partial_0 \alpha^{xx}\, ,
\\
\mathbf{3}: ~&A^{x} \sim A^{x} + \partial_x \alpha^{xx}\, ,
\\
\mathbf{3}' :  ~&\mathbf A^{x'} \sim \mathbf A^{x'} + \partial_y\partial_z \alpha^{xx}\, .
\fe
and
\ie\label{ZNtArep}
\mathbf{1}' \oplus\mathbf{2}: ~&\tilde A_0^{xx'} \sim\tilde A_0^{xx'}+ \partial_0 \tilde\alpha^{xx'}\, ,
\\
\mathbf{3}': ~&\tilde A^{x'} \sim \tilde A^{x'} + \partial_x\tilde  \alpha^{xx'}\, ,
\\
\mathbf{3} :  ~&\tilde {\mathbf A}^{x} \sim \tilde {\mathbf A}^{x} + \partial_y \partial_z \tilde\alpha^{xx'}\, .
\fe
It is important to stress that the fields and the gauge parameters in \eqref{ZNArep} and \eqref{ZNtArep} are the same as in \eqref{eq:ZN_gauge_trans_A} and \eqref{eq:ZN_gauge_trans_tildeA}.  They are not dual to them, and they are not even given by some nontrivial transformations of them.  Instead, they are simply relabeling of the same fields.

The $\bZ_N$ Lagrangian in this presentation is
\ie\label{eq:ZNLag_BF_newbasis}
\mathcal L_E =
\frac{iN}{4\pi} \left[ \sum_i \mathbf A^{i'} ( \partial_\tau \tilde A^{i'} - \partial_i \tilde A_\tau^{ii'} )
+\sum_{i\neq j\neq k\atop \text{cyclic}}  A^{i} ( \partial_\tau \tilde {\mathbf A}^{i} - \partial_j \partial_k \tilde A_\tau^{ii'} )
- \sum_{i\neq j\neq k\atop \text{cyclic}}A_\tau^{ii} ( \partial_i \tilde {\mathbf A}^{i} - \partial_j \partial_k \tilde A^{i'} ) \right]~.
\fe
We will see that this presentation is more natural when we compare the $\bZ_2$  theory with the X-cube models in Section \ref{Xcuberel}.

In this presentation, it will be convenient to shift the dimension 2 gauge fields $\mathbf A^{x'}$ and $\tilde {\mathbf A}^{x}$ by derivatives of the dimension 1 gauge fields $A^i$ and $\tilde A^{i'}$ as follows:
\ie\label{scrA}
&\mathscr{A}^{x'}  =  \mathbf A^{x'}+\partial_y A^{z} +\partial_zA^{y}~, \\
&\mathscr{\tilde A}^{x}  =\tilde  {\mathbf A}^{x}+\partial_y \tilde A^{z'} +\partial_z \tilde A^{y'} ~,\\
\fe
Note that $\mathscr{A}^{x'},\mathscr{A}^{y'},\mathscr{A}^{z'}$ transform  in the $\mathbf{3}'$ of $S_4$.   However, these three components do not form a representation of $G$.
Similarly  $\mathscr{\tilde A}^{x},\mathscr{\tilde A}^{y},\mathscr{\tilde A}^{z}$ transform  in the $\mathbf{3}$ of $S_4$, but they do not form a representation of $G$.

In terms of this basis of gauge fields, the gauge transformations are
\ie
&A_0^{xx} \sim A_0^{xx} +\partial_0 \alpha^{xx}~,\\
&A^x \sim A^x+\partial_x\alpha^{xx}~,\\
&\mathscr{A}^{x'} \sim \mathscr{A}^{x'} +\partial_y\partial_z (\alpha^{xx}+\alpha^{yy}+\alpha^{zz})~,\\
&\tilde A_0^{xx'} \sim \tilde  A_0^{xx'} +\partial_0\tilde \alpha^{xx'}~,\\
&\tilde A^{x'} \sim \tilde A^{x'} +\partial_x \tilde \alpha^{xx'}~,\\
&\mathscr{\tilde A}^{x} \sim \mathscr{\tilde A}^{x} +\partial_y\partial_z (\tilde\alpha^{xx'}+\tilde\alpha^{yy'}+\tilde\alpha^{zz'})~.
\fe
The gauge invariant field strengths are
\ie\label{newfieldstr}
&E^x  =  \partial_0 A^x -\partial_x A_0^{xx}~,\\
&\mathscr{E}^{x'} = \partial_0 \mathscr{A}^{x'}  - \partial_y\partial_z( A_0^{xx} +A_0^{yy}+A_0^{zz})~,\\
&\mathbf B^{xx'}  =\partial_x \mathscr{A}^{x'} -\partial_y \partial_z A^x-\partial_z\partial_x A^y -\partial_x\partial_y A^z~,\\
&\tilde E^{x'}  =  \partial_0 \tilde A^{x'} -\partial_x \tilde A_0^{xx'}~,\\
&\mathscr{\tilde E}^{x} = \partial_0 \mathscr{\tilde A}^{x}  - \partial_y\partial_z(\tilde A_0^{xx'} +\tilde A_0^{yy'}+\tilde A_0^{zz'})~,\\
&\tilde {\mathbf B}^{xx}  =\partial_x \mathscr{\tilde A}^{x} -\partial_y \partial_z\tilde A^{x'}-\partial_z\partial_x \tilde A^{y'} -\partial_x\partial_y \tilde A^{z'}~.
\fe
We can  write the $\bZ_N$ Lagrangian \eqref{eq:ZNLag_BF_newbasis} (in Lorentzian signature) as
\ie\label{eq:ZNLag_BF_newbasis2}
\mathcal{L}&=
\frac{N}{4\pi} \sum_i \left(
-A^{ii}_0 \tilde {\mathbf B}^{ii}
+
\mathscr{ A}^{i'}   \tilde {E}^{i'}
+
 {A}^{i}\mathscr{ \tilde{E}}^{i}\right)~.
\fe
While this expression for the $\bZ_N$ Lagrangian makes only the $S_4\times \bZ_2^C$ symmetry manifest rather than the full $G\times \bZ_2^C$ symmetry, it will be convenient when we compare the checkerboard model with two copies of the X-cube model in Section \ref{sec:2XC}.

Finally, we can exchange the $S_4$ representations between the $A$ and $\tilde A$ gauge fields in each of the above two presentations.

\subsection{Fractons and lineons as defects}

Let us start with gauge-invariant defects at a fixed point in space and extend in time.
Similar to \eqref{Aupperlower}, we define
\ie
&A_0^{XY}  =\frac 12 \left(  A_0^{x} + A_0^{y} -A_0^{z}\right)\,,\quad &&\tilde A_0^{XY}  =\frac 12 \left(\tilde   A_0^{x'} +\tilde  A_0^{y'} -\tilde A_0^{z'}\right)\,,\\
&A_0^{YZ}  =\frac 12 \left(  A_0^{y} + A_0^{z} -A_0^{x}\right)\,,\quad &&\tilde A_0^{YZ}  =\frac 12 \left(\tilde   A_0^{y'} + \tilde A_0^{z'} -\tilde A_0^{x'}\right)\,,\\
&A_0^{ZX}  =\frac 12 \left(  A_0^{z} + A_0^{x} -A_0^{y}\right)\,,\quad &&\tilde A_0^{ZX}  =\frac 12 \left(  \tilde A_0^{z'} + \tilde A_0^{x'} -\tilde A_0^{y'}\right)\,.\\
\fe
In terms of $A^{IJ}_0$, the  defects for $A$ are
\ie\label{fracton}
&f^{XY}\equiv \exp \left[  i \int_{-\infty}^\infty dt~ A_0^{XY} \right]\,,\\
&f^{YZ}\equiv \exp \left[  i \int_{-\infty}^\infty dt~ A_0^{YZ} \right]\,,\\
&f^{ZX}\equiv \exp \left[  i \int_{-\infty}^\infty dt~ A_0^{ZX} \right]\,,
\fe
and three more for $\tilde A$.
These six defects cannot move by themselves -- they are fractons.
Note that there are factors of $1/2$ in the exponents when we express these properly quantized defects in terms  of the $A_0^i$ basis.
We can create two more fractons by considering the combination,
\ie\label{comp_fracton}
f\equiv \exp \left[  i \int_{-\infty}^\infty dt\, \left( A_0^{XY} + A_0^{YZ} + A_0^{ZX} \right) \right],
\fe
and similarly for $\tilde A$.

The composite defects
\ie\label{chline}
&l^x \equiv \exp \left[  i \int_{-\infty}^\infty dt~  (A_0^{XY} +A_0^{ZX}) \right] \,,\\
&l^y\equiv \exp \left[  i \int_{-\infty}^\infty dt~  (A_0^{XY} +A_0^{YZ}) \right] \,,\\
&l^z\equiv \exp \left[  i \int_{-\infty}^\infty dt~  (A_0^{ZX} +A_0^{YZ}) \right] \,,
\fe
by contrast can move in the $x$, $y$, and $z$  direction, respectively.
These  defects are lineons.
The motion of the $x$-lineon is captured by the defects
\ie
&\exp \left[  i \int_{{\cal C}^x \in (t,x)} \left(  dt~  (A_0^{XY} +A_0^{ZX})    +dx~A^{xx}\right) \right] \,,\\
\fe
where ${\cal C}^x$ is a spacetime curve in $(t,x)$, and similarly for the lineons in the other directions.
Similarly there are three lineons for the $\tilde A$ gauge fields.

While a single fracton cannot move, a pair of them separated in the $z$ direction can move collectively in the $x,y$ directions:
\ie\label{fractonplanon}
\exp&\left[ i \int_{z_1}^{z_2} dz \int_{{\cal C} }\Big(    \partial_z A_0^{XY} \, dt
+ \frac12 ( \mathbf A^{yy'}+\partial_z A^{xx} -\partial_x A^{zz})\,dx  +   \frac12 ( \mathbf A^{xx'}+\partial_z A^{yy}-\partial_y A^{zz}) \, dy        \Big)\right] ~,
\fe
where $\mathcal{C}$ is a curve in $(t,x,y)$.
These dipoles of fractons form planons.
The factors of $1/2$ are consistent with the periodicities \eqref{eq:period_phi} of our gauge parameters.
Similarly we have the planons for the $\tilde A$ gauge fields.

The defect is topological in $\cal C$.
When we deform $\cal C$ to ${\cal C}'$, the ratio of the two defects is:
\ie
\exp \Bigg[\frac{i}{2} \int_{z_1}^{z_2}  dz \int_{\cal S}
&\bigg( (\mathbf E^{yy'}+\partial_zE^{xx}-\partial_xE^{zz}) \, dt dx
\\
&+ (\mathbf E^{xx'}+\partial_zE^{yy}-\partial_yE^{zz}) \, dt  dy + (\mathbf B^{x'}-\mathbf B^{y'}) \, dx  dy\bigg)
\Bigg]\,.
\fe
where $\mathcal S$ is the surface bounded by $\mathcal C$ and $\mathcal C'$. Since the field strengths vanish on-shell, this operator is trivial and therefore the defect is topological in $\cal C$.

Let us consider a pair of $x$-lineons separated in the $z$ direction.  They can move collectively on the $xy$-plane:
\ie\label{lineonplanon}
\exp\left[ i
\int_{z_1} ^{z_2} dz  \int _{{\cal C}}
\Big(
\partial_z  ( A_0^{XY}   + A_0^{XZ}  ) \, dt
+\partial_z A^{xx} \, dx
+\mathbf A^{xx'} \, dy
\Big)
\right]~,
\fe
where $\mathcal{C}$ is a curve in $(t,x,y)$.
This dipole of lineons also forms a planon.  Similarly we have the planons for the $\tilde A$ gauge fields.
The defect is topological in $\cal C$.
When we deform $\cal C$ to ${\cal C}'$, the ratio of the two defects is:
\ie
\exp\left[i \int_{z_1}^{z_2}  dz \int_{\cal S}
\left(
\partial_z E^{xx} \, dt dx
+ \mathbf E^{xx'}  \, dt  dy
+ \mathbf B^{x'} \, dx  dy
\right)
\right]\,.
\fe
where $\mathcal S$ is the surface bounded by $\mathcal C$ and $\mathcal C'$. As above, since this depends only on the field strengths, this ratio is trivial on-shell.

\subsection{Global symmetries} \label{sec:ZN_global_sym}

There are two $\mathbb Z_N$ global symmetries.  The first $\mathbb Z_N$ electric symmetry is generated by
\ie
&W^{xx}(y,z)=\exp \left[ i\oint dx~ A^{xx} \right]~,
\\
&W^{xx'}_{y\,-}(x,y_1,y_2)=\exp \left[ \frac{i}{2} \int_{y_1}^{y_2} dy \oint dz~  \left( \mathbf A^{xx'} - \partial_y A^{zz} \right) \right]~,
\fe
and the second $\mathbb Z_N$ electric symmetry is generated by
\ie
&\tilde W^{xx}_{y\,-}(x,y_1,y_2)=\exp \left[ \frac i2 \int_{y_1}^{y_2} dy \oint dz~ \left( \tilde {\mathbf A}^{xx} -\partial_y \tilde A^{zz'} \right) \right]~,
\\
&\tilde W^{xx'}(y,z)=\exp \left[ i\oint dx~ \tilde A^{xx'} \right]~.
\fe
They satisfy the commutation relations
\ie
&W^{xx}(y,z) \tilde W^{xx}_{y\,-}(x,y_1,y_2) = e^{2\pi i/N}\tilde W^{xx}_{y\,-}(x,y_1,y_2) W^{xx}(y,z)~,\quad\text{if }~y_1<y<y_2~,
\\
&W^{xx'}_{y\,-}(x,y_1,y_2) \tilde W^{xx'}(y,z) = e^{2\pi i/N} \tilde W^{xx'}(y,z) W^{xx'}_{y\,-}(x,y_1,y_2)~,\quad\text{if }~y_1<y<y_2~,
\fe
and they commute otherwise. Their $N$-th powers are the identity operator.

Since $\mathbf B^{x'} = 0$, we have
\ie
\partial_y \partial_z \oint dx~ A^{xx} = \oint dx~ \partial_x \mathbf A^{xx'} = 0~,
\fe
and hence, the operator $W^{xx}$ factorizes as
\ie\label{eq:ZN_constraint_1}
W^{xx}(y,z) = W^{xx}_y(y) W^{xx}_z(z)~.
\fe
Only the product of zero modes of $W^{xx}_y$ and $W^{xx}_z$ is physical, so there is a gauge ambiguity
\ie\label{eq:ZN_gauge_1}
W^{xx}_y(y) \sim \eta^x W^{xx}_y(y)~,\qquad W^{xx}_z(z) \sim (\eta^x)^{-1} W^{xx}_z(z)~,
\fe
where $\eta^x$ is a $\mathbb Z_N$ phase. Since $\mathbf {B}^{x'}+ \mathbf {B}^{z'}=0$, we have
\ie
\partial_x \int_{y_1}^{y_2} dy \oint dz~ \left( \mathbf {A}^{xx'} - \partial_y A^{zz} \right) = \int_{y_1}^{y_2} dy \oint dz~ \partial_z\left(  \partial_y  A^{xx} - \mathbf A^{zz'} \right) = 0~,
\fe
hence, the operator $W^{xx'}_{y\,-}$ is independent of $x$. Moreover, it satisfies
\ie\label{eq:ZN_constraint_2}
W^{xx'}_{y\,-}(0,\ell^y) = W^{xx'}_{z\,-}(0,\ell^z)~.
\fe
Similarly, $\tilde W^{xx}_{y\,-}$ is independent of $x$, and satisfies
\ie\label{eq:ZN_constraint_3}
\tilde W^{xx}_{y\,-}(0,\ell^y) = \tilde W^{xx}_{z\,-}(0,\ell^z)~,
\fe
and $\tilde W^{xx'}$ factorizes as
\ie\label{eq:ZN_constraint_4}
\tilde W^{xx'}(y,z) = \tilde W^{xx'}_y(y) \tilde W^{xx'}_z(z)~,
\fe
where $\tilde W^{xx'}_y$ and $\tilde W^{xx'}_z$ have a gauge ambiguity
\ie\label{eq:ZN_gauge_2}
\tilde W^{xx'}_y(y) \sim \eta^{yz} \tilde W^{xx'}_y(y)~,\qquad \tilde W^{xx'}_z(z) \sim (\eta^{yz})^{-1} \tilde W^{xx'}_z(z)~,
\fe
where $\eta^{yz}$ is a $\mathbb Z_N$ phase.

In addition to the discrete exotic global symmetries above, there are also various ordinary discrete global symmetries, including the rotation group $S_4$ and the three $\bZ_2^{(i)}$ symmetries.

Let us start with the action of the 90 degree rotation in the $xy$-plane on the defects:
\ie
&f \rightarrow f^{YZ}~,\quad &&f^{XY} \rightarrow ( f^{ZX})^{-1} \,,\quad &&f^{YZ} \rightarrow (f^{XY})^{-1}~,\quad &&f^{ZX} \rightarrow f~,
\\
&\tilde f \rightarrow (\tilde f^{YZ})^{-1}~,\quad &&\tilde f^{XY} \rightarrow  \tilde f^{ZX} \,,\quad &&\tilde f^{YZ} \rightarrow \tilde f^{XY}~,\quad &&\tilde f^{ZX} \rightarrow \tilde f^{-1}~,
\\
\fe
where the fracton defects $f^{IJ}, f$ are defined in \eqref{fracton} and \eqref{comp_fracton}, and similarly for $\tilde f^{IJ}, \tilde f$.

Next, we consider the action of the $\mathbb Z_2^{(i)}$ symmetry on the defects in the $\bZ_N$ theory.
We will focus on $\bZ_2^{(z)}$, while the discussions for $\bZ_2^{(x)}, \bZ_2^{(y)}$ are similar.
The lineons are invariant under $\bZ_2^{(z)}$, whereas the fractons are permuted as
\ie
\bZ_2^{(z)}:~~&f \rightarrow f^{XY}~,\quad &&f^{XY} \rightarrow  f \,,\quad &&f^{YZ} \rightarrow (f^{ZX})^{-1}~,\quad &&f^{ZX} \rightarrow (f^{YZ})^{-1}~,
\\
&\tilde f \rightarrow \tilde f^{XY}~,\quad &&\tilde f^{XY} \rightarrow  \tilde f \,,\quad &&\tilde f^{YZ} \rightarrow (\tilde f^{ZX})^{-1}~,\quad &&\tilde f^{ZX} \rightarrow( \tilde f^{YZ})^{-1}~.
\fe

The self-duality \eqref{dualitytran} on the lattice becomes  a $\bZ_4^D$ global symmetry in the continuum.
It acts on the gauge fields as
\ie\label{dualityA}
\bZ_4^D:~~&A_0^i \to - \tilde A_0^{i'} \,,~~&&A^{ii}\to -\tilde A^{ii'}\,,~ ~&& \mathbf A^{ii'} \to -\tilde {\mathbf A}^{ii}\,, \\
                  &\tilde A_0^{i'} \to A_0^i \,, ~~&& \tilde A^{ii' }\to  A^{ii}\,,~ ~&&\tilde {\mathbf A}^{ii}\to \mathbf A^{ii'}\,.
\fe
This global symmetry  acts on the defects as
\ie
\bZ_4^D:~~&f \rightarrow \tilde f^{-1}~,~~ &&f^{XY} \rightarrow  (\tilde f^{XY})^{-1}  \,,~~ &&f^{YZ} \rightarrow (\tilde f^{YZ})^{-1}~,~~ &&f^{ZX} \rightarrow (\tilde f^{ZX})^{-1}~,
\\
&\tilde f \rightarrow f~,~~ &&\tilde  f^{XY} \rightarrow  f^{XY}  \,,~~ &&\tilde f^{YZ} \rightarrow f^{YZ}~,~~ &&\tilde f^{ZX} \rightarrow f^{ZX}~.
\fe
This order 4 action squares to the generator of the charge conjugation symmetry $\bZ_2^C$, which is the diagonal $\bZ_2$ subgroup of $\bZ_2^{(x)}\times \bZ_2^{(y)}\times\bZ_2^{(z)}$.
Furthermore, this $\bZ_4^D$ commutes with  all the  three $\bZ_2^{(i)}$ symmetries.

Let us discuss the symmetry group in more detail.  In addition to the time reversal and parity symmetry, the symmetry group of the system is denoted by $S$.  It includes as a subgroup $\mathbb{Z}_2^C\times G \subset S$, which we discussed above.  The action of the duality group $\bZ_4^D$ is added as follows.  We have an exact sequence
\ie\label{dualitys}
1\to \mathbb{Z}_2^C\times G \to S \to \bZ_2^D \to 1~.
\fe
The nontrivial duality transformation \eqref{dualityA} appears in the quotient $ S/(\mathbb{Z}_2^C\times G )$ because its square is the charge conjugation element in $\mathbb{Z}_2^C$.  Focusing on these elements, \eqref{dualitys} is the familiar sequence $1\to \mathbb{Z}_2^C\to \mathbb{Z}_4^D \to \bZ_2^D \to 1$.   Our temporal gauge fields $(A_0^i, \tilde A_0^{i'})$ and our spatial gauge fields $(A^{ii},\tilde A^{ii'})$ and $(\mathbf A^{ii'},\tilde {\mathbf A}^{ii})$ are combined into three six-dimensional irreducible representations of $S$.

The case of $N=2$ is special because the fracton defects are their own inverse.
In particular, this implies that the charge conjugation symmetry $\bZ_2^C$ acts trivially.  This has two important consequences: first, $\bZ_4^D$ does not act faithfully and becomes  $\bZ_2^D$ and second, $\bZ_2^D$ commutes with $G$.   As a result, in this case, the global symmetry that acts faithfully (in addition to time reversal and parity) is $S/\mathbb{Z}_2^C=G\times \bZ_2^D$.

\subsection{Ground state degeneracy}

\bigskip\centerline{\it Higgs presentation}\bigskip

Let us compute the ground state degeneracy on a spatial $3$-torus starting from the Higgs presentation \eqref{eq:ZNLag_Higgs}. The equations of motion state that all the fields can be expressed in terms of $\phi^i$, and the solution space reduces to
\ie
\left\{ \phi^i~\right| \left. \phi^i \sim \phi^i + N \alpha^i  \right\}~.
\fe
All but the winding modes of $\phi^i$ can be gauged away. Among the winding modes, those whose winding charges are multiples of $N$ are gauged away by the winding modes of $\alpha^i$. So only winding modes with winding charges that are nonzero modulo $N$ are physical. On a lattice with $L^i$ sites in the $i$ direction, the winding charges are labelled by $4L^x+4L^y+4L^z-6$ integers. Therefore, there are $N^{4L^x+4L^y+4L^z-6}$ physical winding modes.
This  reproduces the results in \cite{Vijay:2016phm,Shirley:2019xnp} when $N=2$.

\bigskip\bigskip\centerline{\it $BF$-type presentation}\bigskip

We can also compute the ground state degeneracy on a spatial $3$-torus using the $BF$-type presentation \eqref{eq:ZNLag_BF_oldbasis}. In the temporal gauge $A_0^i=0$ and $\tilde A^{i'}_0=0$, the phase space is
\ie
\{ A^{ii},\mathbf A^{ii'},\tilde A^{ii'}, \tilde {\mathbf A}^{ii}~\vert~ \mathbf B^{i'}=0,\tilde {\mathbf B}^{i}=0 \}/\{\text{gauge transformations in \eqref{eq:ZN_gauge_trans_A} and \eqref{eq:ZN_gauge_trans_tildeA}}\}~,
\fe
where we mod out by time-independent gauge transformations. The solutions up to gauge transformation are
\ie
&A^{xx}  =  \frac{1}{\ell^x} f^{xx}_y(y) + \frac{1}{\ell^x} f^{xx}_z(z)\,,
\\
&\mathbf A^{xx'}  = \frac{2}{\ell^z} {\mathbf f}^{xx'}_y(y) + \frac{2}{\ell^y} {\mathbf f}^{xx'}_z(z)\,,
\\
&\tilde A^{xx'}  =  \frac{1}{\ell^x} \tilde f^{xx'}_y(y) + \frac{1}{\ell^x} \tilde f^{xx'}_z(z)\,,
\\
&\tilde {\mathbf A}^{xx}  = \frac{2}{\ell^z} \tilde {\mathbf f}^{xx}_y(y) + \frac{2}{\ell^y} \tilde {\mathbf f}^{xx}_z(z)\,,
\fe
and similarly for the other directions. There is a gauge ambiguity due to zero modes,
\ie\label{eq:ZN_gauge_amb}
&f^{xx}_y(t,y) \sim f^{xx}_y(t,y) + \ell^x c^{xx}(t)\,,\qquad &&f^{xx}_z(t,z) \sim f^{xx}_z(t,z) - \ell^x c^{xx}(t)\,,
\\
&{\mathbf f}^{xx'}_y(t,y) \sim {\mathbf f}^{xx'}_y(t,y) + \ell^z \mathbf c^{xx'}(t)\,,\qquad &&{\mathbf f}^{xx'}_z(t,z) \sim {\mathbf f}^{xx'}_z(t,z) - \ell^y \mathbf c^{xx'}(t)\,,
\\
&\tilde f^{xx'}_y(t,y) \sim \tilde f^{xx'}_y(t,y) + \ell^x \tilde c^{xx'}(t)\,,\qquad &&\tilde f^{xx'}_z(t,z) \sim \tilde f^{xx'}_z(t,z) - \ell^x \tilde c^{xx'}(t)\,,
\\
&\tilde {\mathbf f}^{xx}_y(t,y) \sim \tilde {\mathbf f}^{xx}_y(t,y) + \ell^z \tilde {\mathbf c}^{xx}(t)\,,\qquad &&\tilde {\mathbf f}^{xx}_z(t,z) \sim \tilde {\mathbf f}^{xx}_z(t,z) - \ell^y \tilde {\mathbf c}^{xx}(t)\,,
\fe
and similarly in other directions. Let us define gauge invariant variables
\ie
&\bar {\mathbf f}^{xx'}_y(t,y) = {\mathbf f}^{xx'}_y(t,y) + \frac{1}{\ell^y} \oint dz~ {\mathbf f}^{xx'}_z(t,z)\,,
\\
&\bar {\mathbf f}^{xx'}_z(t,z) = {\mathbf f}^{xx'}_z(t,z) + \frac{1}{\ell^z} \oint dy~ {\mathbf f}^{xx'}_y(t,y)\,.
\\
&\check {\mathbf f}^{xx}_y(t,y) = \tilde {\mathbf f}^{xx}_y(t,y) + \frac{1}{\ell^y} \oint dz~ \tilde {\mathbf f}^{xx}_z(t,z)\,,
\\
&\check {\mathbf f}^{xx}_z(t,z) = \tilde {\mathbf f}^{xx}_z(t,z) + \frac{1}{\ell^z} \oint dy~ \tilde {\mathbf f}^{xx}_y(t,y)\,.
\fe
They satisfy the constraints
\ie\label{eq:ZN_constraints}
&\oint dy~ \bar {\mathbf f}^{xx'}_y(t,y) = \oint dz~ \bar {\mathbf f}^{xx'}_z(t,z)\,,
\\
&\oint dy~ \check {\mathbf f}^{xx}_y(t,y) = \oint dz~ \check {\mathbf f}^{xx}_z(t,z)\,.
\fe

The periodicities of $f^{ii}_j$ and $\bar {\mathbf f}^{ii'}_j$ are discussed in Section \ref{sec:elec_modes}, and the periodicities of $\tilde f^{ii'}_j$ and $\check {\mathbf f}^{ii}_j$ can be obtained similarly.

The effective Lagrangian is
\ie
L_{eff} = \frac{N}{2\pi} \left[ \oint dy~ \left(f^{xx}_y \partial_0 \check {\mathbf f}^{xx}_y + \bar {\mathbf f}^{xx'}_y \partial_0 \tilde f^{xx'}_y\right) +\oint dz~ \left(f^{xx}_z \partial_0 \check {\mathbf f}^{xx}_z + \bar {\mathbf f}^{xx'}_z \partial_0 \tilde f^{xx'}_z\right) \right]~,
\fe
and similarly in other directions. Let us focus on the following part of the Lagrangian
\ie
\frac{N}{2\pi} \oint dy~ \left(f^{zz}_y \partial_0 \check {\mathbf f}^{zz}_y + \bar {\mathbf f}^{xx'}_y \partial_0 \tilde f^{xx'}_y \right)=\frac{N}{2\pi} \oint dy~ \left( F^{zz}_y \partial_0 \check {\mathbf F}^{zz}_y + \bar {\mathbf F}^{xx'}_y \partial_0 \tilde F^{xx'}_y \right)~,
\fe
where we define the new variables
\ie
F^{zz}_y(y) &= \frac12 \left( f^{zz}_y(y) + f^{zz}_y(\ell^y) \right) - \int_y^{\ell^y} dy'~ \bar {\mathbf f}^{xx'}_y(y')~,
\\
\bar {\mathbf F}^{xx'}_y(y) &= \bar {\mathbf f}^{xx'}_y(y) - \frac12 \partial_y f^{zz}_y(y)~,
\\
\tilde F^{xx'}_y(y) &= \frac12 \left( \tilde f^{xx'}_y(y) + \tilde f^{xx'}_y(0) \right)  + \int_0^y dy'~ \check {\mathbf f}^{zz}_y(y')~,
\\
\check {\mathbf F}^{zz}_y(y) &=  \check {\mathbf f}^{zz}_y(y) - \frac12  \partial_y \tilde f^{xx'}_y(y) ~.
\fe
These upper-case variables satisfy the same gauge ambiguities \eqref{eq:ZN_gauge_amb} and constraints \eqref{eq:ZN_constraints} as their lower-case cousins. However, their periodicities are now independent of each other. The effective Lagrangian in terms of these new variables is
\ie
L_{eff} = \frac{N}{2\pi} \left[ \oint dy~ \left(F^{xx}_y \partial_0 \check {\mathbf F}^{xx}_y + \bar {\mathbf F}^{xx'}_y \partial_0 \tilde F^{xx'}_y\right) +\oint dz~ \left(F^{xx}_z \partial_0 \check {\mathbf F}^{xx}_z + \bar {\mathbf F}^{xx'}_z \partial_0 \tilde F^{xx'}_z\right) \right]~,
\fe
and similarly in other directions.

On a lattice with lattice spacing $a$, using the gauge ambiguity of $F^{xx}_i$ and $\tilde F^{xx'}_i$, we can fix $F^{xx}_z(\hat z = L^z) = 0$ and $\tilde F^{xx'}_z(\hat z = L^z) = 0$. The remaining variables have periodicity $F^{xx}_i\sim F^{xx}_i+2\pi$ and $\tilde F^{xx'}_i\sim \tilde F^{xx'}_i +2\pi$. On the other hand, the constraints of $\bar {\mathbf F}^{xx'}_i$ and $\check {\mathbf F}^{xx}_i$ determine $\bar {\mathbf F}^{xx'}_z(\hat z = L^z)$ in terms of other $\bar {\mathbf F}^{xx'}_i$'s, and $\check {\mathbf F}^{xx}_z(\hat z = L^z)$ in terms of the other $\check {\mathbf F}^{xx}_i$'s. The remaining variables have periodicity $\bar {\mathbf F}^{xx'}_i\sim \bar {\mathbf F}^{xx'}_i+\frac{2\pi}{a}$ and $\check {\mathbf F}^{xx}_i \sim \check {\mathbf F}^{xx}_i + \frac{2\pi}{a}$. Hence, there are $L^y+L^z-1$ pairs of variables $(F^{xx}_i , \check {\mathbf F}^{xx}_i)$ and $L^y+L^z-1$ pairs of variables $(\tilde F^{xx'}_i, \bar {\mathbf F}^{xx'}_i)$. Each pair leads to an $N$-dimensional Hilbert space. Combining the modes from the other directions, we find the ground state degeneracy to be $N^{4L^x+4L^y+4L^z-6}$.

\bigskip\bigskip\centerline{\it Ground state degeneracy from global symmetries}\bigskip

The ground state degeneracy can also be understood from the $\mathbb Z_N$ global symmetries. On a lattice with lattice spacing $a$, due to \eqref{eq:ZN_constraint_1} and \eqref{eq:ZN_gauge_1}, there are $L^y + L^z -1$ symmetry operators $W^{xx}_y$ and $W^{xx}_z$ along $x$ direction. Similarly, due to \eqref{eq:ZN_constraint_3}, there are $L^y+L^z-1$ symmetry operators $\tilde W^{xx}_{y\,-}$ and $\tilde W^{xx}_{z\,-}$ along $x$ direction. The commutation relations between these operators are isomorphic to $L^y+L^z-1$ copies of the clock and shift algebra $AB = e^{2\pi i/N}BA$ and $A^N=B^N=1$. The isomorphism is given by
\ie
&A_{\hat y} = W^{xx}(\hat y, 1)~,\quad &&B_{\hat y} = \tilde W^{xx}_{y\,-}(\hat y,\hat y+1)~,\quad &&\hat y = 1,\ldots, L^y~,
\\
&A_{\hat z} = W^{xx}(1,\hat z)W^{xx}(1,1)^{-1}~,\quad &&B_{\hat z} = \tilde W^{xx}_{z\,-}(\hat z,\hat z+1)~,\quad &&\hat z = 2,\ldots,L^z~.
\fe
Similarly, due to \eqref{eq:ZN_constraint_2}, \eqref{eq:ZN_constraint_4} and \eqref{eq:ZN_gauge_2}, we have $L^y+L^z-1$ more copies of the clock and shift algebra in the $x$ direction
\ie
&A_{\hat y} = W^{xx'}_{y\,-}(\hat y,\hat y+1)~,\quad &&B_{\hat y} = \tilde W^{xx'}(\hat y, 1)~,\quad &&\hat y = 1,\ldots, L^y~,
\\
&A_{\hat z} = W^{xx'}_{z\,-}(\hat z,\hat z+1)~,\quad &&B_{\hat z} = \tilde W^{xx'}(1,\hat z)\tilde W^{xx'}(1,1)^{-1}~,\quad &&\hat y = 2,\ldots,L^z~.
\fe
Combining all the directions, we have $4L^x+4L^y+4L^z-6$ copies of the clock and shift algebras, which force the ground state degeneracy to be $N^{4L^x+4L^y+4L^z-6}$.

\section{Relation to some anisotropic theories} \label{sec:anisotropic}

In this section, we relate the anisotropic models discussed in Section \ref{anisI} to the $\phi$-theory, $A$-theory, and $\bZ_N$-theory in Section \ref{sec:tetrphitheory}, Section \ref{sec:tetrgaugetheory}, and Section \ref{ZNgaugetheory}, respectively.

For each of these anisotropic theories, we can take three copies of them, one for the $x$ direction, one for $y$, and one for $z$.
The resulting decoupled theory is then related to their counterparts in the previous sections by coupling to certain $\bZ_2$ gauge fields.
This gives another realization of the $\bZ_N$ checkerboard model of Section  \ref{ZNgaugetheory}.
We will see that this presentation will be especially useful when we compare the $\bZ_N$ checkerboard model with the X-cube model.

\subsection{The $\phi$-theories and the $U(1)$ gauge theories}\label{sec:anisophiA2}

We now take three copies of anisotropic $\phi$-theories \eqref{eq:aniso_phi_periodI}, one for the $x$ direction, one for $y$ and one for $z$. We will denote these fields by $\phi^x,\phi^y,\phi^z$.
This decoupled theory has the same Lagrangian as the $\phi$-theory \eqref{eq:Lag_generic} (ignoring the $K_1,K_2$ terms), but the fields have different periodicities.
More specifically, the fields here have  independent $2\pi$ periodicities \eqref{anisoperiodicityI}, while those in Section \ref{sec:tetrphitheory} have correlated periodicities as in  \eqref{eq:period_phi}.

We now discuss two ways to relate one model to another by gauging a certain global symmetry.

In the first option, we gauge the $\mathbb{Z}_2$ momentum symmetry that acts on the fields as
\ie
&\phi^x(t,x,y,z)\sim \phi^x(t,x,y,z)+\pi n^{zx}(y)+\pi n^{xy}(z)~,
\\
&\phi^y(t,x,y,z)\sim \phi^y(t,x,y,z)+\pi n^{xy}(z)+\pi n^{yz}(x)~,
\\
&\phi^z(t,x,y,z)\sim \phi^z(t,x,y,z)+\pi n^{yz}(x)+\pi n^{zx}(y)~,
\fe
where $n^{xy}(z),n^{yz}(x),n^{zx}(y)\in\mathbb{Z}$ are integer-valued functions.
This is a $(\mathbf{2},\mathbf{3}')$ tensor symmetry of \cite{paper2,paper3,Gorantla:2020xap}.
The gauging of this momentum symmetry introduces more identifications in the space of $\phi^i$.
As a result, $\phi^x,\phi^y,\phi^z$ have correlated periodicities as in \eqref{eq:period_phi} with $w^i(x^i),m^i(x^i)\in\frac{1}{2}\mathbb{Z}$.
We therefore arrive at the Lagrangian \eqref{eq:Lag_generic} by identifying $\bar\mu _0 =4 \mu_0,~\bar \mu = 4\mu ,\bar\mu_1 =4\mu_1$.

In the second option, we gauge the diagonal $\mathbb{Z}_2$  winding symmetry, which is a $(\mathbf{3}',\mathbf{1})$ dipole symmetry in \cite{paper2,paper3,Gorantla:2020xap}.
This amounts to removing some identifications of $\phi^i$.
Explicitly, this gauging can be implemented by adding the term
\ie
&{1\over 2\pi} \hat C \left[  \partial_0 (\phi^x+\phi^y+\phi^z)  + 2\partial_0 \phi^{XYZ} \right]+{1\over2\pi}\sum_{i<j} \hat C^{ij}_0 \left[ \partial_i\partial_j (\phi^x+\phi^y+\phi^z)  +2  \partial_i\partial_j\phi^{XYZ} \right]~,
\fe
where  $(\hat C^{ij}_0,\hat C)$ are Lagrange multipliers in the $(\mathbf{3}',\mathbf{1})$ of $S_4$, which are the gauge fields discussed in \cite{Gorantla:2020xap}.  The field $\phi^{XYZ}$ is a scalar with $2\pi$ periodicity:
\ie\label{eq:phiXYZ_period}
&\phi^{XYZ}(t,x,y,z)\sim \phi^{XYZ}(t,x,y,z) +2\pi n^x(x)+2\pi n^y(y)+2\pi n^z(z)~,\\
&n^x(x),n^y(y),n^z(z)\in\mathbb{Z}\,.
\fe
Indeed, this is the same as the periodicity of $\phi^{XYZ}$ in Section \ref{sec:XYtetra}.
Integrating out $(\hat C^{ij}_0,\hat C)$ constrains $\phi^x,\phi^y,\phi^z$ to have the same correlated periodicities as in \eqref{eq:period_phi} with $w^i(x^i),m^i(x^i)\in\mathbb{Z}$.
We therefore arrive at the Lagrangian \eqref{eq:Lag_generic} by identifying $\bar\mu _0 = \mu_0,~\bar \mu = \mu ,\bar\mu_1 =\mu_1$.

Similarly, we  take three copies of the anisotropic $A$-theories \eqref{anisoAI}, one for $x$, one for $y$ and one for $z$.
We will denote these fields by $ (A_0^{ii} , A^{i},\mathbf A^{i'})$ and their gauge parameters by $\alpha^{ii}$.
This decoupled theory has a $G\times \bZ_2^C$ symmetry, and we assign the $S_4$ representations for these fields as in \eqref{ZNArep}.
See more discussions in Section \ref{sec:Adiscrete}.

This decoupled theory  is similar to \eqref{eq:A_Lag} (ignoring the $\kappa_1,\kappa_2$ terms), but the gauge parameters in the two theories have different periodicities.
More specifically, the gauge parameters  here have  independent periodicities, while those in Section \ref{sec:tetrgaugetheory} have correlated periodicities \eqref{alphaperiodicity}.

Again,  we can map one model to another by coupling to a $\bZ_2$ gauge field in two different ways.

In the first option, we gauge the diagonal $\mathbb{Z}_2$ electric symmetry, which is a $(\mathbf{3}',\mathbf{1})$ dipole global symmetry of \cite{paper2,paper3,Gorantla:2020xap}.
This is dual to gauging the $\bZ_2$ winding symmetry for the decoupled $\phi^i$ theory above.
The gauging of this electric symmetry adds more identifications to the space of the gauge parameters.
As a result, the gauge parameters are the same as \eqref{eq:period_phi} with $w^i(x^i),m^i(x^i)\in\frac{1}{2}\mathbb{Z}$.
We therefore arrive at the Lagrangian \eqref{eq:Lag_generic} by identifying $(\bar g _{e1})^2 =( g_{e1})^2/4,~(\bar g_{e2})^2 = (g_{e2})^2/4,(\bar g_m)^2  =(g_m)^2/4$.

In the second option, we gauge the $\mathbb{Z}_2$ magnetic symmetry, which is a $(\mathbf{ 2},\mathbf{3}')$  tensor symmetry  of \cite{paper2,paper3,Gorantla:2020xap}.
Specifically, we would like to impose a constraint such  that the gauge parameters for the combinations $A_0^{xx}+A_0^{yy} +A_0^{zz}$ and  $ \mathscr{A}^{i'}$ have  $4\pi$ as opposed to $2\pi$ periodicities.
Here $\mathscr{A}^{x'}= \mathbf A^{x'} +\partial_y A^z +\partial_z A^y$ is defined in \eqref{scrA} and its  field strengths    are defined in \eqref{newfieldstr}.
That is,  we would like to impose
\ie\label{A2a}
&A_0^{xx}+A_0^{yy} +A_0^{zz} = 2a_0 \,,\\
&\mathscr{A}^{x'}  = 2a_{yz}  ,~~~\mathscr{A}^{y'} =2a_{zx},~~~\mathscr{A}^{z'}= 2a_{xy}\,,
\fe
for some gauge fields $(a_0, a_{ij})$ with $2\pi $ periodicities in the $(\mathbf{1},\mathbf{3}')$ of $S_4$.
This constraint is implemented by adding the terms:\footnote{Let us draw an analogy with a similar operation on an ordinary  $U(1)$ gauge theory in $d$ spacetime dimensions. (See \cite{Seiberg:2010qd}, for closely related constructions.) Let $A^{(1)}$ be  an ordinary  one-form gauge field with gauge transformation $A^{(1)} \sim A^{(1)}+d\alpha^{(0)}$, where $\alpha^{(0)} \sim \alpha^{(0)}+2\pi$.   Gauging a $\bZ_2$ electric 1-form symmetry adds more identifications and reduces the periodicity of $\alpha^{(0)}$ by a factor of 2, i.e., $\alpha^{(0)} \sim \alpha^{(0)}+\pi$. On the other hand, gauging a $\bZ_2$ magnetic $(d-3)$-form symmetry removes some identifications and increases the periodicity of $\alpha^{(0)}$ by a factor of 2, i.e., $\alpha^{(0)}\sim \alpha^{(0)}+4\pi$. This can be implemented by adding the term
\ie
{1\over 2\pi } \hat a^{(d-2)}\wedge  ( dA^{(1)} - 2da^{(1)})
\fe
to the Lagrangian.  Here $\hat a^{(d-2)}$ is a Lagrange multiplier enforcing the constraint $dA^{(1)} = 2da^{(1)}$ for some one-form gauge field $a^{(1)}$. }
\ie\label{eq:A_condition}
&  {1\over 2\pi} \hat a^{[xy]z} _0 \left[  (\mathbf B^{xx'} - \mathbf B^{yy'}) - 2(\partial_x a_{yz} - \partial_y a_{xz}) \right]
-{1\over 2\pi} \hat a_{yz} \left[\mathscr{E}^{x'}- 2(\partial_0 a_{yz}-\partial_y\partial_z a_0)\right]+\cdots
\fe
to the Lagrangian. Here the ellipsis represents similar terms associated with the other directions.
The    Lagrange multipliers $(\hat a^{[ij]k} _0 ,\hat a_{ij})$ are in the $(\mathbf{ 2},\mathbf{3}')$ of $S_4$.
Note that $a$ and $\hat a$ are the gauge fields participating in the continuum field theory for the X-cube model \cite{paper3}, which we  review in Appendix \ref{sec:X-cube}.

This is dual to gauging the $\bZ_2$ momentum symmetry for the decoupled $\phi^i$ theory above.

The gauging of the magnetic symmetry removes some identifications in the space of the gauge parameters $\alpha^i$.
As a result, the gauge parameters have correlated periodicities, as in \eqref{eq:period_phi} with $w^i(x^i),m^i(x^i)\in\mathbb{Z}$.
We therefore arrive at the Lagrangian \eqref{eq:Lag_generic} by identifying $(\bar g _{e1})^2 =( g_{e1})^2,~(\bar g_{e2})^2 = (g_{e2})^2,(\bar g_m)^2  =(g_m)^2$.

\subsection{The checkerboard model and the anisotropic model}\label{sec:relation}

We start with three copies of anisotropic $\mathbb{Z}_N$ theories \eqref{anisoZNI}, one for the $x$ direction, one for $y$ and one for $z$.
We will denote these gauge fields by $( A_0^{ii} , A^i , \mathbf A^{i'})$ and $(\tilde A_0^{ii'} , \tilde A^{i'} ,\tilde {\mathbf A}^i)$, and their gauge parameters by $\alpha^{ii}$ and $\tilde \alpha^{ii'}$, in the same notations as \eqref{ZNArep} and \eqref{ZNtArep}.
The Lagrangian is
\ie\label{decoupledZN}
\frac{N}{2\pi} \sum_i \left(-
A^{ii}_0 \tilde {\mathbf B}^{ii}
+
\mathscr{A}^{i'}   \tilde {E}^{i'}
+
 {A}^{i}\mathscr{ \tilde{E}}^{i}\right)~.
\fe
Up to an overall factor of two, this sum of decoupled theories has the same Lagrangian  as   the $\bZ_N$ checkerboard model \eqref{eq:ZNLag_BF_newbasis2}.
However, they are different in their global properties.
More specifically, the gauge parameters   in this decoupled theories have independent periodicities, while those  in \eqref{eq:ZNLag_BF_newbasis2} have correlated periodicities \eqref{alphaperiodicity}.

Similar to Section \ref{sec:anisophiA2}, we can map one model to the other by  coupling to certain $\bZ_2$ gauge fields.

One approach is to start with the decoupled theories with even $N  $ and gauge the $\bZ_2$ subgroup of the two electric $\bZ_N$ global symmetries.
After gauging, the gauge parameters have the periodicities in \eqref{alphaperiodicity} with $w^i(x^i),m^i(x^i)\in\frac{1}{2}\mathbb{Z}$.
To normalize the gauge fields  and the gauge parameters in the same way as in Section \ref{ZNgaugetheory}, we then rescale all of our gauge fields as $A\to A/2$ and $\tilde A\to \tilde A/2$. We therefore end up with the $\bZ_{N/2}$ checkerboard model  in \eqref{eq:ZNLag_BF_newbasis}.

\begin{table}[h!]
\begin{align*}
\left.\begin{array}{|c|c|c|}
\hline
&&\\
 ~~\text{original theory: 3 anisotropic } ~~& \text{couple $A$ to} &~~ \text{gauge $\bZ_2$ electric} ~~\\
~~\text{$\bZ_N$ gauge theories} ~~&~~ \text{$\bZ_2$ gauge fields \eqref{eq:A_condition}} ~~&\text{symmetry for $A$}\\
&&\\
 \hline &&\\
 \text{couple $\tilde A$ to}  & N\in \mathbb{Z} & N\in2 \mathbb{Z} \\
\text{$\bZ_2$ gauge fields \eqref{eq:A_condition}} & \bZ_{2N} ~\text{checkerboard}&  \bZ_{N} ~\text{checkerboard}\\
&&\\
 \hline &&\\
 \text{gauge $\bZ_2$ electric} & N\in2 \mathbb{Z} & N\in 2\bZ\\
\text{symmetry for $\tilde A$} & \bZ_{N} ~\text{checkerboard} & \bZ_{N\over 2} ~\text{checkerboard}\\
&&\\
 \hline \end{array}\right.
\end{align*}
\caption{Relations between three copies of the $\bZ_N$ anisotropic models and the   checkerboard model.  Starting from the former, we can couple a $\bZ_2$ gauge fields to each of $A$ and $\tilde A$ in two different ways as discussed in Section \ref{sec:anisophiA2}.  One of them is gauging a $\bZ_2$ electric symmetry, while the other is given in \eqref{eq:A_condition}. This gives four different ways to obtain the checkerboard model. In some of the options, the integer $N$ of the anisotropic model has to be even as indicated above.}\label{tbl:4options}
\end{table}

Alternatively, we can start with the decoupled $\bZ_N$ theories with even $N$, and gauge the electric symmetry for $\tilde A$ and couple $A$ to  $\bZ_2$ gauge fields as in \eqref{eq:A_condition}.  In this way we end up with  a $\bZ_N$ checkerboard model.  Similarly, we can exchange the role of $A$ and $\tilde A$.

Another option is to start with $N\in \bZ$ and couple the three anisotropic $\bZ_N$ theories to gauge fields as in \eqref{eq:A_condition} for both $A$ and $\tilde A$:
\ie\label{eq:checkerboard_anisotropic}
&\frac{N}{2\pi} \sum_i \left(-
A^{ii}_0 \tilde {\mathbf B}^{ii}
+
\mathscr{ A}^{i'}   \tilde {E}^{i'}
+
 {A}^{i}\mathscr{ \tilde{E}}^{i}\right)\\
 & +{1\over 2\pi} \hat a^{[xy]z} _0 \left[  (\mathbf B^{xx'} - \mathbf B^{yy'}) - 2(\partial_x a_{yz} - \partial_y a_{xz}) \right]
-{1\over 2\pi} \hat a^{yz} \left[\mathscr{E}^{x'}- 2(\partial_0 a_{yz}-\partial_y\partial_z a_0)\right]+\cdots\\
  &-  {1\over 2\pi} \hat b^{z(xy)} _0 \left[  (\tilde {\mathbf B}^{xx} - \tilde {\mathbf B}^{yy}) - 2 (\partial_x b_{x} -\partial_y b_{y} )\right]
  + {1\over 2\pi} \hat b^x \left[\tilde{\mathscr{E}}^{x}- 2 (\partial_0 b_{x} -\partial_y \partial_z b_0^{(xyz)} )\right]+\cdots
\fe
where we have introduced the gauge fields $b,\hat b$ whose $S_4$ representations differ from $a,\hat a$ by tensoring with $\mathbf{1}'$.
(See Section \ref{Xcuberel} for more details.)
The ellipses  stand for the similar terms associated with the other directions in space.
These couplings to the $\bZ_2$ gauge fields $a$ and $b$ in the second and the third lines above make the gauge parameters have the same  periodicities as in \eqref{alphaperiodicity} with $w^i(x^i),m^i(x^i)\in\mathbb{Z}$.
Comparing with \eqref{eq:ZNLag_BF_newbasis2}, we therefore arrive at the $\bZ_{2N}$ checkerboard model.

We summarize these four different ways of obtaining the checkerboard model from three decoupled anisotropic models in Table \ref{tbl:4options}.

\section{Relation to the $\mathbb Z_2$ X-cube model}\label{Xcuberel}

\subsection{Relation to two copies of the $\bZ_2$  X-cube model}\label{sec:2XC}

We now show that  the $\bZ_2$ checkerboard model is equivalent to two copies of the $\bZ_2$ X-cube models in the continuum.

 Many of  the expressions below  contain several equations related by cyclically permuting $x,y,z$.
 In some of these expressions, we will write only one of the  equations to avoid cluttering.
In some other expressions, the  ellipses  stand for   the terms associated with the other directions.

We will use the presentation of the $\bZ_2$ checkerboard model in \eqref{eq:checkerboard_anisotropic} with $N=2$.
In this presentation, the gauge fields have independent periodicities.
It can be rewritten as
\ie\label{2LXC}
& {\cal L}_{XC} (N=2; \hat a_0 ^{i(jk)} , \hat a^{ij} ,a_0 ,a_{ij})+ {\cal L}_{XC} (N=2; \hat b_0 ^{[jk]i} , \hat b^{k} ,b^{(xyz)}_0 ,b_{k})\\
&+\frac{1}{2\pi} \left[ -
(A^{xx}_0 + \hat b_0^{[yz]x} )( \tilde {\mathbf B}^{xx}-\partial_x \partial_y  \hat a^{xy} -\partial_x \partial_z \hat a^{zx}  -\partial_y\partial_z \hat a^{yz})\right.\\
&\qquad \quad  \left.
+
\mathscr{ A}^{x'}   (   \tilde {E}^{x'}  + \partial_0 \hat a^{yz}  -\partial_x \hat a_0^{x(yz)} )
+
 ( {A}^{x} + \hat b^{x}) \mathscr{ \tilde{E}}^{x}\right] +\cdots
 \fe
 where ${\cal L}_{XC}$, defined in \eqref{LXC}, is the Lagrangian of the X-cube model \cite{Slagle:2017wrc,paper3}.
Since all the gauge fields have  independent periodicities, we can perform the following field redefinition to shift the gauge fields $A,\mathscr{A}, \tilde A, \mathscr{\tilde A}$ by $a,\hat a, b, \hat b$ as
\ie
&A_0^{xx} \to A_0^{xx} - \hat b_0^{[yz]x} \,,~~~A^x \to A^x - \hat b^{x } \,,~~~\mathscr{A}^{x'}\to \mathscr{A}^{x'}\,,\\
&\tilde A_0^{xx'} \to\tilde A_0^{xx'} - \hat a_0^{x(yz)} \,,~~~\tilde A^{x'} \to\tilde A^{x'} -\hat a^{yz} \,,~~~\mathscr{\tilde A}^{x}\to \mathscr{\tilde A}^{x}\,,\\
\fe
and similarly for the fields associated with the other directions.
Note that these field redefinitions are $S_4$ covariant.
In terms of the shifted gauge fields, the Lagrangian becomes
\ie
& {\cal L}_{XC} (N=2; \hat a_0 ^{i(jk)} , \hat a^{ij} ,a_0 ,a_{ij})+ {\cal L}_{XC} (N=2; \hat b_0 ^{[jk]i} , \hat b^{k} ,b^{(xyz)}_0 ,b_{k})\\
&+ \frac{1}{2\pi} \left[-
A^{xx}_0 \tilde {\mathbf B}^{xx}+
\mathscr{ A}^{x'}      \tilde {E}^{x'}
+
  {A}^{x} \mathscr{ \tilde{E}}^{x}\right] +\cdots
 \fe
 Again, all the fields have  independent periodicities. The second line is just three decoupled trivial anisotropic theories \eqref{anisoZNI}.
 We have thus arrived at the Lagrangian of two $\bZ_2$ X-cube models.\footnote{Let us draw an analogy with an equivalence between two ordinary TQFTs in 2+1 dimensions.	  Consider the Lagrangian
 	\ie\label{L2}
 	{\cal L} = {2\over 4\pi } \tilde A^1 dA^1 +{2\over 4\pi } \tilde A^2 dA^2~.
 	\fe
 	where the gauge parameters of these gauge fields have correlated periodicities $\alpha^I \sim\alpha^I+2\pi n^I$ with $n^I\in \bZ$ and $n^1+n^2\in 2\bZ$, and similarly for $\tilde\alpha^I$. This TQFT ${\cal L}$ is the analog of the $\bZ_2$ checkerboard model, where the gauge parameters have similar correlated periodicities  \eqref{alphaperiodicity}.
 	
 	Similar to \eqref{eq:checkerboard_anisotropic}, the theory can be rewritten as
	 	\ie
 	{\cal L} =  {1\over 2\pi } \tilde A^1 dA^1 +{1\over 2\pi } \tilde A^2 dA^2
 	- {1\over 2\pi} \hat a \left[  d(A^1+ A^2 ) -2da  \right] - {1\over 2\pi} \hat b \left[  d(\tilde A^1+\tilde A^2 ) -2db  \right]\,.
 	\fe
 	Unlike in \eqref{L2}, the gauge parameters of all these one-form gauge fields now have independent $2\pi$ periodicities. After some field redefinitions, the Lagrangian can rewritten as two copies of $\bZ_2$ gauge theories.   	
 }

The duality $\bZ_2^D$ of the checkerboard model becomes the $\bZ_2$ symmetry that exchanges the two copies of the X-cube model.

Note that this derivation of the equivalence between the $\bZ_2$ checkerboard model and two $\bZ_2$ X-cube models does not extend to higher $N$.
In fact, in Section \ref{sec:unequal}, we will show that the two models are inequivalent for $N>2$.
For even $N>2$, the presentation \eqref{eq:checkerboard_anisotropic} says that the $\bZ_{N}$ checkerboard model is equivalent to three $\bZ_{N/2}$ anisotropic models coupled to two $\bZ_2$ X-cube models. When $N=2$, this reduces to the above equivalence.

\bigskip\bigskip\centerline{\it Map between the fractons and the lineons}\bigskip

We have just established the off-shell equivalence between the $\bZ_2$ checkerboard model and two copies of the X-cube model.   Even though it follows from this off-shell equivalence, it is nice to check explicitly how the defects are mapped between these models.  In this map we can use the equations of motion to simplify the presentation of the defects.

Similar to \eqref{Aupperlower}, we first define
\ie
&A_0^{XY}  =\frac 12 \left(  A_0^{xx} + A_0^{yy} -A_0^{zz}\right)\,,~~&&\tilde A_0^{XY}  =\frac 12 \left(\tilde   A_0^{xx'} +\tilde  A_0^{yy'} -\tilde A_0^{zz'}\right)\,,\\
&A_0^{YZ}  =\frac 12 \left(  A_0^{yy} + A_0^{zz} -A_0^{xx}\right)\,,~~&&\tilde A_0^{YZ}  =\frac 12 \left(\tilde   A_0^{yy'} + \tilde A_0^{zz'} -\tilde A_0^{xx'}\right)\,,\\
&A_0^{ZX}  =\frac 12 \left(  A_0^{zz} + A_0^{xx} -A_0^{yy}\right)\,,~~&&\tilde A_0^{ZX}  =\frac 12 \left(  \tilde A_0^{zz'} + \tilde A_0^{xx'} -\tilde A_0^{yy'}\right)\,.\\
\fe
This basis for the temporal gauge fields will be more natural for the  defects.

The map for the fractons is
\ie
 f  \equiv\exp \left[  i \int_{-\infty}^\infty dt~ (A_0^{XY} + A_0^{YZ} + A_0^{ZX}) \right] &\longleftrightarrow~ \exp \left[  i \int_{-\infty}^\infty dt~ a_0 \right]~,
\\
 f^{XY}  \equiv\exp \left[  i \int_{-\infty}^\infty dt~ A_0^{XY} \right]&\longleftrightarrow~ \exp \left[  i \int_{-\infty}^\infty dt~ (a_0 + \hat b_0^{[xy]z}) \right]~,
\\
\tilde f\equiv \exp \left[  i \int_{-\infty}^\infty dt~ (\tilde A_0^{XY} + \tilde A_0^{YZ} + \tilde A_0^{ZX}) \right]  &\longleftrightarrow~ \exp \left[  i \int_{-\infty}^\infty dt~ b_0^{(xyz)} \right]~,
\\
 \tilde f^{XY}\equiv\exp \left[  i \int_{-\infty}^\infty dt~ \tilde A_0^{XY} \right]  &\longleftrightarrow~ \exp \left[  i \int_{-\infty}^\infty dt~ (b_0^{(xyz)} + \hat a_0^{z(xy)}) \right]~.
\fe
The maps for the other fractons $f^{YZ},f^{ZX},\tilde f^{YZ},\tilde f^{ZX}$ are obtained by cyclic permutations.
The map for the lineons is
\ie\label{lineonmap}
 l^x  \equiv\exp \left[  i \int_{-\infty}^\infty dt~  (A_0^{XY} + A_0^{ZX}) \right] &\longleftrightarrow~ \exp \left[  i \int_{-\infty}^\infty dt~  \hat b_0^{[yz]x} \right]~,
\\
\tilde l^x \equiv \exp \left[  i \int_{-\infty}^\infty dt~  (\tilde A_0^{XY} + \tilde A_0^{ZX}) \right]  &\longleftrightarrow~ \exp \left[  i \int_{-\infty}^\infty dt~  \hat a_0^{x(yz)} \right]~.
\fe
The maps for the other lineons $l^y,l^z,\tilde l^y,\tilde l^z$ are obtained by cyclic permutations.

As a consistency check,  we show that the defects from the two theories obey the same fusion relations.
The fractons satisfy
\ie
f^{XY} f^{YZ} f^{ZX} = f~,\qquad \tilde f^{XY} \tilde f^{YZ} \tilde f^{ZX} = \tilde f~.
\fe
The lineons satisfy
\ie
l^x l^y l^z = 1~,\qquad \tilde l^x \tilde l^y \tilde l^z = 1~.
\fe
Finally, we have
\ie
l^x= f^{YZ} f~,\qquad \tilde l^x = \tilde f^{YZ} \tilde f
\fe

\subsection{$\bZ_N$ checkerboard model with $N>2$}\label{sec:unequal}

In the previous section, we have shown that the $\bZ_2$ checkerboard model  is equivalent to two copies of the $\bZ_2$ X-cube models in the continuum.
We now show that their $\bZ_N$ versions are \textit{not} equivalent for $N>2$ by showing that   the fusion rules are different in these two theories.

Let us consider the lineons in the two theories.
In two copies of the $\bZ_N$ X-cube models, we have the following lineons:
\ie
&l_{(a)}^i \equiv  \exp \left[i  \int_{-\infty}^\infty dt~ \hat a_0^{ i(jk)}\right]\,,\\
&l_{(b)}^i\equiv \exp\left[i  \int_{-\infty}^\infty dt~ \hat b_0^{ [jk]i}\right]\,.
\fe
They obey
\ie\label{3lineon}
&l_{(a)}^x l_{(a)}^yl_{(a)}^z=1\,,\\
&l_{(b)}^x l_{(b)}^yl_{(b)}^z=1\,.
\fe
as well as $(l_{(a)}^i)^N  =(l_{(b)}^i)^N  =1$.

On the other hand, the lineons of the $\bZ_N$ checkerboard model are (see \eqref{chline})
\ie\label{chlinec}
&l^x \equiv \exp \left[  i \int_{-\infty}^\infty dt~  (A_0^{XY} +A_0^{ZX}) \right] \,,\\
&l^y\equiv \exp \left[  i \int_{-\infty}^\infty dt~  (A_0^{XY} +A_0^{YZ}) \right] \,,\\
&l^z\equiv \exp \left[  i \int_{-\infty}^\infty dt~  (A_0^{ZX} +A_0^{YZ}) \right] \,.
\fe

Let us try to match the lineons in the two theories.  Any map between them should be of the form
\ie\label{proposemap}
l^i _{(a)} =  (l^i )^{n_i} (\tilde l^i )^{\tilde n_i}
\fe
with some integers $n_i, \tilde n_i$.  (There is a similar map for $l^i_{(b)}$, but we will not need it.)

Using $\exp\left[iN  \int_{-\infty}^\infty dt~  A_0^{IJ} \right]=\exp\left[iN  \int_{-\infty}^\infty  dt~\tilde A_0^{IJ} \right] =1$, the relations  \eqref{3lineon}, \eqref{chlinec}, and \eqref{proposemap} are consistent only if
\ie
n_i + n_j =0~{\rm mod}~ N\,,~~~~\tilde n_i +\tilde n_j =0~{\rm mod} ~N~,
\fe
for $i\neq j$.  This implies that
\ie
2n_i =2\tilde n_i = 0~{\rm mod}~ N\,.
\fe
Since $(l^i)^N=(\tilde l^i)^N=1$, \eqref{proposemap} leads to $(l^i_{(a)})^2=1$, which is false unless $N=2$.
Hence the $\bZ_N$ checkerboard model is inequivalent to two copies of the $\bZ_N$ X-cube models when $N>2$.

\section*{Acknowledgements}

We thank F.~Burnell, M.~Cheng, T.~Devakul, M.~Hermele, K.~Slagle and Y.~You for helpful discussions. PG was supported by Physics Department of Princeton University.  HTL was supported
by a Centennial Fellowship and a
Charlotte Elizabeth Procter Fellowship from Princeton University and Physics Department of Princeton University. The work of NS was supported in part by DOE grant DE$-$SC0009988.  NS and SHS were also supported by the Simons Collaboration on Ultra-Quantum Matter, which is a grant from the Simons Foundation (651440, NS).
SHS  thanks the Department  of Physics at National Taiwan University for its hospitality while this work was being completed.
Opinions and conclusions expressed here are those of the authors and do not necessarily reflect the views of funding agencies.

\appendix

\section{Some useful group theory}\label{usefulgroup}

\subsection{Review of the representation theory of the cubic group $S_4$}\label{sec:app_reps_S4}

The symmetry group of the cubic lattice (modulo translations) is the \textit{cubic group}, which consists of 48 elements.
We will focus on the group of orientation-preserving symmetries of the cube, which is isomorphic to the permutation group of four objects $S_4$.

The irreducible representations of $S_4$ are the trivial representation $\mathbf{1}$, the sign representation $\mathbf{1}'$, a two-dimensional irreducible representation $\mathbf{2}$, the standard representation $\mathbf{3}$, and another three-dimensional irreducible representation $\mathbf{3}'$.
The representation $\mathbf{3}'$ is the tensor product of the sign representation and the standard representation, $\mathbf{3}' = \mathbf{1}' \otimes \mathbf{3}$.

It is convenient to embed $S_4\subset SO(3) $ and decompose the $SO(3)$ irreducible representations in terms of $S_4$ representations.  The first few are
\ie\label{rep}
&SO(3)~~  &\supset ~~&S_4~\\
&~~~\mathbf{1} &=~~&~\mathbf{1}~\\
&~~~\mathbf{3} &=~~&~\mathbf{3}~\\
&~~~\mathbf{5} &=~~&~\mathbf{2}\oplus \mathbf{3}' ~\\
&~~~\mathbf{7} &=~~&~\mathbf{1}'\oplus \mathbf{3} \oplus \mathbf{3}' ~\\
&~~~\mathbf{9} &=~~&~\mathbf{1}\oplus \mathbf{2} \oplus \mathbf{3} \oplus \mathbf{3}'~
\fe

We will label the components of $S_4$ representations using $SO(3)$ vector indices as follows.
The three-dimensional standard representation $\mathbf{3}$ of $S_4$  carries an $SO(3)$ vector index $i$, or equivalently, an antisymmetric pair of indices $[jk]$.\footnote{We will adopt the convention that  indices in the square brackets are antisymmetrized, whereas indices in the parentheses are symmetrized. For example, $A_{[ij]} = -A_{[ji]}$ and $A_{(ij)} = A_{(ji)}$.}
Similarly, the irreducible representations of $S_4$ can be expressed in terms of the following tensors:
\ie\label{eq:SFindices}
&~\mathbf{1} &:~~&~S & &\\
&~\mathbf{1}'&:~~&~T^{(xyz)}&\,\\
&~\mathbf{2} &:~~&~B_{[ij]k}&\,,~~~i\neq j\neq k& \quad ,  ~B_{[ij]k}+B_{[jk]i}+B_{[ki]j}=0\\
&&&~B_{i(jk)}&\,,~~~i\neq j\neq k&\quad ,~B_{i(jk)}+B_{j(ki)} +B_{k(ij)}=0\\
&~\mathbf{3} &:~~&~ V^i& & \\
&~\mathbf{3}'&:~~&~E_{ij}&\,,~~~i\neq j ~~~~~~& \quad,\  E_{ij}=E_{ji}\\
&&&~E^{k'}&
\fe
In the above we have two different expressions, $B_{[ij]k}$ and $B_{i(jk)}$,  for the irreducible representation  $\mathbf{2}$ of $S_4$.
In the first expression, $B_{[ij]k}$ is the component of $\mathbf{2}$ in the tensor product $\mathbf{3}\otimes \mathbf{3}  =\mathbf{1}\oplus \mathbf{2}\oplus\mathbf{3}\oplus\mathbf{3'}$.
In the second expression, $B_{i(jk)}$ is the component of $\mathbf{2}$ in the tensor product $\mathbf{3}\otimes \mathbf{3'}  =\mathbf{1'}\oplus \mathbf{2}\oplus\mathbf{3}\oplus\mathbf{3'}$. We also consider fields in the  reducible representations $\mathbf{1}\oplus \mathbf{2}$ and $\mathbf{1}'\oplus \mathbf{2}$ of $S_4$, for which we use the following tensors:
\ie
\mathbf{1}\oplus \mathbf{2} ~:& ~ \phi^{ii}\\
\mathbf{1}'\oplus \mathbf{2} ~:& ~ \phi^{ii'}
\fe
where there is no constraint on the sum of the three components.
Note that in the embedding of $S_4\subset SO(3)$, the reducible representation $\mathbf{1}\oplus \mathbf{2}$  is embedded into the diagonal components of a symmetric tensor in the $\mathbf{1}\oplus\mathbf{5}$ of $SO(3)$, and hence the notation $\phi^{ii}$.

In most of this paper, the indices $i,j,k$ in every expression are not equal, $i\neq j\neq k$ (see \eqref{eq:SFindices} for example).
We will often use $x,y,z$ both as coordinates and as the indices of a tensor.

\subsection{Review of the symmetry of the FCC lattice}\label{app:fcc}

We will discuss the orientation-preserving symmetry of the FCC lattice.
Let the shortest distance between  a pair of sites in the FCC lattice be $a/\sqrt{2}$.

The FCC lattice has translation symmetry generated by $a/2$ in two of the three directions:
\ie
&T^{xy} : ~(x,y,z) \mapsto (x+\frac a2 , y+\frac a2 , z)\,,\\
&T^{zx} : ~(x,y,z) \mapsto (x+\frac a2 , y , z+\frac a2)\,,\\
&T^{yz} : ~(x,y,z) \mapsto (x , y+\frac a2 , z+\frac a2)\,.
\fe
An interesting subgroup of these translations is generated by
\ie\label{cubictranslation}
&T^{x} : ~(x,y,z) \mapsto (x+a , y , z)\,,\\
&T^{y} : ~(x,y,z) \mapsto (x , y+a , z)\,,\\
&T^{z} : ~(x,y,z) \mapsto (x , y, z+a)\,.
\fe

Next, we discuss the rotation symmetry on the lattice.
We will label a counterclockwise 90 degree space rotation $R^k _{(x,y,z)}$ by the direction $k$ of its rotation axis and a point $(x,y,z)$ on the rotational axis.
The FCC lattice has the following 90 degree  rotation symmetry:
\ie\label{R}
R^x_{(0,0,0)} \,, ~~R^y_{(0,0,0)}\,,~~R^z_{(0,0,0)} \,.
\fe

The FCC lattice has other symmetries  that can be generated by $T^{ij}$ and $R^k_{(0,0,0)}$.
These include a 90 degree rotation around the face center
\ie
&R^x _{(0,\frac a2, \frac a2 )} =  T^{yz}R^x_{(0,0,0)} (T^{yz})^{-1}\,,\\
&R^y _{(\frac a2,0, \frac a2 )} =  T^{zx}R^y_{(0,0,0)} (T^{zx})^{-1}\,,\\
&R^z _{(\frac a2, \frac a2,0 )} =  T^{xy}R^z_{(0,0,0)} (T^{xy})^{-1}\,.
\fe
and the following composition of a translation and a rotation:
\ie\label{r}
&r^x=  (T^x)^{1\over2} R^x_{(0, \frac a4, \frac a4)}   = T^{xy} R^x_{(0,0,0)}\,,\\
&r^y=  (T^y)^{1\over2} R^y_{(\frac a4,0, \frac a4)}   = T^{yz} R^y_{(0,0,0)}\,,\\
&r^z=  (T^z)^{1\over2} R^z_{(\frac a4, \frac a4,0)}   = T^{zx} R^z_{(0,0,0)}\,,\\
\fe
where $(T^z)^{1\over2}:  (x,y,z)\mapsto (x,y,z+\frac a2)$.
Note that neither $(T^z)^{1\over2}$ nor $R^z_{( \frac a4, \frac a4, 0)} $  is a symmetry of the FCC lattice by itself, but their composition is.
Similarly, $(T^z)^{1\over2} (R^z_{(\frac a4, \frac a4,0)})^{-1}   = T^{yz} (R^z_{(0,0,0)})^{-1}$.

We conclude that the orientation-preserving symmetry of the FCC lattice is generated by the three translations $T^{xy},T^{yz},T^{zx}$ and the 90 degree rotations $R^k_{(0,0,0)}$.
These generate the space group F432.

Let $ G$ be the quotient  of the space group  F432 by the translations $T^x,T^y,T^z$.  It is
\ie
G= (\bZ_2\times \bZ_2) \rtimes S_4\,.
\fe
Note that $G$ is different from the point group of F432, which is known to be $S_4$. The latter is the quotient group of F432 by the translations $T^{xy},T^{yz},T^{zx}$, or equivalently,
\ie
S_4= {G\over \bZ_2\times \bZ_2}~.
\fe
We will discuss more about $G$ as an abstract finite group in Appendix \ref{app:G}.

Let $\bar T^{ij}$ and $\bar R^k_{(0,0,0)}$ be the group elements of $G$ corresponding to $ T^{ij}$ and $ R^k_{(0,0,0)}$ in F432. Note that $\bar R^k_{(0,0,0)} = \bar R^k _{(a\hat x,a\hat y, a\hat z)}$ for any $\hat x,\hat y,\hat z\in \bZ$.
Below we work out the action of these symmetries on the fields living on the sites.
For example, these could be the $\phi^{IJ}$ fields in Section \ref{sec:XYtetra}.
 $\bar T^{zx}$ acts as
\ie
\bar T^{zx} :~&\phi^{XYZ} \rightarrow \phi^{ZX} \,,\\
&\phi^{XY}\rightarrow \phi^{YZ}\,,\\
&\phi^{ZX}\rightarrow \phi^{XYZ}\,,\\
&\phi^{YZ}\rightarrow \phi^{XY} \,,
\fe
$\bar R^z_{(0,0,0)}$ acts as
\ie\label{barR}
\bar R^z_{(0,0,0)} :~&\phi^{XYZ} \rightarrow \phi ^{XYZ}\,,\\
&\phi^{XY}\rightarrow \phi^{XY}\,,\\
&\phi^{ZX}\rightarrow \phi^{YZ}\,,\\
&\phi^{YZ}\rightarrow \phi^{ZX} \,,\\
\fe
The composition $\bar r ^z\equiv \bar T^{zx} \bar R^z_{(0,0,0)}$ acts as
\ie\label{barr}
\bar r^z \equiv \bar T^{zx} \bar R^z_{(0,0,0)} :~&\phi^{XYZ} \rightarrow \phi^{ZX} \,,\\
&\phi^{XY}\rightarrow \phi^{YZ}\,,\\
&\phi^{ZX}\rightarrow \phi^{XY}\,,\\
&\phi^{YZ}\rightarrow \phi ^{XYZ}\,,
\fe

The four fields $\phi^{XYZ},\phi^{XY},\phi^{ZX},\phi^{YZ}$ form a 4-dimensional reducible representation of $G$.
The sum $\phi^{XYZ}+ \phi^{XY} +\phi^{ZX} +\phi^{YZ}$ is in a trivial representation of $G$.  Therefore,
the subspace
\ie
\phi^{XYZ}+ \phi^{XY} +\phi^{ZX} +\phi^{YZ}=0\,
\fe
preserves the $G$ symmetry.
We can then solve $\phi^{XYZ}$ in terms of
the remaining three $\phi^{IJ}$, which form an irreducible representation of $G$.
Let us define
\ie
&\phi^x = \phi^{XY}+\phi^{ZX}\,,\\
&\phi^y = \phi^{XY}+\phi^{YZ}\,,\\
&\phi^z = \phi^{YZ}+\phi^{ZX}\,.
\fe

The group $G$ has several $S_4$ subgroups.
Let $S_4^{(1)}$ be the $S_4$ subgroup of $G$ generated by $\bar r^x ,\bar r^y,\bar r^z$ from \eqref{r}.
For example, $\bar r^z$  acts as
\ie
\bar r^z:~~&\phi^x(t,x,y,z) \to \phi^y(t,y,-x,z)\,,\\
&\phi^y(t,x,y,z) \to - \phi^x(t,y,-x,z)\,,\\
&\phi^z(t,x,y,z) \to -\phi^z(t,y,-x,z)\,.
\fe
Therefore $\phi^i$ is in the $\mathbf{3}'$ of $S_4^{(1)}$, and hence, $\partial_i \phi^i$ (no sum over $i$) and $\partial_i \partial_j\phi^k$ (with $i\ne j\ne k$)  are in the $\mathbf{1}'\oplus\mathbf{2}$ and   $\mathbf{1}\oplus\mathbf{2}$, respectively.

Alternatively, consider a different $S_4$ subgroup of $G$ generated by $\bar R^x_{(0,0,0)} ,\bar R^y_{(0,0,0)},\bar R^z_{(0,0,0)}$.
We will denote this subgroup as $S_4^{(2)}$.
For example, $\bar R^z$ acts on $\phi^i$ as
\ie
\bar R^z_{(0,0,0)}:~~&\phi^x(t,x,y,z) \to \phi^y(t,y,-x,z)\,,\\
&\phi^y(t,x,y,z) \to \phi^x(t,y,-x,z)\,,\\
&\phi^z(t,x,y,z) \to \phi^z(t,y,-x,z)\,.
\fe
Therefore $\phi^i$ is in the $\mathbf{1}\oplus \mathbf{2}$ of $S_4^{(2)}$,
and hence, $\partial_i \phi^i$ (no sum over $i$) and $\partial_i \partial_j\phi^k$ (with $i\ne j\ne k$) are in the $\mathbf{3}$ and   $\mathbf{3}'$, respectively.

In addition to $G$, our system also has a charge conjugation internal $\bZ_2^C$ symmetry generated by  $\phi^i\to -\phi^i$ for all $i$.  Unlike the other symmetries discussed here, which are the symmetries of the FCC lattice, $\bZ_2^C$ commutes with the symmetries of the lattice. Just as $G$ has several different $S_4$ subgroups, $G\times \bZ_2^C$ leads to more options.
We can compose the order 4 elements of $S_4^{(1)}$ with the generator of $\bZ_2^C$  to turn $\phi^i$ into $\mathbf{3}$.  Then, $\partial_i \phi^i$ (no sum over $i$) and $\partial_i \partial_j\phi^k$ (with $i\ne j\ne k$) are in the $\mathbf{1}\oplus\mathbf{2}$ and   $\mathbf{1}'\oplus\mathbf{2}$, respectively.
We will denote this new $S_4$  subgroup as $S_4^{(1)C}$.

Alternatively, we can compose the order 4 elements of $S_4^{(2)}$ with the generator of $\bZ_2^C$  to turn $\phi^i$ into  $\mathbf{1}'\oplus \mathbf{2}$, and hence, $\partial_i \phi^i$ (no sum over $i$) and $\partial_i \partial_j\phi^k$ (with $i\ne j\ne k$) are in the $\mathbf{3}'$ and   $\mathbf{3}$, respectively.
We will denote this new $S_4$ subgroup as $S_4^{(2)C}$.

\subsubsection{More details about  $G=(\mathbb{Z}_2\times \mathbb{Z}_2)\rtimes S_4$}\label{app:G}

Above we have identified the relevant finite group symmetry $G$ for our models and how it is realized on the FCC lattice.
Here we provide more details on the representation theory of $G$ as an abstract finite group.

The orientation-preserving symmetry group of the FCC lattice modulo translations of the minimal cubic sublattice \eqref{cubictranslation} is
\begin{equation}
G=(\mathbb{Z}_2\times \mathbb{Z}_2)\rtimes S_4~,
\end{equation}
which consists of 96 elements. The group $G$ can be generated by 4 generators $\{s,t,a,b\}$
\begin{equation}
G=\{s,t,a,b|s^2=t^3=(st)^4=1, sas=a, sbs=ab, tat^{-1}=b, tbt^{-1}=ab\}~.
\end{equation}
In terms of their actions on the FCC lattice, $s$ is a 180 degree rotation around the $(110)$ axis, $t$ is a 120 degree rotation around the $(111)$ axis, $a$ is a translation along $xy$ plane, $b$ is the translation along $yz$ plane.

The group $G$ has 12 $S_4$ subgroups. They are divided into three conjugacy classes of subgroups. Each of these classes have four $S_4$ subgroups that are related by internal automorphisms. $G$ has an $S_3$ outer automorphism that permutes the 3 conjugacy classes of $S_4$ subgroups.

The group $G$ has 10 irreducible representations: the trivial representation $\mathbf{1}$, another one-dimensional representation $\mathbf{1}'$, a two-dimensional representation $\mathbf{2}$, six three-dimensional representation $\mathbf{3}_i$, $\mathbf{3}_i'$ with $i=1,2,3$ and a six-dimensional representation $\mathbf{6}$. The representation $\mathbf{3}_i'$ is the tensor product $\mathbf{3}_i' = \mathbf{1}' \otimes \mathbf{3}_i$. The $S_3$ outer automorphism permutes the representation $\mathbf{3}_i$ and $\mathbf{3}_i'$ of different subscripts.

The character table of the group $G$ is \cite{GAP4}
\begin{center}
\begin{tabular}{|c|c|c|c|c|c|c|c|c|c|c|c|c|c|}
\hline
representative & $\{1\}$  & $\{stst\}$ 
  & $\{st\}$ & $\{s\}$ & $\{t\}$  & $\{a\}$ 
  &  $\{sb\}$  & $\{ststb\}$ & $\{sta\}$   & $\{ststa\}$ \\
\hline
order & 1 & 2 & 4 & 2 & 3 & 2 & 4 & 2 & 4  & 2 \\
\hline
size & 1 & 3 & 12 & 12 & 32 & 3 & 12 & 3 & 12  & 6 \\
\hline
$\mathbf{1}$ &  1 & 1 &1 &1 &1 & 1 & 1 &1 &1 &1   \\
\hline
$\mathbf{1}'$ & 1 & 1 &-1 &-1 &1 & 1 & -1& 1 & -1  &1 \\
\hline
$\mathbf{2}$ & 2 & 2 &0 &0 &-1 & 2 & 0 & 2 & 0  & 2\\
\hline
$\mathbf{3}_1$ & 3 & -1 &1 &-1 &0 &3  &-1 & -1 & 1  &-1 \\
\hline
$\mathbf{3}'_1$ & 3 & -1 &-1 &1 &0 &3 &1 & -1 & -1  &-1 \\
\hline
$\mathbf{3}_2$ & 3 & -1 &1 &-1 &0 &-1 &1 & 3 &-1  &-1 \\
\hline
$\mathbf{3}'_2$ & 3 & -1 &-1 &1 &0 & -1 &-1 & 3 & 1&-1 \\
\hline
$\mathbf{3}_3$ & 3 & 3 &-1 &-1 &0 &-1 &1 &-1 & 1  &-1 \\
\hline
$\mathbf{3}'_3$ & 3 & 3 &1 &1 &0 &-1 &-1&-1&-1&-1 \\
\hline
$\mathbf{6}$ & 6 & -2 & 0 & 0 & 0 &-2 &0 &-2&0&2 \\
\hline
\end{tabular}
\end{center}

We can choose one $S_4\subset G$ and then the representations of $G$ can be decomposed into irreducible representations of the $S_4$ subgroup as
\begin{equation}
\begin{aligned}
&\mathbf{1}=\mathbf{1},\quad \mathbf{1}'=\mathbf{1}',\quad \mathbf{2}=\mathbf{2},\quad \mathbf{6}=\mathbf{3}\oplus\mathbf{3}',
\\
& \mathbf{3}_1=\mathbf{3},\quad \mathbf{3}_1'=\mathbf{3}',\quad \mathbf{3}_2=\mathbf{3},\quad \mathbf{3}_2'=\mathbf{3}',\quad\mathbf{3}_3=\mathbf{1}'\oplus\mathbf{2},\quad\mathbf{3}'_3=\mathbf{1}\oplus\mathbf{2}~.
\end{aligned}
\end{equation}
With a different $S_4\subset G$ subgroup that is related by an outer-automorphism, the representations of $G$ decompose as
\begin{equation}
\begin{aligned}
&\mathbf{1}=\mathbf{1},\quad \mathbf{1}'=\mathbf{1}',\quad \mathbf{2}=\mathbf{2},\quad \mathbf{6}=\mathbf{3}\oplus\mathbf{3}',
\\
& \mathbf{3}_1=\mathbf{3},\quad \mathbf{3}_1'=\mathbf{3}',\quad \mathbf{3}_2=\mathbf{1}'\oplus\mathbf{2},\quad \mathbf{3}_2'=\mathbf{1}\oplus\mathbf{2},\quad\mathbf{3}_3=\mathbf{3},\quad\mathbf{3}'_3=\mathbf{3}'~.
\end{aligned}
\end{equation}

The following are some useful nontrivial decompositions of tensor products of irreducible representations of $G$
\begin{center}
\begin{tabular}{|c|c|c|c|c|c|}
\hline
$\otimes$ & $\mathbf{1}'$ & $\mathbf{2}$ & $\mathbf{6}$ & $\mathbf{3}_i$ \\
\hline
$\mathbf{1}'$ & $\mathbf{1}$ & $\mathbf{2}$ & $\mathbf{6}$  & $\mathbf{3}_i'$\\
\hline
$\mathbf{2}$ &  & $\mathbf{1} \oplus \mathbf{1}' \oplus \mathbf{2}$ & $\mathbf{6} \oplus \mathbf{6}$ & $\mathbf{3}_i \oplus \mathbf{3}_i'$ \\
\hline
$\mathbf{3}_i$ & & & $\mathbf{3}_j \oplus \mathbf{3}_j' \oplus \mathbf{3}_k \oplus \mathbf{3}_k' \oplus \mathbf{6}$ & $\mathbf{1} \oplus \mathbf{2} \oplus \mathbf{3}_i \oplus \mathbf{3}_i'$ \\
\hline
$\mathbf{3}_j$ & & & & $\mathbf{3}_k' \oplus \mathbf{6}$ \\
\hline
\end{tabular}
\end{center}
where $i\ne j\ne k$. Finally, we have
\ie
\mathbf{6} \otimes \mathbf{6} = \mathbf{1} \oplus \mathbf{1}' \oplus (2\times \mathbf{2}) \oplus \mathbf{3}_1 \oplus \mathbf{3}_1' \oplus \mathbf{3}_2 \oplus \mathbf{3}_2' \oplus \mathbf{3}_3 \oplus \mathbf{3}_3' \oplus (2\times \mathbf{6})~.
\fe

\section{Review of the continuum field theory of the X-cube model} \label{sec:X-cube}

Here we review the continuum field theory description of the X-cube model in \cite{Slagle:2017wrc,paper3}.\footnote{Other related presentations of this continuum field theory are discussed in \cite{Slagle:2018swq,Slagle:2020ugk,layers}.}

The continuum theory of (3+1)d $\mathbb Z_N$ X-cube model consists of two gauge fields
\ie\label{a1}
\mathbf{1}&:~ a_0\sim a_0+\partial_0\gamma~,\\
\mathbf{3}'&:~ a_{ij}\sim a_{ij}+\partial_i\partial_j\gamma~,
\fe
and
\ie\label{a2}
\mathbf{2}&:~\hat{a}_0^{i(jk)}\sim \hat{a}_0^{i(jk)}+\partial_0\hat{\gamma}^{i(jk)}~,\\
\mathbf{3}'&:~\hat{a}_{ij}\sim \hat{a}_{ij}+\partial_k\hat{\gamma}^{k(ij)}~.
\fe
The Lagrangian of this model is
\ie\label{LXC}
&\mathcal L_{XC}(N;\hat a_0^{i(jk)} , \hat a^{ij},  a_0 , a_{ij})\\
&  = {N\over 2\pi} \left[
\hat a_0^{[xy]z}  ( \partial_x a_{yz} - \partial_y a_{zx} )
-\hat a^{xy} (\partial_0 a_{xy} - \partial_x\partial_y a_0 )
+\text{cyclic permutations}
\right]\,.
\fe

We can relabel our gauge fields using the representation of a different $S_4$ subgroup, which is related to the original $S_4$ subgroup by an outer automorphism of the global symmetry group $\mathbb Z_2^C \times S_4$.
Explicitly, we rename the fields as $a_0\to b_0^{(xyz)}, a_{ij} \to b_{k}, \hat a_0^{i(jk)} \to \hat b_0^{[jk]i} , \hat a^{ij} \to \hat b^{k}$, where here and below the indices $i,j,k$ are always cyclically ordered.
The fields and gauge parameters in this new presentation are in the following $S_4$ representations:
\ie\label{b1}
\mathbf{1}'&:~ b_0^{(xyz)}\sim b_0^{(xyz)}+\partial_0\beta^{(xyz)}~,\\
\mathbf{3}&:~ b_{k}\sim b_{k}+\partial_i\partial_j\beta^{(xyz)}~,
\fe
and
\ie\label{b2}
\mathbf{2}&:~\hat{b}_0^{[ij]k}\sim \hat{b}_0^{[ij]k}+\partial_0\hat{\beta}^{[ij]k}~,\\
\mathbf{3}&:~\hat{b}^{k}\sim \hat{b}^{k}+\partial_k\hat{\beta}^{[ij]k}~.
\fe
The two presentations differ by tensoring with the $\mathbf{1}'$ representation of $S_4$.
We also define $\hat a_0^{[ij]k}  \equiv \frac 13 (\hat a_0^{i(jk)} - \hat a_0^{j(ki)} )$ and $\hat b_0^{k(ij)}  \equiv \frac 13 (\hat b_0^{[jk]i} - \hat b_0^{[ki]j} )$.

It is important to stress that the fields and the gauge parameters in \eqref{a1}, \eqref{a2} and \eqref{b1}, \eqref{b2}   are not dual to each other, and they are not even related by some nontrivial transformations.  Instead, they are simply relabeling of the same fields.

\bibliographystyle{JHEP}
\bibliography{checkerboard}

\end{document}